\documentclass[acmsmall]{acmart}

\AtBeginDocument{%
  }

\setcopyright{acmlicensed}
\copyrightyear{2018}
\acmYear{2018}
\acmDOI{XXXXXXX.XXXXXXX}

\acmJournal{JACM}
\acmVolume{37}
\acmNumber{4}
\acmArticle{111}
\acmMonth{8}

\usepackage{multirow}
\usepackage{caption}
\usepackage{subcaption}
\usepackage{tcolorbox}
\usepackage{enumitem}
\usepackage{amsmath}

\newcommand\scalemath[2]{\scalebox{#1}{\mbox{\ensuremath{\displaystyle #2}}}}

\def\figref#1{Figure~\ref{fig:#1}}

\def\tabref#1{Table~\ref{tab:#1}}

\def\tabref#1{Table~\ref{tab:#1}}

\newcommand{\revision}[1]{\textcolor{black}{ #1}}

\begin{document}

\title{Unveiling Code Pre-Trained Models: Investigating Syntax and Semantics Capacities}

\author{Wei Ma}
\affiliation{
	\institution{Nanyang Technological University}
	\country{Singapore}
}
\email{ma_wei@ntu.edu.sg}

\author{Shangqing Liu}
\authornote{Corresponding author.}
\affiliation{
	\institution{Nanyang Technological University}
	\country{Singapore}
}
\email{liu.shangqing@ntu.edu.sg}

\author{Mengjie Zhao}
\affiliation{
	\institution{Ludwig Maximilian University of Munich}
 \city{Munich}
	\country{Germany}
}
\email{mzhaolmu@gmail.com}

\author{Xiaofei Xie}
\affiliation{
	\institution{Singapore Management University}
	\country{Singapore}
}
\email{xfxie@smu.edu.sg}

\author{Wenhan Wang}
\affiliation{
	\institution{University of Alberta}
 \city{Edmonton}
	\country{Canada}
}
\email{wenhan12@ualberta.ca}

\author{Qiang Hu}
\affiliation{
	\institution{The University of Tokyo}
        \city{Tokyo}
	\country{Japan}
}
\email{qianghu0515@gmail.com}

\author{Jie Zhang}
\affiliation{
	\institution{Noah’s Ark Lab, Huawei}
 \city{Xi'An}
	\country{China}
}
\email{clark.zhang@huawei.com}

\author{Yang Liu}
\affiliation{
	\institution{Nanyang Technological University}
	\country{Singapore}
}
\email{yangliu@ntu.edu.sg}

\renewcommand{\shortauthors}{W. Ma, S. Liu and M. Zhao et al.}

\begin{abstract}
Significant progress has been made in code intelligence through code models that embed knowledge about programming languages. Past research has examined how well these models grasp code syntax, yet their understanding of code semantics still needs to be explored. Moreover, current analyses typically link the number of edges in an abstract syntax tree (AST) with syntax distance. They also frequently necessitate reducing the high-dimensional space of deep learning models to a lower dimension, potentially leading to inaccuracies. We extensively analyze seven code models to investigate how code models represent code syntax and semantics. This includes four prominent code pre-trained models (CodeBERT, GraphCodeBERT, CodeT5, and UnixCoder) and three large language models (StarCoder, CodeLlama, and CodeT5+). We have developed four probing tasks to evaluate the models' abilities to learn code syntax and semantics. These tasks focus on reconstructing code syntax and semantic structures—such as Abstract Syntax Trees (AST), Control Flow Graphs (CFG), Control Dependency Graphs (CDG), and Data Dependency Graphs (DDG)—within the models' representation spaces. These structures are fundamental to understanding code.
Additionally, we explore the role of syntax tokens in each token representation and the extended dependencies among code tokens. Furthermore, we examine the distribution of attention weights concerning code semantic structures. Through detailed analysis, our results emphasize the strengths and weaknesses of various code models in mastering code syntax and semantics. The findings reveal that these models are proficient in grasping code syntax, effectively capturing the relationships and roles of syntax tokens. However, their ability to encode code semantics shows more variability. CodeT5 and CodeBERT excel at capturing control and data dependencies, whereas UnixCoder performs less effectively. We also find that large language models (LLMs) do not generally outperform pre-trained models significantly.
Interestingly, the shallower layers of LLMs demonstrate better performance compared to their deeper layers. Our analysis of attention weights indicates that different attention heads are specialized for distinct roles in encoding code semantics. Our research underscores the necessity for further improvements in code models to enhance their ability to learn code semantics effectively. This study enriches our understanding of the capabilities of code models in analyzing syntax and semantics. Our findings offer valuable insights for future code model enhancements, helping optimize their application across a range of code-related tasks.
\end{abstract}

\maketitle

\section{Introduction}
Many code models~\cite{kanade2019pre, feng2020codebert, guo2020graphcodebert, lu2021codexglue, svyatkovskiy2020intellicode,buratti2020exploring, karampatsis2020scelmo, wang2021codet5, ahmad2021unified, liu2023contrabert} have been proposed to greatly advance the development of code intelligence. \revision{Many software engineering approaches are based on fine-tuning these pre-trained models, such as code clone detection, vulnerability detection and code completion. Recently, large language models~(LLM) have proven to have emergent abilities~\cite{wei2022emergent} that pre-trained models do not possess. This technology makes it possible to generate software automatically, such as MetaGPT~\cite{hong2023metagpt}. }
Despite the fact that these models have been proven to be efficient on various code-related tasks, a fundamental issue still remains unresolved for these code models \revision{ about how they understand code}. \revision{Significantly, recent works~\cite{10.1609/aaai.v37i12.26739,10.1145/3428230,10.1145/3591227} indicate that code models cannot give reasonable results if token replacement or insertion tricks are used. Regarding the code models, we should consider deeply their ability to learn the basic characteristics of the code, ``\textit{What kind of code knowledge can these code models learn?}''. }
A program consists of syntax features (e.g., AST) and semantic information (e.g., data dependency); as a result, this issue can be further decomposed into ``\textit{Can code models capture program syntax well?}'' and ``\textit{What kind of program semantics can code models learn?}''. \revision{Prior studies have begun to explore the questions raised, especially the first sub-question. However, a deeper understanding of the knowledge gained by code models remains elusive. While research by \citet{10.1145/3510003.3510050} and \citet{lopez2022ast} shows that code pre-trained models can grasp program syntax, their analysis does not extend to program semantics. It is also important to note that these studies have two assumptions:1) the number of links between nodes is related to code syntax; 2) a linear relationship exists between the representation of code in high-dimensional and low-dimensional spaces. However, these assumptions have some limitations: 1) the number of links is not necessary to be related to code syntax. 2) the smaller distance in the raw code text does not mean the syntax closeness in the representation space. 
Furthermore, \citet{troshin2022probing} examine both code syntax and semantics through various tasks. However, this work follows a similar approach to the earlier studies when analyzing syntax. It also falls short of thoroughly investigating the various semantics inherent in programming. More importantly, there are no studies on the transformer decoder, which is the architecture of most LLMs, such as the StarCoder~\cite{li2023starcoder} and Llama~\cite{roziere2023code} families.}

\revision{These questions are significant for us to review the applications of code models in software engineering. If the code models can understand the code almost perfectly, the outputs of the code models can be trusted, and automated software generation is possible. } To address the aforementioned challenges, in this paper, we comprehensively investigate four widely adopted code pre-trained models: CodeBERT~\cite{feng2020codebert} (Encoder-only), GraphCodeBERT~\cite{guo2020graphcodebert} (Encoder-only), CodeT5~\cite{wang2021codet5} (Encoder-decoder) and UnixCoder~\cite{guo2022unixcoder} (UniLM-style). \revision{We also include three large language models (LLMs), StarCoder~\cite{li2023starcoder}, CodeLlama~\cite{roziere2023code} and CodeT5+~\cite{wang2023codet5+}.}

Four probing tasks are \revision{employed} to analyze the capabilities of the models in learning code syntax and semantics. Specifically, we utilize two syntax probing tasks, namely \textit{syntax node pair prediction} and \textit{token syntax tagging prediction}. Both tasks aim to manipulate Abstract Syntax Trees (AST) to assess the capabilities of pretrained models in learning code syntax since AST carries all the syntax information of the code. Syntax node pair prediction aims to determine whether vector representations of two syntax-close spans exhibit syntactic similarity while token syntax tagging prediction aims to identify whether the vector representation captures the syntax rule of the individual token. \revision{Syntax node pair prediction is to recover the  AST structure from the vector representation, and token syntax tagging prediction is to assign the syntax role to each code token in the representation space. Both of the syntax information are significant for code models to understand the code syntax at the global and local levels.
The intention behind the two syntax tasks is that the code-token vector representation should keep the syntax properties that exist in the code: the syntax relationship between code tokens, and the syntax property of each code token.}
In addition to syntax analysis, we further design two semantic probing tasks, namely \textit{semantic relation prediction} and \textit{semantic propagation prediction}. Both tasks are designed to investigate the extent to which code models can effectively learn various aspects of code semantics. \revision{Semantic relation prediction is to recover the significant code semantic structures in the representation space, including Data Dependency Graph~(DDG), Control Dependency Graph~(CDG) and Control Flow Graph~(CFG). These semantic structures can represent code execution and states inside. Semantic propagation prediction is to see if we can observe the long dependency relationship in the vector space because a variable can be declared in the first statement but is used in the end.} 
Lastly, we performed the statistical analysis for the attention weights to comprehensively understand the role and attention distribution in learning the semantics of the code.

\revision{Our syntax analysis shows that 1). code models effectively capture syntactic relationships between token pairs and this property is easier to observe in the shallow hidden layers, indicating a strong understanding of code syntax at a structural level; 2). code models are proficient at identifying the syntax roles of individual tokens and this property is easier to observe in the deep hidden layers. However, code pre-trained models show superior performance than LLMs for syntax tagging, indicating the syntax characteristics are more difficult to observe in the code representation from LLMs. The ease of being observed is not directly related to the performance of the model on downstream tasks, as it depends on many factors, such as data quality and fine-tuning methods. However, it can indicate the difficulty of how to build and tune a good model for the downstream task based on LLM. In practice, people have tried and even found that the performance of traditional small models is better than that of large models in some cases~\cite{dou2023towards,huging_llm_comparing}. }

\revision{
Our semantic analysis demonstrates varying effectiveness in the ability of code models to predict semantic relationships. The current models provide some nuanced understanding of code semantics, such as data dependencies, but also highlight areas for improvement, particularly in capturing complex semantic structures. Compared with the performance of syntactic tasks, the performance of the code models in semantics is relatively low. For CodeT5+, the performance difference between the encoder and decoder is very large. We think this may be caused by the different working mechanisms of the encoder and decoder. The encoder focuses more on the nesting of global information. The decoder is more focused on generating the next token based on the previous text. This difference is especially obvious when analyzing the attention mechanism. When we conducted attention analysis based on control dependencies, we found that for the decoder architecture, the weight contribution of tokens with control dependencies is smaller than the weight contribution of non-control dependencies. We believe this is because the encoder focuses on encoding global information, but the decoder focuses more on the previous tokens and other non-control-dependent information, resulting in different weight distributions.
}

\revision{Through extensive analysis, our work indicates that code models still need to be improved to learn code syntax and semantics, making these properties more observable, so that reducing the difficulty of building a downstream task model. The improvement should focus on how to integrate code syntax and semantics into code models. This may need novel training strategies that can integrate the whole code structure instead of flattening these structures. Although the model ability is scaled with its size~\cite{52065,kaplan2020scaling}, it is challenging to release its ability which needs a non-trivial extra work to make the features more observable in the representation space for the downstream tasks. In summary, our work has the following contributions:
\begin{itemize}[itemsep=0.5em,topsep=0.5em,leftmargin=*]
    \item \textit{We propose the syntax and semantic probing tasks to analyze the code model's ability to understand syntax and semantics by directly recovering the code syntax and semantic structures ~(AST, CFG, CDG and DDG) from the code representation. We study the distribution of the attention weights related to the code semantics.}
    \item  \textit{We reveal that the syntax relationship is more observable in the shallow hidden layers, while the token-role syntax is more observable in the deep layers. Code models have a superior performance for the syntax representation than the semantic representation. Meanwhile, code models have inferior performance for CFG compared with the other semantic structures, which requires code models to be enhanced to represent CFG semantics. Different hidden layers show the different observable levels for different types of code syntax and semantics.}
    \item \textit{
We first include large language models and show that their performance on probing tasks does not have a massive advantage over pre-trained models, considering their huge amount of parameters. This reflects that the code syntax and semantics are hidden and not obvious in the representation of LLMs.} 
\end{itemize}
}
\revision{
We hope these insights can inspire
researchers to train more powerful code models. When utilizing these models, it is essential to design the workflows guided by an understanding of these models, taking into consideration whether the integration of additional code features is warranted. Although numerous factors, including data quality, the complexity of downstream task models, and feature integration methods influence the model performance in downstream tasks, our work can offer some direction for their models and data design. Models devoid of semantic comprehension require consideration of semantic information enhancement via feature integration when addressing downstream tasks. Our work highlights that it is not mandatory to use the output of the final layer - intermediate layers can potentially express syntax and semantics just as adequately.} All code and data can be found at the  repository~\footnote{\url{https://github.com/Marvinmw/probing_analysis_tosem.git}}.

\section{Motivation}
\label{sec:motivation}
The capabilities of pre-trained code models have been extensively discussed in previous studies \cite{10.1145/3510003.3510050, troshin2022probing,lopez2022ast}. However, their analysis is based on the assumption that closer syntax between two tokens in the code would result in smaller syntax distance, measured by the number of edges between nodes in the AST ~\cite{10.1145/3510003.3510050}, and correspondingly, a smaller distance in vector representation, such as the Euclidean distance encoded by the code model. In other words, the assumption states that ``closer syntax" leads to ``smaller edge distance in AST" which in turn leads to "smaller representation distance." However, first, ``smaller edge distance in AST'' is not necessary to mean ``close syntax''. Second, code vector representation space is a high-dimension space. For example, CodeBERT forms a 768-dimensional space but node distance in AST is a low dimensional space. Therefore, the traditional distance metrics used in these works will not work well due to the curse of high dimensionality~\cite{mirkes2020fractional, 10.5555/645504.656414}. First, the distance mapping from high-dimension space to low-dimension space may not be accurate i.e., the smaller distance in the AST may not definitely represent the smaller euclidean distance in the high-dimension space. Second, a small edge distance between two nodes in an AST does not guarantee a close syntax relationship. \revision{This bias can make their conclusions and approaches not generalized and can hinder our understanding of how code models encode syntax. As well, it wrongly estimated the code syntax similarity and can lead to missing analysis of the syntax structure for code models. The conclusion may not be useful for improving model performance.} We provide two examples for better illustration. 

\begin{figure}[!t]
\centering
\scalebox{0.8}{
\includegraphics[width=0.8\textwidth]{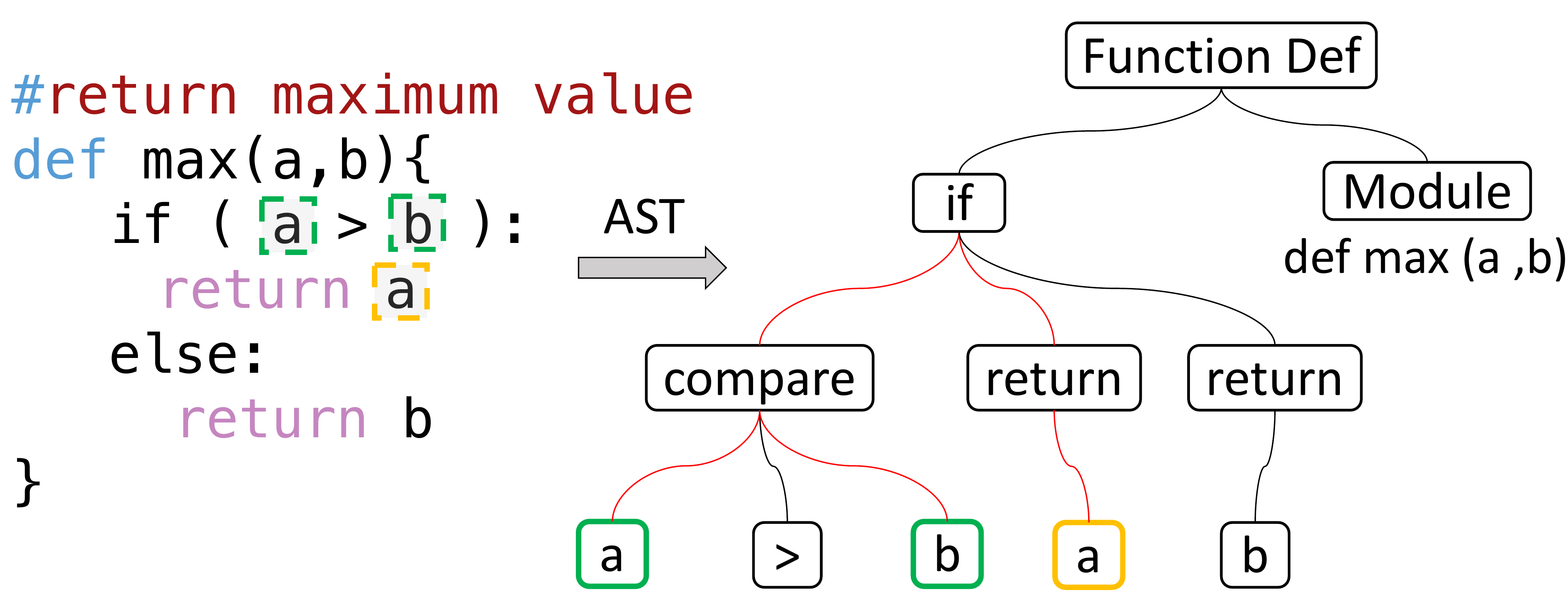}}
\Description[Motivation AST Example]{}
\vspace{-0.9em}
	\caption{A simple code snippet with its AST.}
\label{fig:ast_motivation}
\end{figure}

\begin{figure}[t]
\centering
\scalebox{0.7}{
\includegraphics[width=0.8\textwidth]{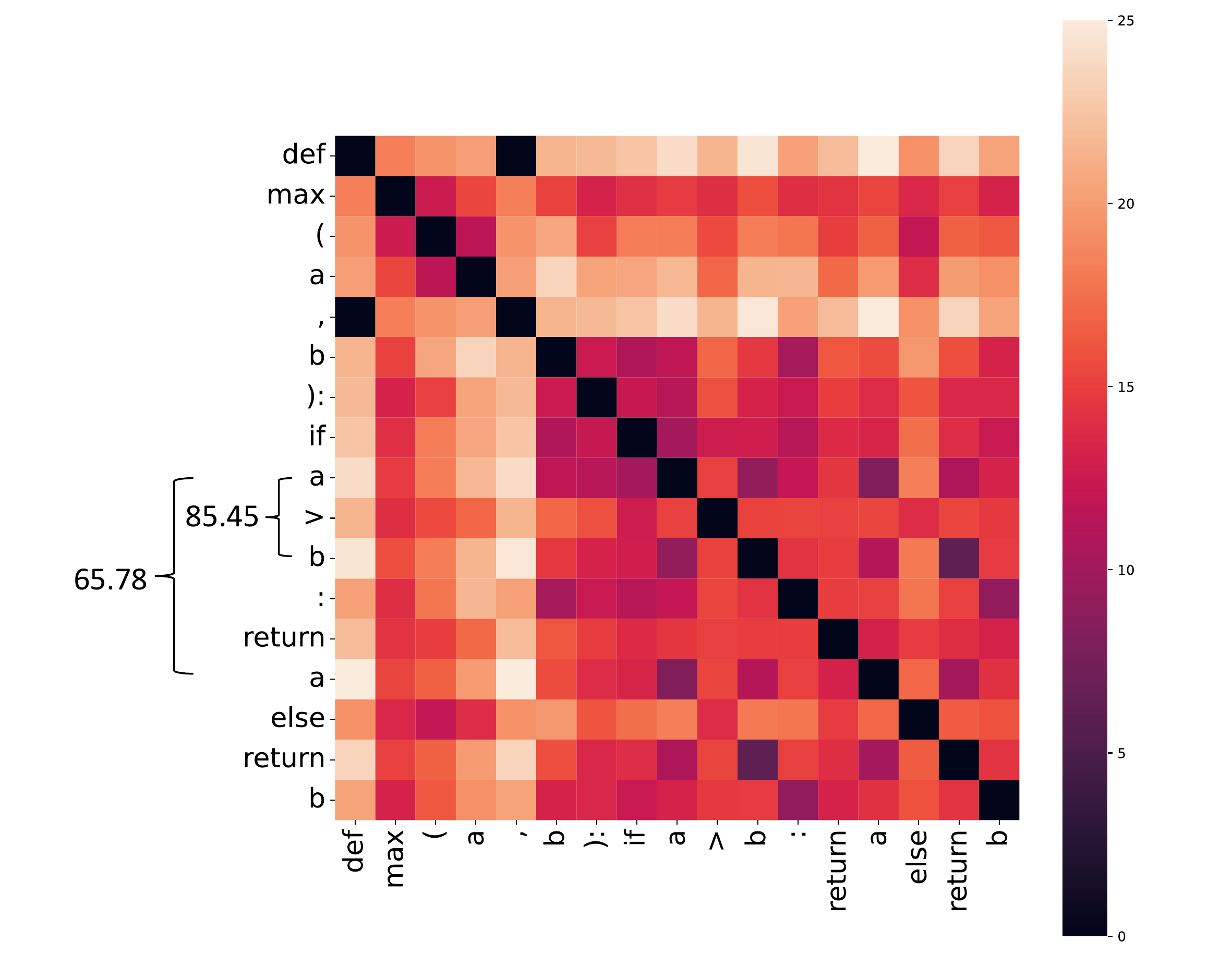}}
\vspace{-0.9em}
	\caption{Euclidean distance of token representations.}
 \Description[Motivation Distance Example]{}
\label{fig:motivation_euclidean}
\end{figure}

\figref{ast_motivation} is the first example and provides a visualization of a function parsed into an AST. We can see that the node distance between the variable ``a'' and ``b'' from the if-condition (marked in green squares) is \textbf{2} hops while the node distance between this variable ``a'' (green square) and the variable ``a'' (orange square) from the return statement is \textbf{4} hops. Hence, we can find that in the low-dimension space, the variable ``a'' from the if-condition is close to the variable ``b'' and they are syntax closer than the variable ``a'' from the return statement. However, the conclusion is opposite in the high-dimension space. We encode this function by CodeBERT and calculate the euclidean distance between the token vector representations for these variables. As shown in \figref{motivation_euclidean}, we can see that the distance from the variable ``a'' (green square) to the variable ``b'' (green square) from if-condition is \textbf{85.45} while the distance from it to the variable ``a'' (orange square) from the return statement is \textbf{65.78}. Hence, through this example, we find that the distances in the low space and high space are not consistent, and they may not be positively related.

For the other example, \revision{\figref{ast_probing_task1} demonstrates that the function argument variable "b" has 4 hops to "return" in
the last statement of this function, and it shares the same distance as the function argument variable "c". However, "b"
should be closer in syntax to "c" since they are function arguments.}

Furthermore, these works mainly focus on exploring the capabilities of code pre-trained models in learning code syntax. In-depth discussions about the learned code semantics (e.g., control/data dependency analysis) are missing. \citet{troshin2022probing} uses some tasks related to semantics while they are not only related to one single code property that is usually required by the probing analysis, the diversity of program semantics is also ignored and the deep analysis is lacked. 
Discussion of code semantics is an essential part and cannot be overlooked as it forms the foundation for many code-related tasks~\cite{drain2021generating, li2022transrepair, zhou2019devign, liu2020retrieval, liu2020atom}. \revision{These works also lack the analysis of LLMs.}

\revision{To overcome the problems above, we need to explore better approaches without the assumption that the number of edges between nodes is related to code syntax. 
Code is a very special data structure. First of all, it has a strict syntax definition; all code must comply with these syntax rules. All names used in the code only need to follow specific naming rules. However, for the convenience of human reading, humans usually annotate the code by names for easy reading and understanding. These syntax rules can be expressed in a structured way by an abstract syntax tree (AST). Based on these defined syntax rules, the code also expresses deeper semantic information. The code has certain functionality, is executable, and implements specific logic. This semantic information includes control flow, data flow, and dependencies. Almost all problems related to the code require this information to solve. The control flow describes the specific logic of the code, expressing the functionality implemented by the code. Data flow and dependencies express the semantic association between different parts of the code. They are very important for code quality, and defect vulnerability detection and repair. We can see that AST, Control Flow Graph (CFG), Control Dependency Graph (CDG), and Data Dependency Graph (DDG) are structured expressions of code syntax and semantics. 
Code tokens have different syntax and semantic relationships or properties based on these structures. These relationships and properties are significant for understanding code and should be represented in the learned vector space of code models.
The ability of a code model is to map the characteristics of the code to vector space. For a good code model, its feature space should keep all the characteristics of the code as much as possible. The probing tasks we design are to reconstruct these data structure relationships in the vector feature space.}

\revision{
Therefore, for syntax analysis, we try to reconstruct AST~(syntax node pair prediction) and predict the syntax label of the code tokens~(token syntax tagging). AST carries all syntax information of the code. Syntax node pair prediction can reflect how code models learn the syntax structure by reconstructing AST in the representation space. Token syntax tagging is to see if the syntax role of each token is encoded in the representation, which is a micro syntax property. Both of them are significant for code models to learn code syntax.  
For semantic analysis, we reconstruct code semantic structure~(semantic propagation prediction) that includes a control dependency graph (CDG), data dependency graph (DDG) and control flow graph (CFG). We also include the control and data dependency with long distances~(semantic propagation prediction). The three semantics structures, CDG, DDG and CFG, are core concepts in program analysis, optimization, toolbars, and various other software engineering tasks. The long dependency is one special characteristic of the program. One can declare a variable at the beginning but use it after hundreds of lines. Understanding them is significant for code models. Details can be found in Section \ref{sec:methodlogy}.
}

\section{Methodology}
\label{sec:methodlogy}
In this section, we introduce our analysis of the probing approaches for code syntax and semantics. 

\subsection{Preliminary Knowledge}
\revision{Early software engineering researchers used feature engineering based on expert empirical knowledge to extract code features as input to machine learning algorithms. Typically, experts design feature extraction rules based on the specific tasks to be solved, such as the number of loops and code complexity. These rule-based features are highly interpretable but limited by the experience of experts and cannot cover the code syntax and semantic structure. In recent years, with the rise of Transformer and the emergence of CodeBERT, software engineering researchers can extract code features based on the encoding ability of pre-trained models, greatly reducing their reliance on expert experience and knowledge. The transformer with the encoder-decoder architecture was proposed in 2017~\cite{vaswani2017attention}. The encoder and the decoder are stacked by multiple encoder and decoder layers, respectively. Both contain complex attention-calculation and allocation mechanisms. The difference between the encoder layer and the decoder layer is that the encoder layer considers the context of the model; the decoder layer only considers the output context of the model. The code pre-trained models will use massive code data that is collected from the Internet such as GitHub and Stack Overflow. Code pre-trained models can be grouped into three groups regarding their network architecture. The first group uses a transformer encoder like CodeBERT. The second group uses a transformer decoder like CodeGPT. The third group uses the transformer encoder-decoder like CodeT5.}

\revision{
Before pre-training, we need to train a tokenizer based on code text to learn how to segment text code, such as the algorithm Byte-Pair Encoding tokenization (BPE)\footnote{https://huggingface.co/learn/nlp-course/en/chapter6/5}. When using Transformer for feature learning, the original text code will be tokenized and we will get a sequence of code tokens. Then, we converted it into the corresponding index in the embedding layer. This completes the conversion from the raw code to the vector. The model is then optimized based on the training paradigm implemented. Compared with traditional pre-train models, LLM is also a transformer architecture but has a huge number of parameters and a huge amount of training data. LLM is aligned with human preference. Since tuning the whole parameters of LLM is very expensive, we usually use LoRA~\cite{hu2022lora}, fewshot learning and prompt tuning to solve the downstream tasks. Some research shows that it has emergence and generalization capabilities~\cite{wei2022emergent}. Researchers in software engineering have found that large models have good code generation and analysis capabilities~\cite{yuan2023no,du2023classeval,fan2023large}.}

\revision{
Feature extraction methods based on code models have a premise: the structural features of the code are included in the feature space of the code model. The current mainstream code model training paradigms are based on masked language modelling ~(MLM), causal language modelling ~(CLM) and other variants of the previous two. To put it simply, MLM means randomly masking code tokens and then recovering them. The difference is that CLM predicts the next token based on the above. These pretraining methods are not adapted for learning code features. In almost all the cases, we covert the code into the text format required by them. The analysis and understanding of the feature space of the code models is a challenge.}

\subsection{Probing Analysis Model}
\begin{figure}[t]
\centering
\scalebox{0.9}{
\includegraphics[width=0.8\textwidth]{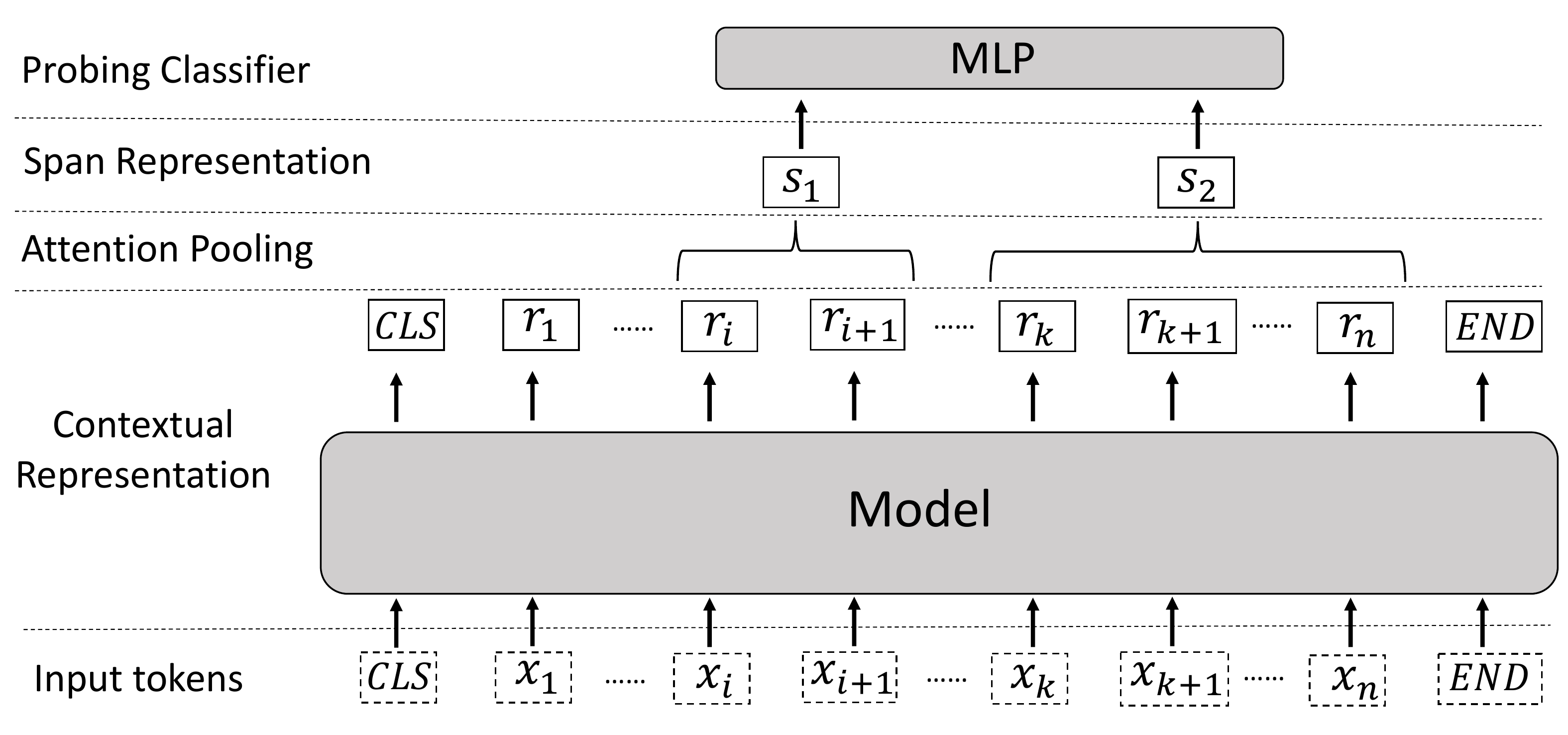}}
\Description[Probing Model Architecture]{}
\vspace{-0.9em}
	\caption{Analysis model.}
\label{fig:probing_model}
\end{figure}

\revision{Probing analysis is a research method used for understanding and evaluating the knowledge and information encoding in language models. This analytical method reveals a model's mastery over specific types of linguistic information during its learning process by designing and applying a series of probing tasks. Probing tasks are usually simple, targeted tasks that are specifically designed to test a model's understanding of a particular linguistic attribute, such as grammatical structure, word meaning, and sentence relations.}
In our study, for all designed probing tasks, we employ the edge probing classifier~\cite{tenney2019you} as depicted in \figref{probing_model}. Consistent with previous \revision{works~\cite{conneau-etal-2018-cram, 10.1145/3357384.3358028,10.1145/3383313.3412249,tenney2019you}} in the probing literature, we keep the parameters of code pre-trained models fixed, \revision{which means they will not be updated during training. For a given input code $x$, we use a tokenizer to tokenize $x$ into a token sequence with two special starting and ending tokens as denoted in the bottom of \figref{probing_model}.}

Initially, we can obtain the contextual representation $r_i$ of \revision{each hidden layer} for each token $x_i$ from the code models. Subsequently, \revision{we extract the specific code token spans  that are associated with the graph or tree node we are interested in, and they are denoted as $s_1$ and $s_2$ in \figref{probing_model}}. One code token span represents one piece of code from one node in AST or the dependency graph. \revision{The length of one token span is various and can contain different numbers of tokens.}
Then, these token spans are then passed through an attention pool, which maps them to a fixed-size vector. \revision{The attention pool is an attention layer that can automatically assign different weights to each token representation in the token span, and then aggregate them.}
Finally, the resulting vector is forwarded to the probing classifier, which is implemented as a multi-layer perceptron (MLP) classifier for classification. The MLP classifier is trainable and is denoted as the symbol $C$. The fixed-size feature vectors of two token spans served as the input for $C$, and the classifier determined whether the two token spans had a syntax or semantic relationship. We employed token spans because code pieces that are related in terms of syntax or semantics are tokenized into two lists of tokens with different lengths, respectively. These token lists are referred to as token spans, which are then converted to a fixed-size vector by the attention pool.

\begin{figure}[!t]
\centering
\scalebox{0.9}{
\includegraphics[width=0.8\textwidth]{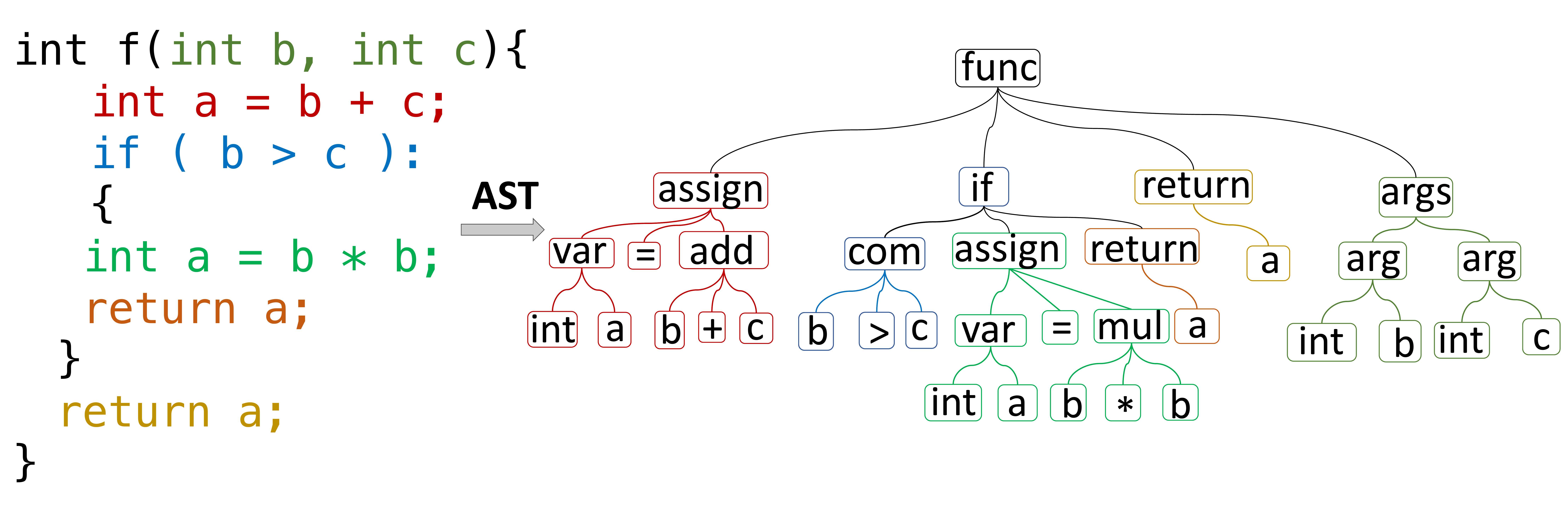}
}
\caption{Syntax Pair Node Prediction.}
\Description[Syntax Pair Node Prediction]{}
\label{fig:ast_probing_task1}
\end{figure}

\subsection{Syntax Probing}\label{sec:syntax-probing}
To address potential inaccuracies resulting from transforming a high-dimensional space to a lower-dimensional space, as well as the possible lack of relationship between distance in the AST and syntax distance as indicated in the motivation, we introduce two syntax probing tasks: syntax-node-pair prediction and token-syntax tagging. These tasks directly predict the attributes of the nodes in the AST. The two tasks complement each other. The first task aims to predict the syntactic-connectivity relationship between code tokens, while the second task focuses on examining the syntactic role of individual code tokens. \revision{Code tokens are the basic code words after the code is tokenized and each node contains few code tokens.} \revision{Through the first task, we can reconstruct AST given the code token representation from the code models. If the code token representations do not contain information about the AST structure, it is not possible to reconstruct the AST. Through the second task, we can find if the code token representations contain the corresponding syntax rules assigned by the programming language rules.}

\subsubsection{Syntax Node Pair Prediction}\label{sec:syntax-node-pair-prediction}
Given a source code, we can parse it to obtain AST and further split this AST into different subtrees.
Each subtree is one syntax expression from the original code and we name it as one syntax unit. Because each syntax unit represents one complete syntax expression from the code, hence the nodes in a unit are syntax-close. This task is designed to predict any pair of nodes in AST belonging to a subtree. \revision{The AST is generated by the syntax parser and contains all the syntax information of the code. For code models, they should retain such a syntax structure in the vector space. If they cannot, it suggests that the model is not capable of effectively encoding code syntax. Via this task, we aim to verify whether the code model can capture and understand this syntax structure from the source code, and subsequently reconstruct the syntax information in the vector space it represents.}
In particular, we present an example for better illustration. As shown in \figref{ast_probing_task1}, there is a function whose corresponding AST is presented on the right hand. For this AST, we split it into different units marked with different colors. For example, the node of ``='' in the red unit, should be syntax close to its left node in this unit i.e., the node of ``int'' and ``a''. Hence, they are labeled with  \textbf{1} as the positive samples. For the nodes of ``int'' and ``a'' in another unit (marked with green), since they belong to another unit, they are labeled with \textbf{0} as the negative samples. 
Formally, this task can be formulated as follows:
 
\[ C(n_0, n_1)=\begin{cases} 
      1 & n_0 \in T_i \cap n_1 \in T_i \\
      0 & n_0 \in T_i \cap n_1 \in T_j 
   \end{cases}
\]
where $T_i$ and $T_j$ ($i \ne j$) are two different syntax units in AST and $n_0$ and $n_1$ are the nodes in the unit. We train a binary probing classifier $C$ to learn whether any pair of nodes is syntax-close based on the node representations computed from the token representation of code models by the attention pool (\figref{probing_model}). 

\subsubsection{Token Syntax Tagging}
\revision{The first task checks the syntactic similarity between different AST nodes. Except for the syntax similarity, all tokens are assigned a syntax role by the programming language rules. To check if code models can learn individual token syntax roles, we propose a multi-class classification task, namely token syntax tagging, and this task requires more better and fine-grained code representation to tag the syntax role of each code token. 
This is designed to challenge and evaluate the depth of understanding of programming languages for code models. A code model that can successfully perform token syntax tagging can be more useful in complex coding applications, such as code generation.}

As each AST node has its node type, a direct idea is to use the node type for tagging. However, these syntax labels are highly abstractive, for example, the variable name is labeled with ``identifier'' by the syntax rule, but it has different syntax meanings in the context. If it is in the function declaration, it is actually an argument for this function; similarly, if it is in the class invocation, it is a class attribute. Hence, we need to design concrete labels for these abstractive tokens. We initially take into account the syntax type of the leaf nodes in the AST. Subsequently, we refer to the types of their parent nodes to construct the concrete labels. Specifically, we design 36 tagging labels for Java and 33 labels for C/C++. %
We filter the labels with low frequency ($fre<200$) for Java250 and POJ-104. The labels we used are depicted in \tabref{lablel_tag}. 

\begin{table*}[!t]
\centering
\caption{The labels of Token Syntax Tagging for Java250 and POJ-104.}
\label{tab:lablel_tag}
\large
\scalebox{0.7}{
\begin{tabular}{llll}
\hline
\multicolumn{4}{c}{Java250} \\ 
modifiers & local\_variable\_declaration & variable\_declarator & formal\_parameters \\
array\_type & dimensions & formal\_parameter & block \\ object\_creation\_expression & argument\_list &
field\_access & integral\_type \\
method\_invocation & while\_statement  & parenthesized\_expression & if\_statement \\
expression\_statement & break\_statement & update\_expression &
assignment\_expression \\
identifier & for\_statement  & binary\_expression &
return\_statement \\ 
array\_creation\_expression & dimensions\_expr & array\_access & ERROR \\ 
unary\_expression & throw\_statement & enhanced\_for\_statement & ternary\_expression \\ 
cast\_expression & generic\_type & type\_arguments & array\_initializer \\
\multicolumn{4}{c}{POJ-104} \\ 
declaration & array\_declarator & function\_definition & parameter\_list \\
parameter\_declaration & compound\_statement & for\_statement  & assignment\_expression \\
binary\_expression & update\_expression  & subscript\_expression & expression\_statement \\
if\_statement & parenthesized\_expression & return\_statement & call\_expression \\
argument\_list &  string\_literal & pointer\_expression & init\_declarator \\
function\_declarator & cast\_expression & type\_descriptor &break\_statement \\
comma\_expression & initializer\_list & char\_literal & pointer\_declarator \\
continue\_statement & while\_statement& field\_expression &sizeof\_expression \\
case\_statement & && \\
\hline
\end{tabular}}
\end{table*}

\begin{figure}[!t]
\centering
\includegraphics[width=0.8\textwidth]{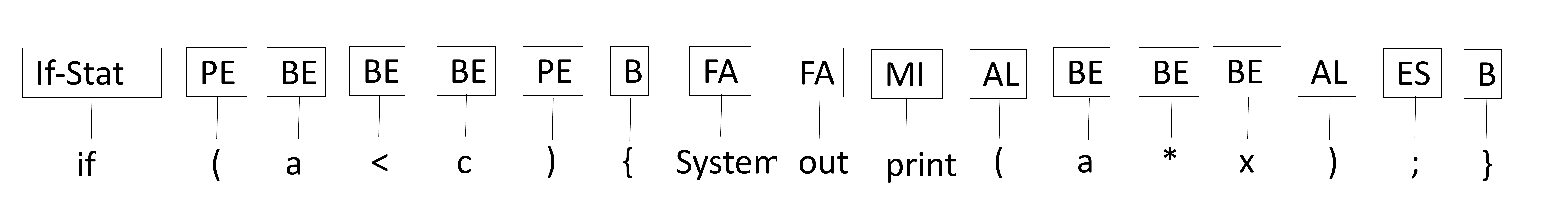}
\caption{Token Syntax Tagging.}
\Description[Token Syntax Tagging]{}
\label{fig:pos_tag}
\end{figure}

We present an example with the corresponding syntax label in \figref{pos_tag}. The detailed information of labels in this example is shown in \tabref{token_tagging_label}. In particular, we can find that the tokens ``('' and ``)'' have different syntax labels in different contexts. The parenthesis ``( )'' is labeled with ``PE'' in the if-condition, while the parenthesis ``( )'' from the method invocation is labeled with ``AL''. We design this task to explore whether these models can learn the code syntax 
properly
from the programming grammar.

\begin{table}[!t]
\centering
\caption{The tagging labels for the tokens used in Figure~\ref{fig:pos_tag}.}
\label{tab:token_tagging_label}
\begin{tabular}{l|l}
\hline
Label                    & Description              \\ \hline
PE                       & Parenthesized expression \\ \hline
BE                       & Binary expression        \\ \hline
B                        & Block                    \\ \hline
FA                       & Field access             \\ \hline
MI                       & Method invocation        \\ \hline
AL                       & Argument list            \\ \hline
ES & Expression statement     \\ \hline
\end{tabular}
\end{table}

\subsection{Semantic Probing}

The previous works mainly focused on analyzing the capacity of code pre-trained models in learning code syntax. However, the analysis of code semantics is also essential. 
\revision{This group evaluates the ability of the code model to understand the meaning and behaviour of code snippets. In this section, we propose two probing tasks to analyze the semantics of the learned code. The first is semantic relation prediction for dependency graphs (CDG, DDG and CFG). Based on the prediction, we can reconstruct the structure of dependency graphs. Dependency graphs depict complex relationships like control and data dependencies, which are crucial for understanding how different parts of a program interact. The second task is to predict the semantic propagation. Understanding control and data flow is a more advanced aspect of semantic analysis, going beyond static analysis to how the program would behave when run. This task assesses the ability of code models to understand the dynamic nature of code execution.}

\subsubsection{Semantic Relation Prediction}\label{sec:semantic_relationship_prediction}
Dependency graphs of a program can well represent code semantics~\citep{10.1145/143062.143156}. Similar to the task of syntax node pair prediction in Section~\ref{sec:syntax-node-pair-prediction}, we also extract the control dependency graph (CDG), data dependency graph (DDG) and control flow graph (CFG) to predict whether the code models can learn code semantics. In the constructed Control Dependency Graph (CDG), Data Dependency Graph (DDG), and Control Flow Graph (CFG), \revision{each node has an attribute with a piece of code that can be tokenized into multiple  tokens. We refer to these code tokens derived from the same node as the \textit{token span}.} We unify these tasks into a meta task namely semantic relation prediction. Formally, this task can be formulated as follows:
 
 \[ 
\scalemath{1}{ C(s_0, s_1)=\begin{cases} 
 1 & \exists e \in G, \quad s_0 \in N_i \cap s_1 \in N_j  \cap \{ N_i \xrightarrow {e} N_j \} \\
      0 & \forall e \in G, \quad s_0 \in N_i \cap s_1 \in N_j \cap \{ N_i \centernot{\xrightarrow {e}} N_j \} 
   \end{cases} }
\]
where $N_i$, $N_j$ is the $i$-th node and $j$-th node in the constructed graph $G\in \{CDG, DDG, CFG\}$ and each node contains one token span from the original code. $s_0$ and $s_1$ denote the corresponding token spans of the original code in nodes $N_i$ and $N_j$ after tokenization. $\xrightarrow {e}$ means that there is an edge between two nodes.

\figref{cdg}, \figref{ddg} and \figref{cfg} demonstrate examples for the three semantic relationships. \figref{cdg} shows one control dependency example. The node $N_4$ is control dependent on the node $N_3$. Based on this fact, we label that the token span from $N_3$ is control dependent on the token span in $N_4$. \figref{ddg} illustrates one data dependency example. The node $N_4$ is data dependent on the node $N_1$. \figref{cfg} shows one example of a control flow graph. We have some execution order facts like that the node $N_2$ is executed after $N_1$ immediately. The token span of $N_2$ has a control-flow relationship with the token span of $N_1$.

 \subsubsection{Semantic Propagation Prediction}
\begin{figure}[!t]
\centering
\begin{subfigure}[b]{0.4\textwidth}
\includegraphics[width=\textwidth]{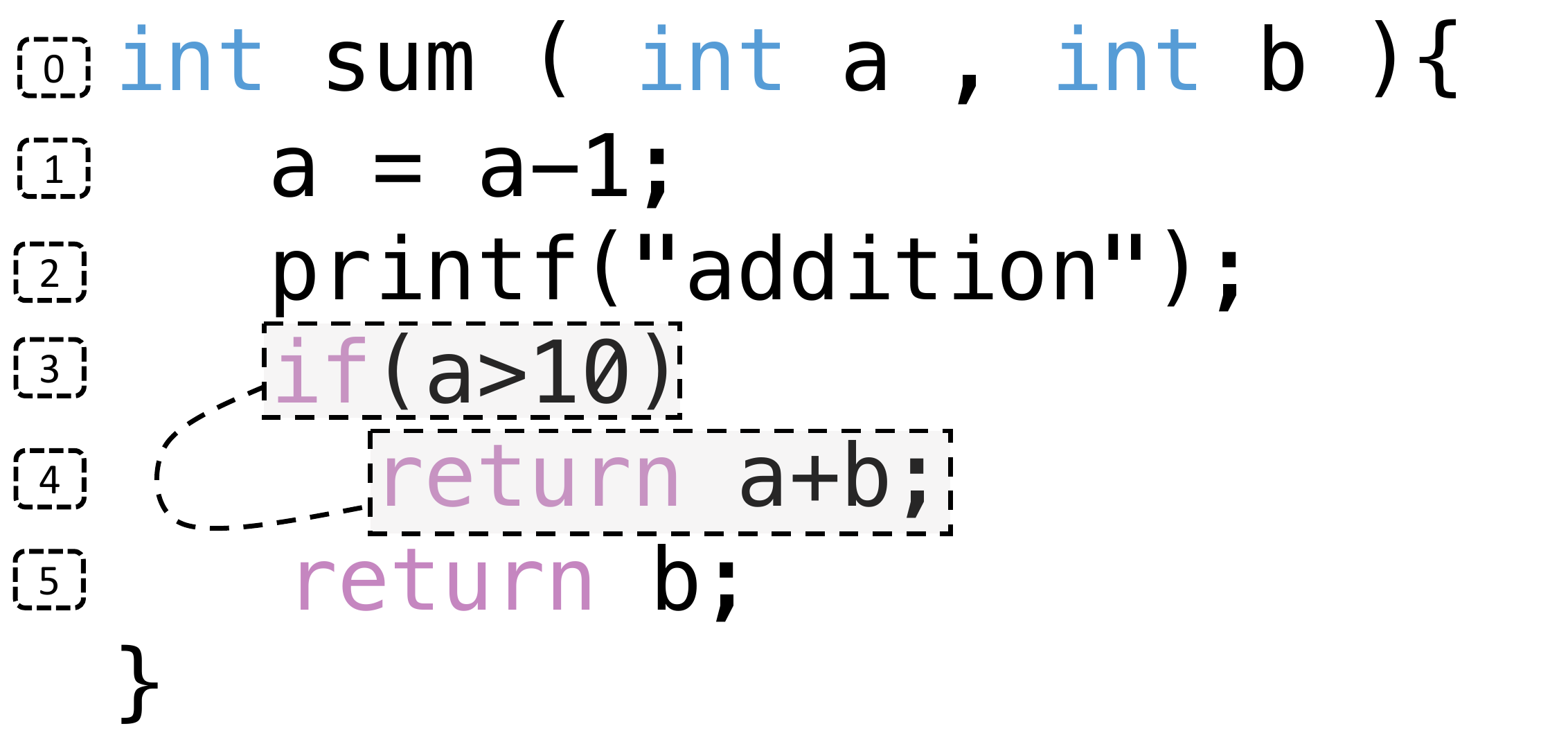}
	\caption{control dependency.}
	\label{fig:cdg}
\end{subfigure}
\begin{subfigure}[b]{0.4\textwidth}
\includegraphics[width=\textwidth]{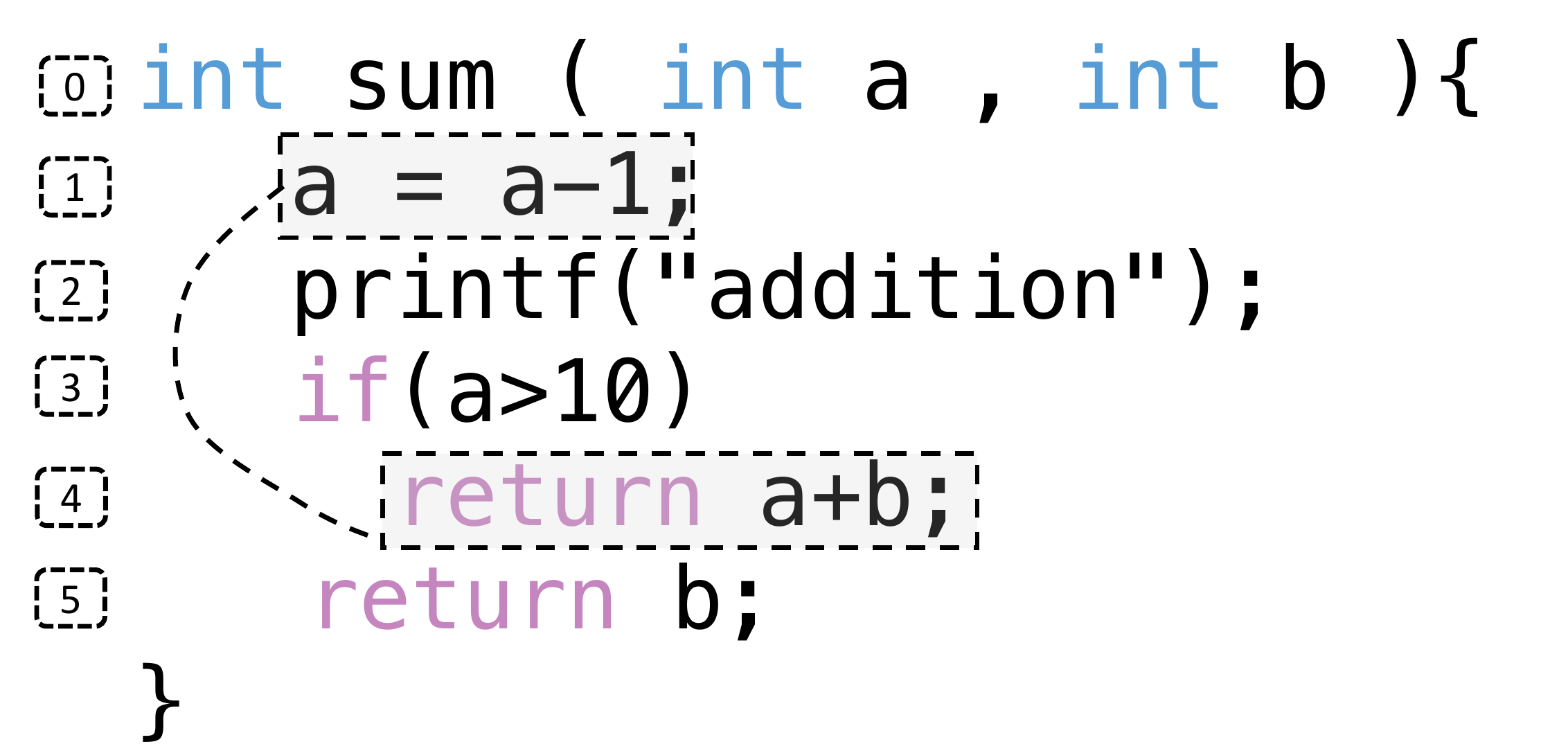}
	\caption{data dependency.}
	\label{fig:ddg}
\end{subfigure}
\begin{subfigure}[b]{0.4\textwidth}
\includegraphics[width=\textwidth]{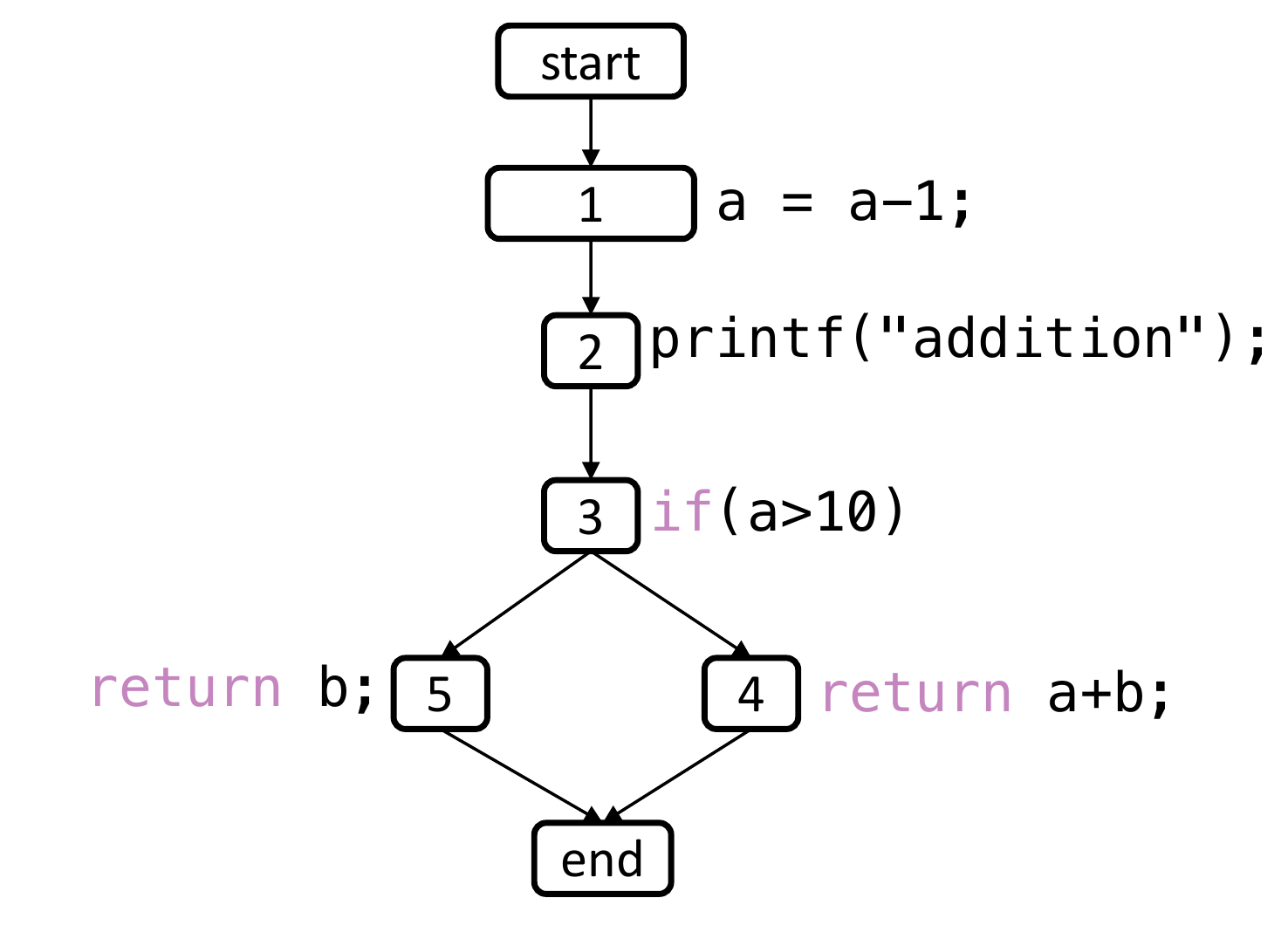}
	\caption{control flow.}
\label{fig:cfg}
\end{subfigure}
\begin{subfigure}[b]{0.4\textwidth}
\includegraphics[width=\textwidth]{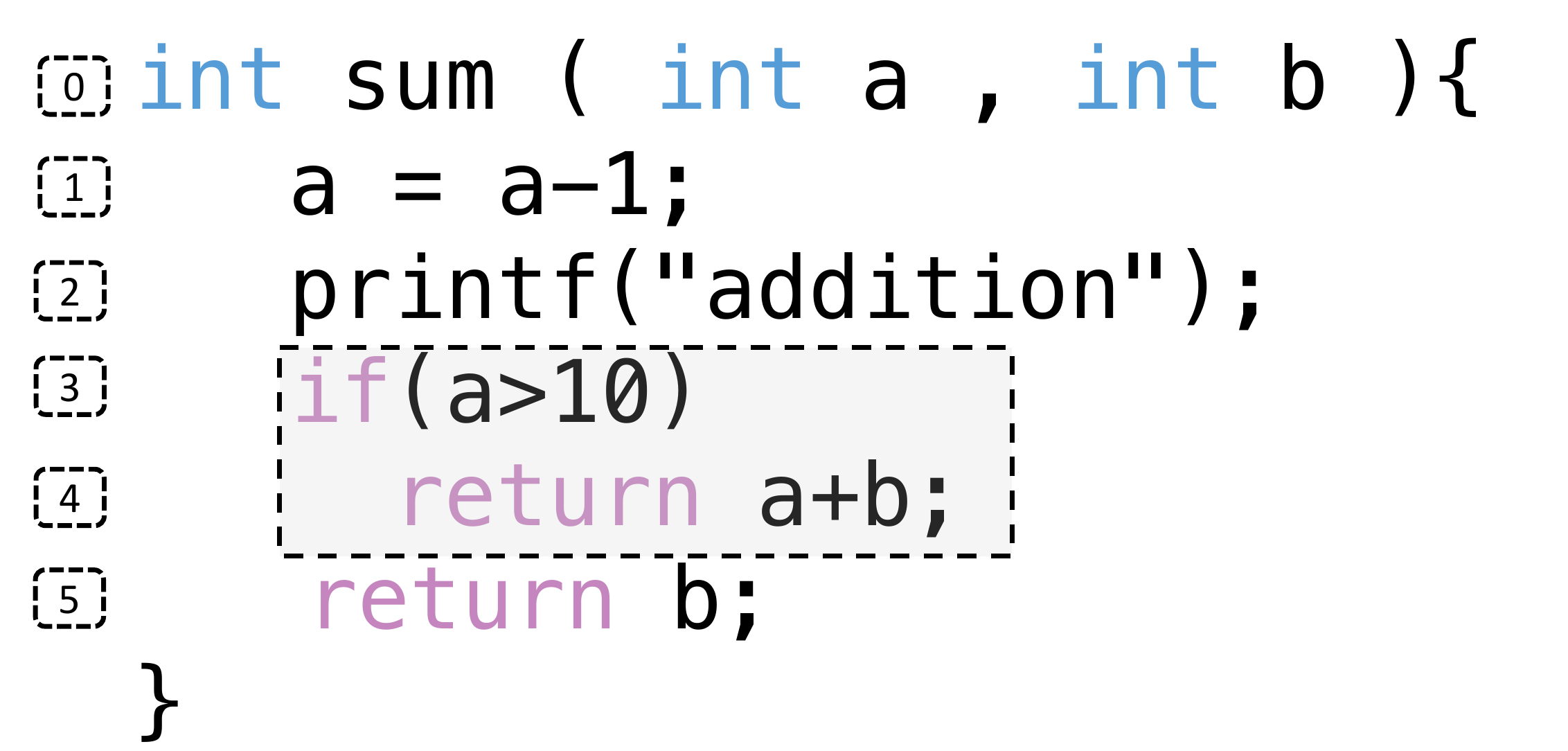}
	\caption{inGraph.}
	\label{fig:inGraph}
\end{subfigure}

\caption{Examples about semantic probing.}
\Description[Examples about semantic probing]{}
\end{figure}

 Data flow information is propagated in the dependency graph. \revision{Tow nodes with the long distance may be data dependent implicitly.} It is a fact that any modification in the dependency graph potentially affects the program output. The implicit dependency flow propagation is one import program semantics. The semantic propagation task (alias \textit{inGraph}) is defined by that
  \[ 
  \scalemath{1}{C(s_0, s_1)=\begin{cases} 
      1 &  \quad s_0 \in N_i \cap  s_1 \in N_j \cap  N_i \in G \cap N_j \in G   \\
      0 & \quad s_0 \in N_i \cap  s_1 \in N_j \cap  N_i \in G \cap N_j \notin G
   \end{cases} }
\] where $\textit{G} \in \{CDG, DDG\}$ , $s_0$ and $s_1$ are the token spans in the nodes of $G$. 
 \figref{inGraph} shows one example. The shadow box highlights the statements with the control dependency relationship. We expect the probing classifier can recognize that the ``printf'' statement is not in the dependency control graph.

\subsection{Attention Analysis}
We conducted an analysis of the attention weights in each self-attention head, based on the program semantic relationships $G\in \{CFG, DDG, CDG\}$ from Section~\ref{sec:semantic_relationship_prediction}. For a given node token span $s_i$ where $i$ is the $i$-th node and a semantic relationship type $G$, we grouped the remaining input tokens into two sets: $R_0$ and $R_1$. $R_1$ which consists of all tokens that have the semantic relationship with $s_i$ while the remaining tokens constitute the set $R_0$. We then divided the attention weights of one attention head related to $s_i$ into two sets, denoted as $W_0$ and $W_1$, based on $R_0$ and $R_1$. \revision{We check the difference of the attention distributions between the two sets. Because we want to know which set makes more contribution to the code representation, we compare the distribution centrality of the two attention sets. If the sum of the attention weights in $W_1$ from a particular attention head is greater than the sum of the attention weights in $W_0$, it indicates that this attention head makes more contribution to learning code semantics and we denote it as the semantic attention head. }%

To determine the statistical significance, we applied the paired t-test with a large sample size for each semantic type $G$. We formulated the null hypothesis $H_0: \mu_d = 0$, where $\mu_d$ represents the true mean difference between $W_1$ and $W_0$. The alternative hypothesis $H_1: \mu_d > 0$ suggests that there is a positive mean difference ($W_1$ is greater than $W_0$). By conducting this test, we can identify if the attention head makes more contribution to learning code semantics or not. This head-level analysis provides a more detailed understanding of self-attention. In total, we analyzed more than 10,000 semantic inputs for the 4 pre-trained code models and \revision{randomly selected 100 semantic inputs for the three LLMs.}

\section{Evaluation Setup}
\label{sec:experiments}
In this section, we introduce the evaluation setup, including data pre-processing, evaluation models, and evaluation metrics. 

\subsection{Dataset and Pre-processing} 
\begin{figure}[!t]
\centering
\includegraphics[width=0.8\textwidth]{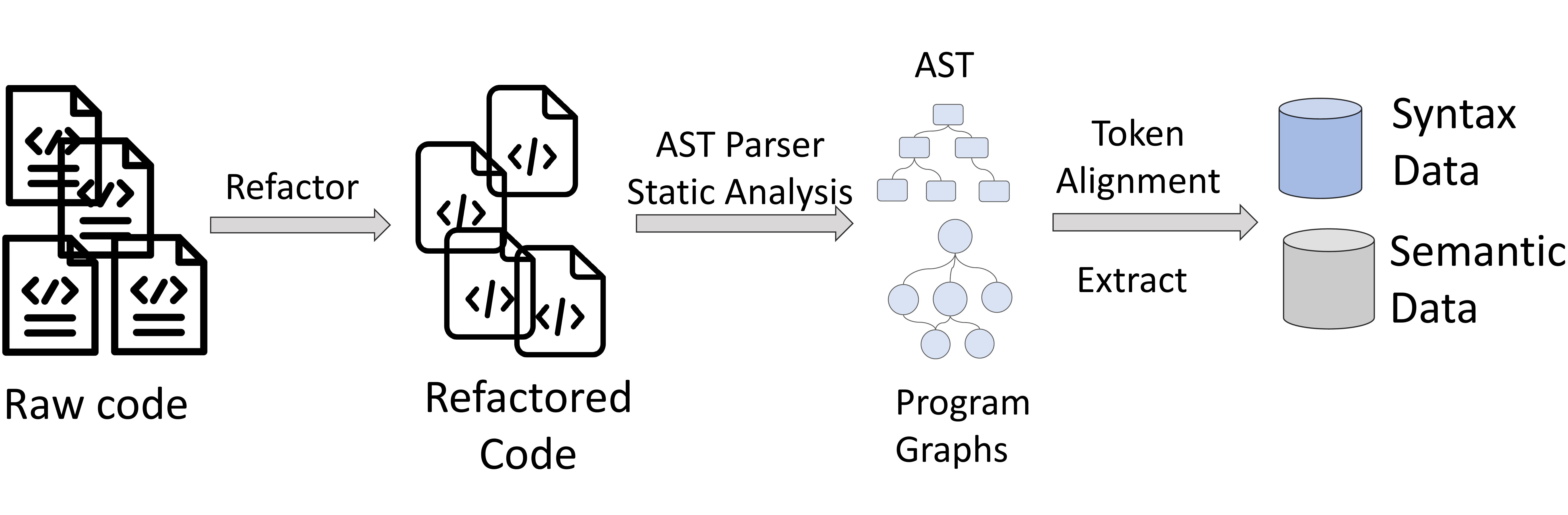}
\caption{The workflow of data pre-processing.}
\label{fig:data_preprocesss}
\Description[The workflow of data pre-processing]{}
\end{figure}

We employ two datasets, namely Java250~\cite{puri2021codenet} and POJ-104~\cite{mou2016convolutional} for our study. The quality of the data has a significant impact on the probing analysis. We follow the pre-processing steps outlined in \figref{data_preprocesss} to generate high-quality syntax and semantic probing data. The code is refactored using the google-java-format tool\footnote{\url{https://github.com/google/google-java-format.git}} and the clang-refactor tool\footnote{\url{https://clang.llvm.org/docs/ClangTools.html}}. Refactoring the code ensures its readability and facilitates token alignment between the graphs and the model input. To ensure the accuracy of the results, we utilize the Joern static analysis framework and the AST parser\footnote{\url{https://joern.io/, https://tree-sitter.github.io/tree-sitter/}} to extract the program graphs and AST. In the case of program graphs, we merge redundant nodes if the code of one node is a subset of its neighbors. We then extract the syntax and semantic relationships among the code tokens and construct the probing datasets.

\subsection{Experimental Models} 
\revision{We chose 7 popular code models with diverse architectures (transformer encoder, transformer decoder, and transformer encoder-decoder) that are widely used in software engineering. The 7 models were built by reputable companies or organizations: Microsoft, Meta, Salesforce, and Huggingface. All of them have a large number of downloads on Huggingface, and their corresponding papers each have more than 150 citations.}
\revision{First,} we conduct experiments using four pre-trained models, CodeBERT~(CB)~\cite{feng2020codebert}, GraphCodeBERT~(GCB)~\cite{guo2020graphcodebert}, UnixCoder~(UC)~\cite{guo2022unixcoder} and CodeT5~(CT)~\cite{wang2021codet5}. CodeBERT and GraphCodeBERT utilize the Transformer encoder, while GraphCodeBERT incorporates additional data flow information. UnixCoder utilizes mask attention matrices with prefix adapters to support encoder, decoder, and encoder-decoder learning. CodeT5 employs the encoder-decoder Transformer architecture. For CodeT5, we consider its encoder, which can be compared with other \revision{pre-trained} models in the same \revision{architecture}. These models have 12 Transformer encoder layers (denoted as $\mathbf{Layer_1}$-$\mathbf{Layer_{12}}$) plus one embedding layer (denoted as $\mathbf{Layer_0}$). We apply our probing tasks to each encoder layer and the embedding layer. This helps us to understand the role of each layer. \revision{Second, we also study 3 large language models~(LLMs), CodeLlama-7b~\cite{roziere2023code} (CL), StarCoder~\cite{li2023starcoder} (SC) and CodeT5+~\cite{wang2023codet5+} (CodeT5p-770m) (CT5P+) using the Java dataset. CodeLlama-7b has 32 hidden layers and StarCoder has 40 hidden layers. Both of them use the transformer decoder. CodeT5p-770m has 24 hidden encoder layers and 24 hidden decoder layers. It is an open code large language model based on the encoder-decoder structure for the CodeT5+ family~\cite{wang2023codet5+}.} All experiments are repeated 3 times with different random seeds. 
 
\subsection{Evaluation Metrics}
We utilize Matthew's correlation coefficient~(MCC)~\cite{MATTHEWS1975442} as our evaluation metric. MCC is one reliable alternative to F1-score for binary classification~\cite{chicco2020advantages, 10.1145/3383219.3383232}, and it considers the whole confusion matrix. \revision{MCC can be computed from the confusion matrix, $$MCC = \frac{TP \times TN - FP \times FN}{\sqrt{(TP + FP) \cdot (TP + FN) \cdot (TN + FP) \cdot (TN + FN)}}
$$, where TP is true positive, TN is true negative, FP is false positive and FN is false negative. }
For token syntax tagging, it is a multiple classifications and we also show its \revision{macro F1 score, $$\text{Macro F1} = \frac{1}{N} \sum_{i=1}^{N} \frac{2 \times \text{Precision}_i \times \text{Recall}_i}{\text{Precision}_i + \text{Recall}_i}
$$, where N is the number of classes and i is the $i_{th}$ class. We selected the macro F1 score as our evaluation metric due to its inherent property of assigning equal significance to every class within our dataset.} As all experiments are repeated by 3 times, we use the mean values for evaluation.

\section{Experimental Results}
\label{sec:results}
In this section, we present the  results of analyzing code syntax and semantics.

\subsection{Syntax Analysis}
\subsubsection{Syntax Pair Node Prediction}\label{sec:syntax-pair-node-prediction}
\begin{figure}[!t]
     \centering
     \begin{subfigure}[b]{0.45\textwidth}
         \centering
         \includegraphics[width=\textwidth]{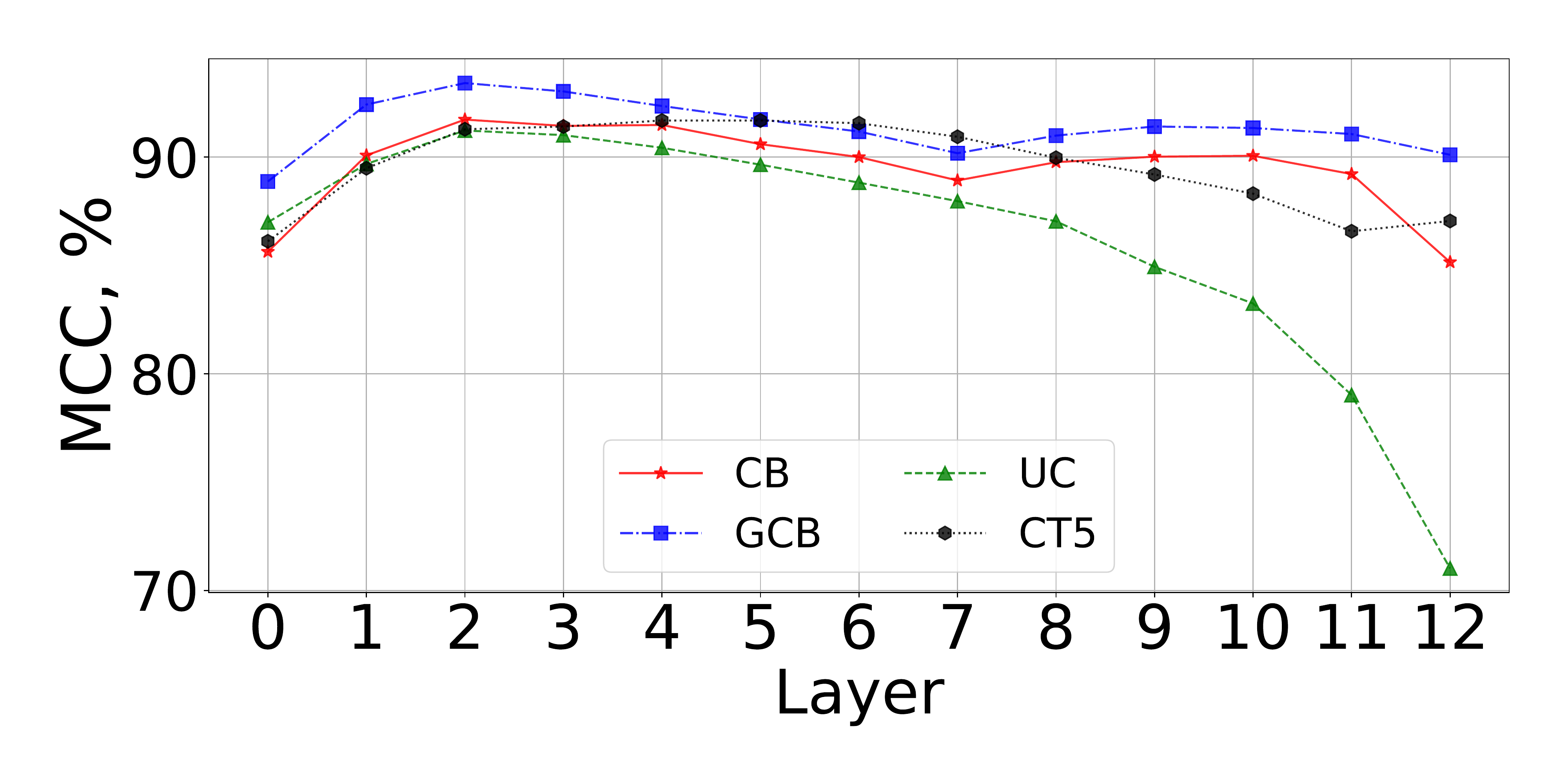}
          \vspace{-2em}
         \caption{Java250-AST.}
         \label{fig:ast_org_java250}
     \end{subfigure} 
     \begin{subfigure}[b]{0.45\textwidth}
         \centering
         \includegraphics[width=\textwidth]{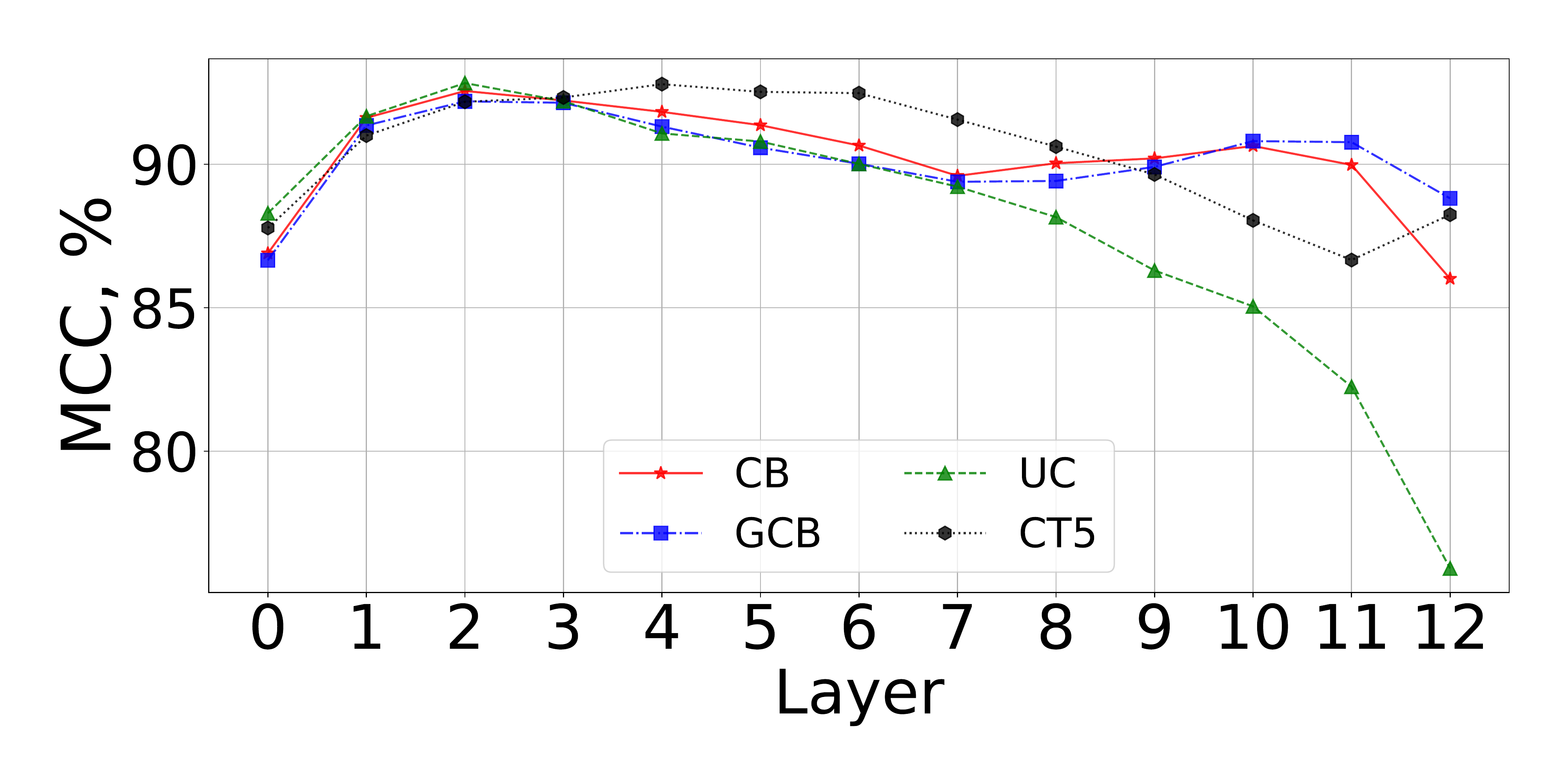}
          \vspace{-2em}
        \caption{POJ-104-AST.}
         \label{fig:ast_org_poj}
     \end{subfigure} \\
          \begin{subfigure}[b]{0.5\textwidth}
         \centering
         \includegraphics[width=\textwidth]{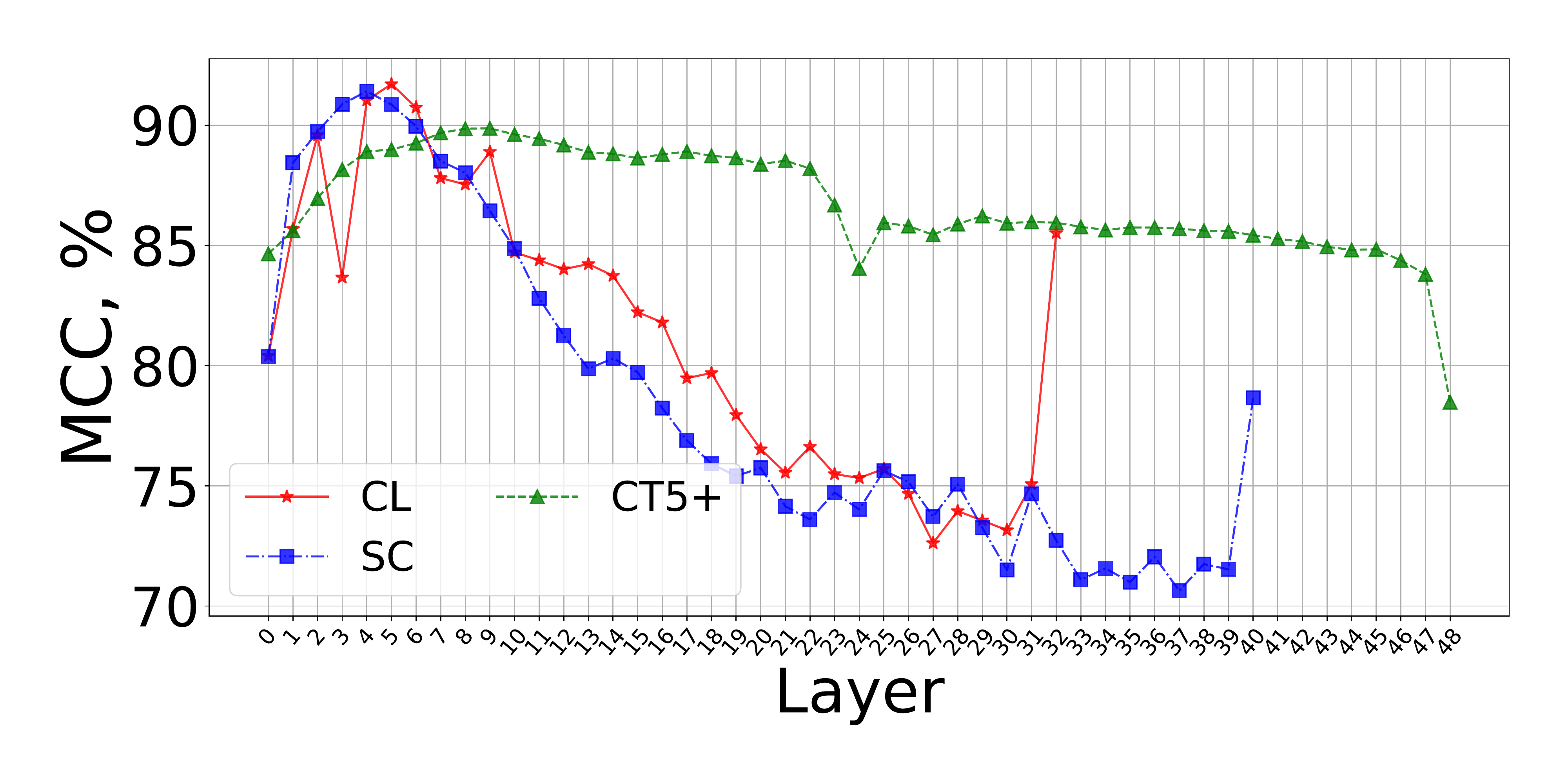}
          \vspace{-2em}
         \caption{Java250-AST of LLM.}
         \label{fig:ast_org_java250_llm}
     \end{subfigure} \\
     \vspace*{-0.9em}
        \caption{Performance~(MCC) about Java250 and POJ-104 for Syntax Pair Node prediction~(AST).}
        \Description[Performance about Java250 and POJ-104 for Syntax Pair Node prediction~(AST)]{}
\end{figure}

\figref{ast_org_java250} and \figref{ast_org_poj} present the results of the probing classifiers for syntax pair node prediction on Java250 and POJ-104 respectively, \revision{for pre-trained code models, CodeBert~(CB), GraphCodeBert~(GCB), UnixCoder~(UC) and CodeT5~(CT5)}. These results are derived from the hidden representations provided by the code models. The X-axis denotes the layer index, while the Y-axis signifies the performance. $Layer_0$ corresponds to the embedding layer, while $Layer_1$-$Layer_{12}$ represent the Transformer layers. 
We can find that overall these models are able to accurately depict the syntactic relations among tokens. For example, the MCC gets over 70\% in different layers across different models. \revision{All pre-trained models achieve their best performance between the shallow $Layer_2$ and $Layer_4$. With the depth of the layers, the performance decreases gradually.} To be more specific, GraphCodeBERT (GCB) exhibits slightly better performance than the other models on Java250, as evidenced in \figref{ast_org_java250}. However, when it turns to POJ-104, the phenomenon is more complex. CodeT5 demonstrates superior performance compared with other models from $Layer_4$ to $Layer_8$, however, when the layer reaches 12, GraphCodeBERT achieves the best performance. Despite UnixCoder and CodeT5 employing more intricate pre-training strategies, leveraging larger datasets, and having a greater number of trainable parameters, their relative advantage is minimal. \revision{\figref{ast_org_java250_llm} shows MCC scores for CodeLlama, StarCoder and CodeT5+. An interesting observation is that the three large language models achieve the best performance in the shallow layers $Layer_4$ and $Layer_5$, which is similar to the code pre-trained models. But for StarCoder and CodeLlama, the performance increased a lot in the last layer which is different from others.} \revision{We can conclude that the syntax relationship between code tokens is easier to observe in the representation spaces of the shallow hidden layers. The decrease in performance of deeper representations on this task is an interesting phenomenon.
Although deep hidden layers are supposed to contain information from shallower layers, their representation space is more complex, encoding more other code attributes like the token-syntax role, so the syntax-token relationship may be less salient.}

 \subsubsection{Token Syntax Tagging}\label{sec:token-syntax-tagging}
\begin{figure}[!t]
     \centering
 \begin{subfigure}[b]{0.45\textwidth}
         \centering
         \includegraphics[width=\textwidth]{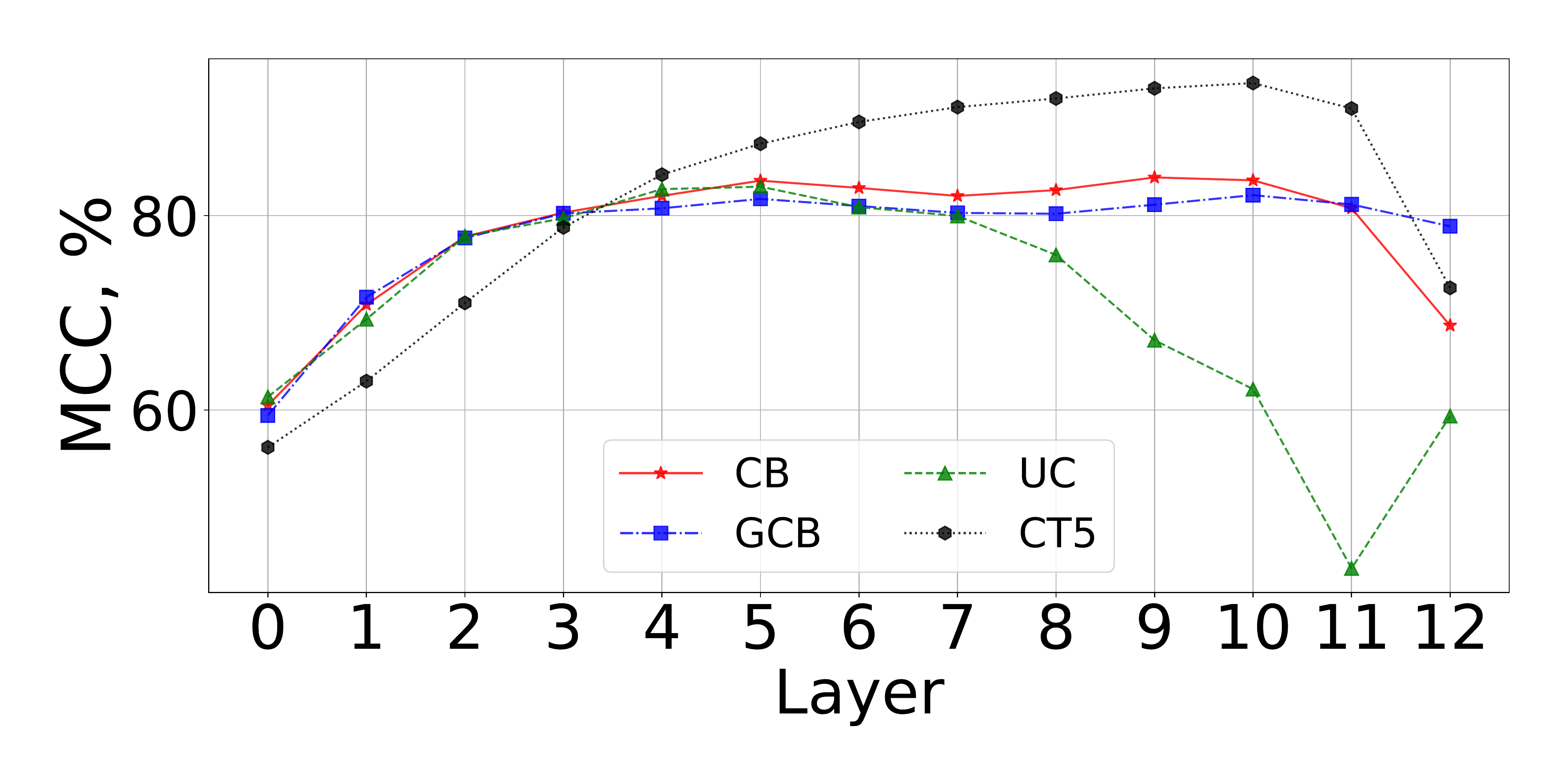}
          \vspace{-2em}
         \caption{Java250-Tagging~(MCC).}
         \label{fig:tagging_org_java250}
     \end{subfigure} 
     \begin{subfigure}[b]{0.45\textwidth}
         \centering
         \includegraphics[width=\textwidth]{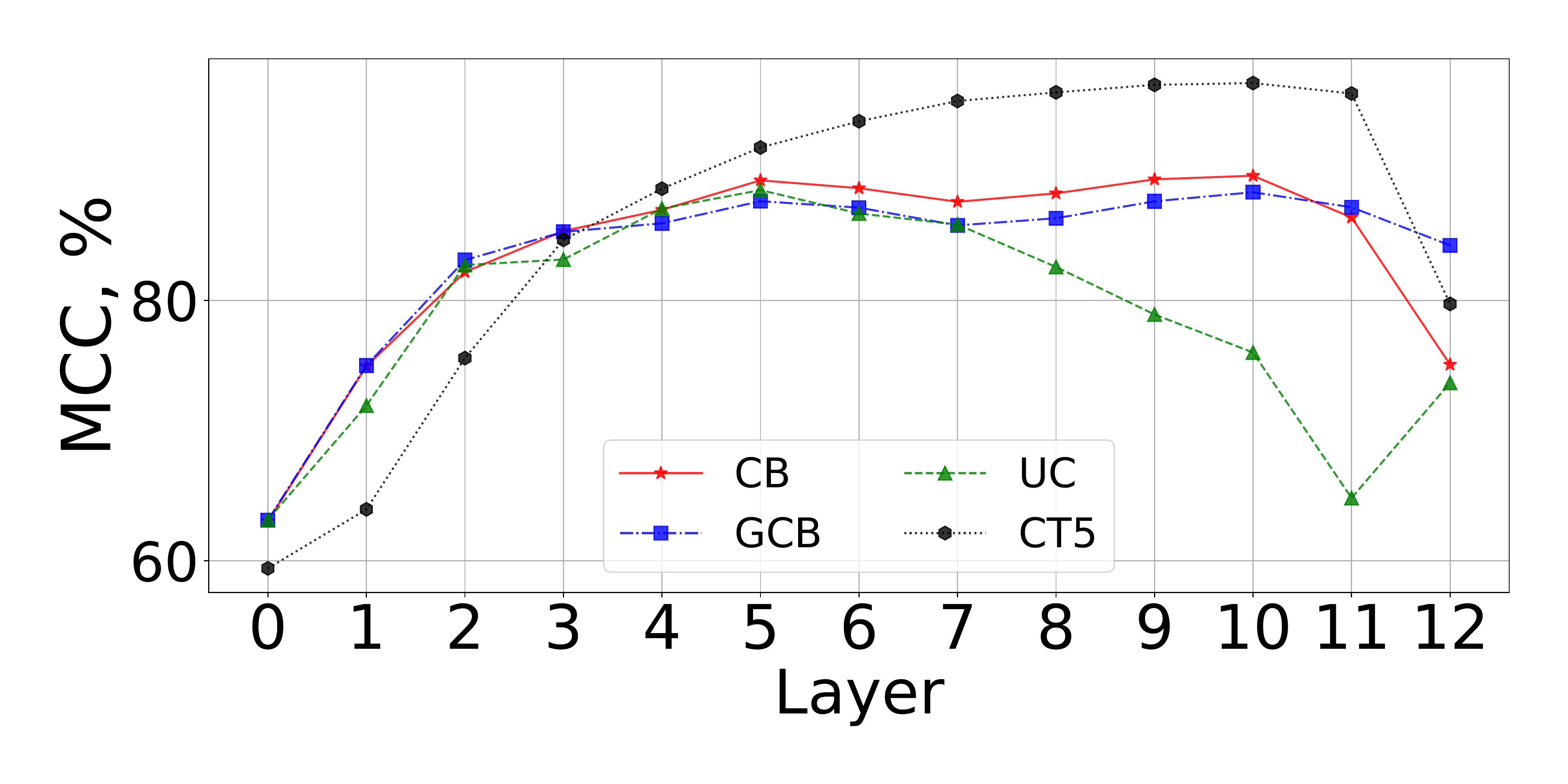}
          \vspace{-2em}
        \caption{POJ-104-Tagging~(MCC).}
         \label{fig:tagging_org_poj}
     \end{subfigure}\\
     \begin{subfigure}[b]{0.45\textwidth}
         \centering
         \includegraphics[width=\textwidth]{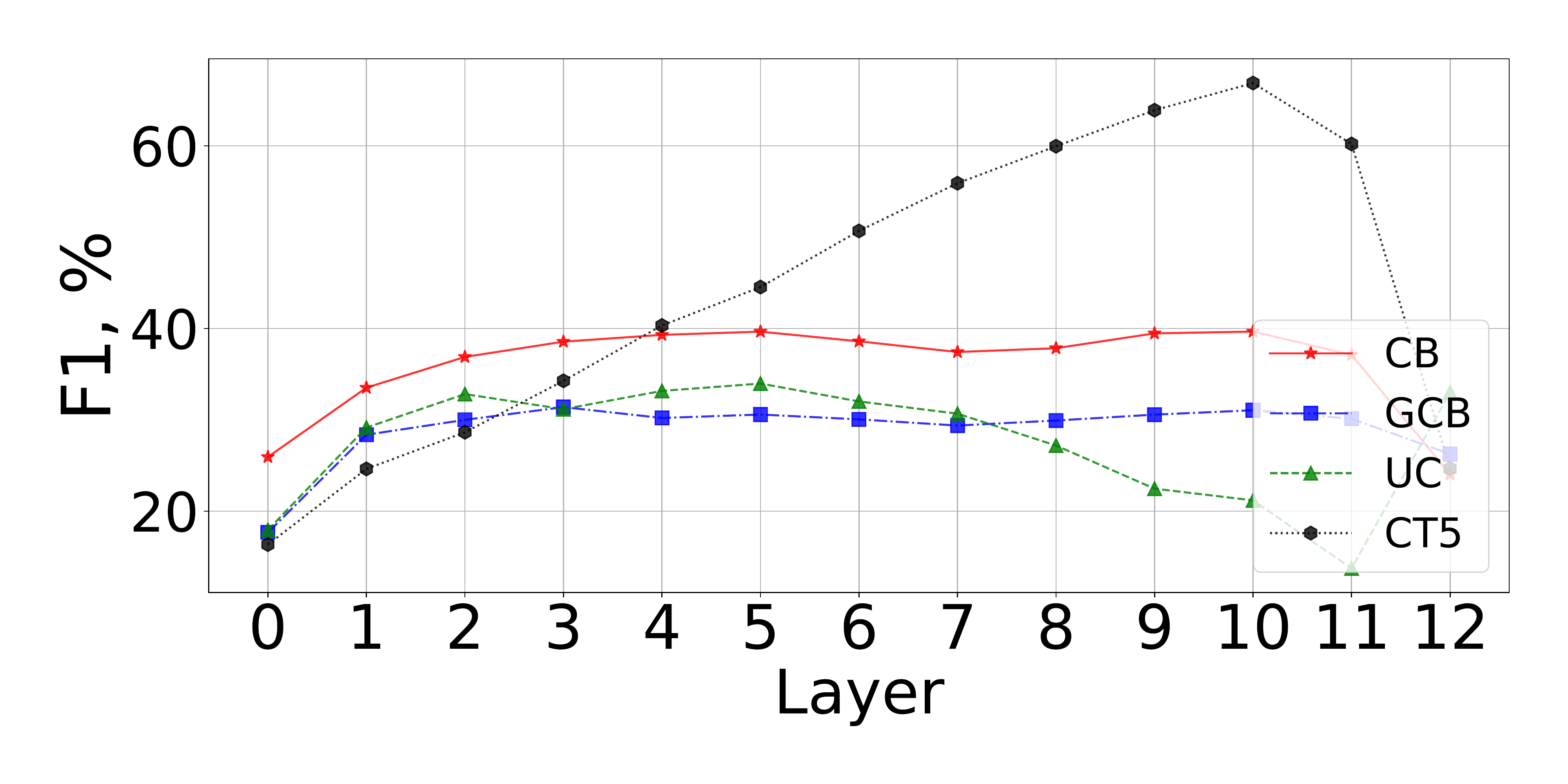}
          \vspace{-2em}
         \caption{Java250-Tagging~(F1).}
         \label{fig:tagging_org_java250_f1}
     \end{subfigure} 
     \begin{subfigure}[b]{0.45\textwidth}
         \centering
         \includegraphics[width=\textwidth]{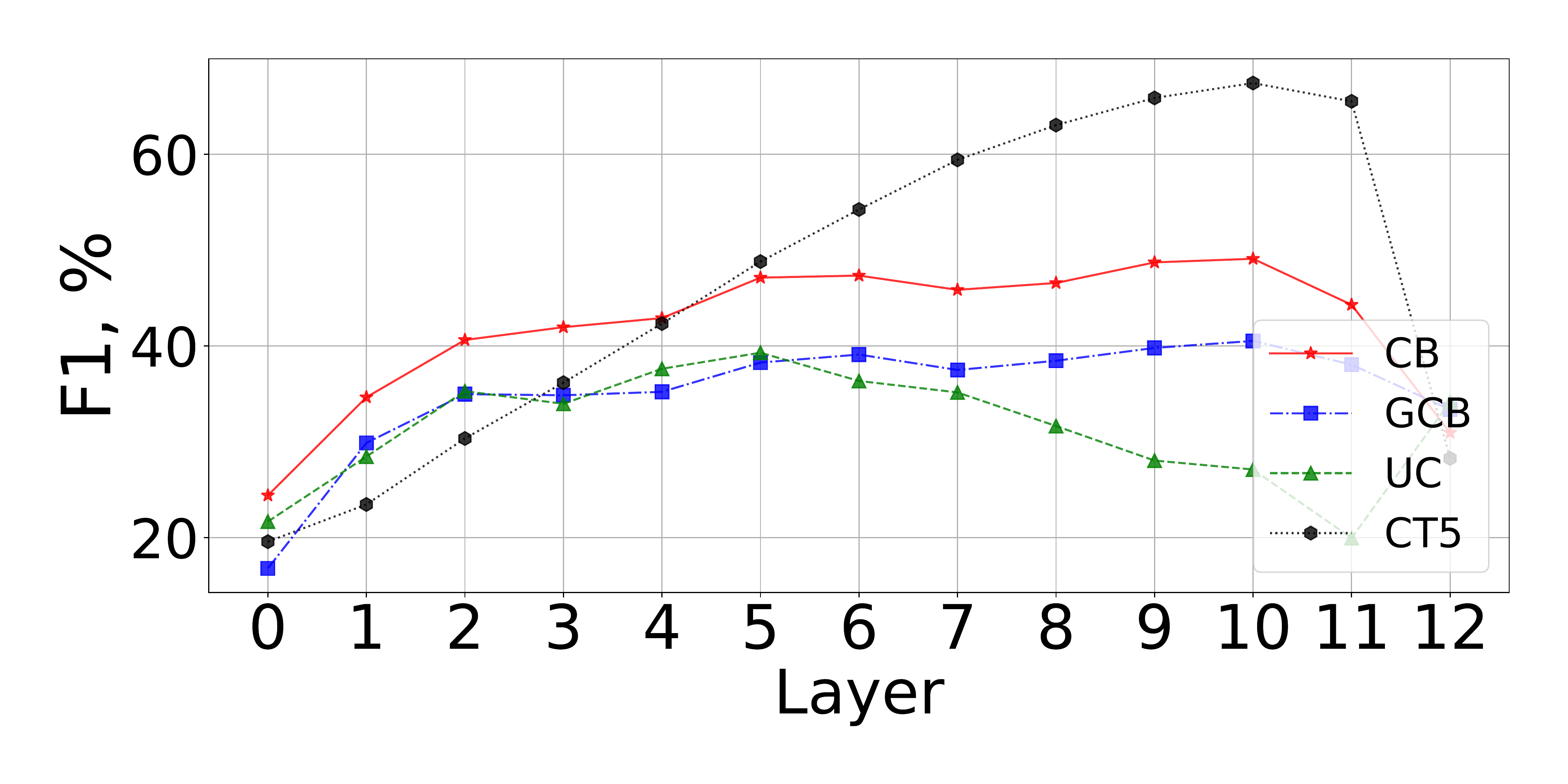}
          \vspace{-2em}
        \caption{POJ-104-Tagging~(F1).}
         \label{fig:tagging_org_poj_f1}
     \end{subfigure}\\
     \vspace*{-0.9em}
        \caption{Performance about Java250 and POJ-104 for Token Syntax Tagging.}
        \Description[Performance about Java250 and POJ-104 for Token Syntax Tagging.]{}
\end{figure}

\begin{figure}[!t]
     \centering
     \begin{subfigure}[b]{0.45\textwidth}
         \centering
         \includegraphics[width=\textwidth]{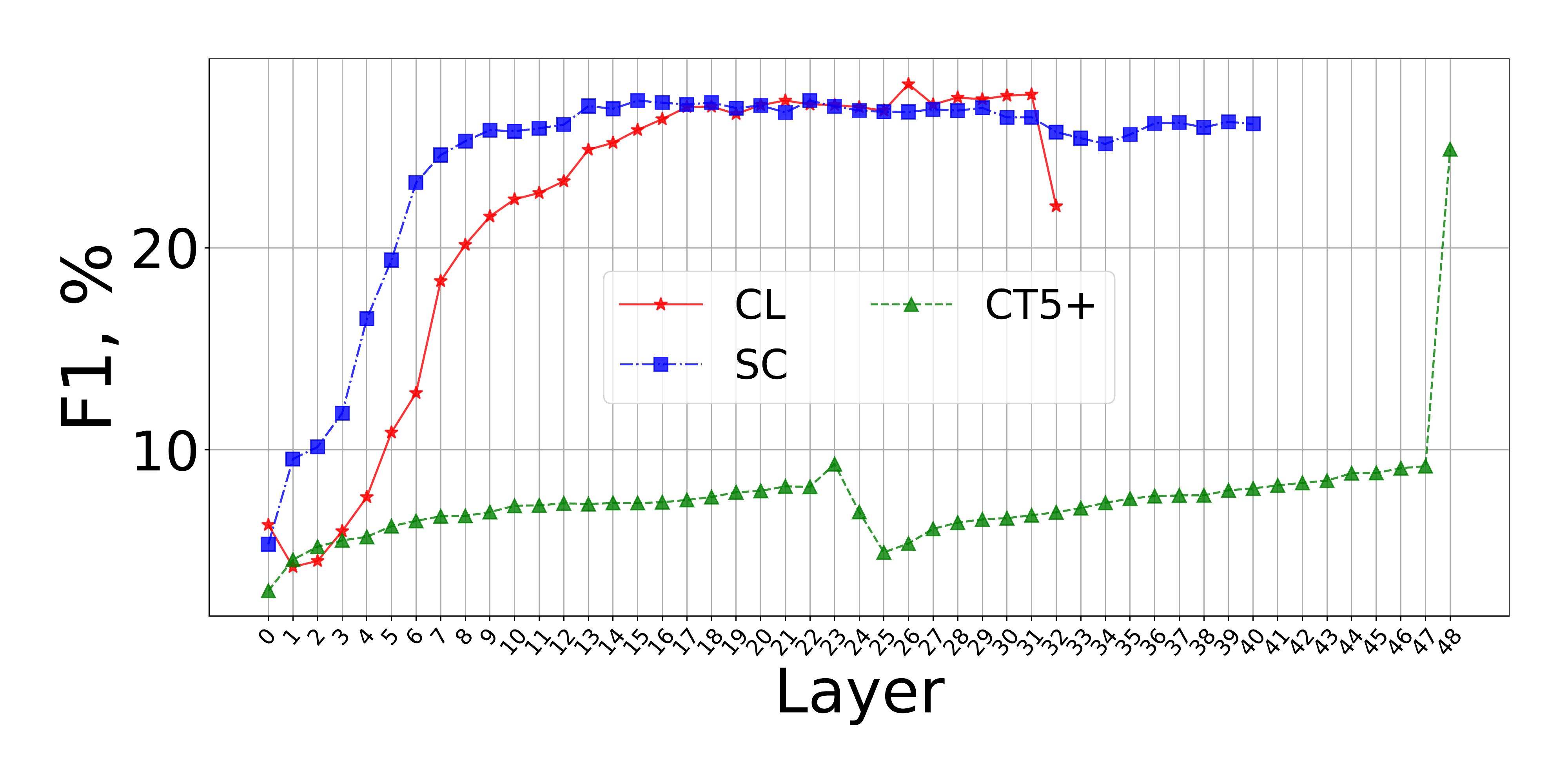}
          \vspace{-2em}
        \caption{Java250-Tagging LLM~(F1).}
         \label{fig:ast_org_java250_f1_llm}
     \end{subfigure} 
    \begin{subfigure}[b]{0.45\textwidth}
         \centering
         \includegraphics[width=\textwidth]{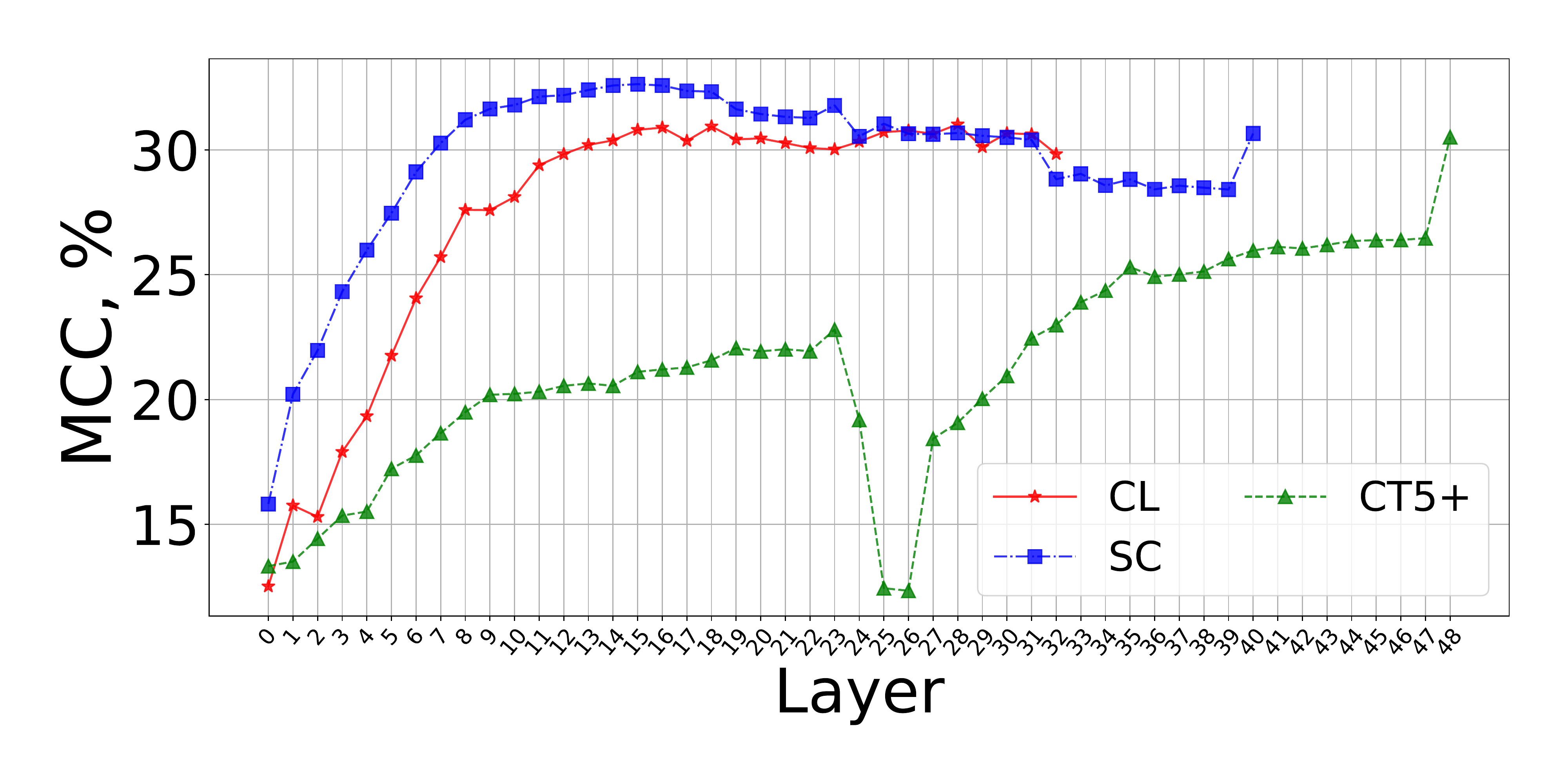}
          \vspace{-2em}
        \caption{Java250-Tagging LLM~(MCC).}
         \label{fig:ast_org_java250_mcc_llm}
     \end{subfigure} 
     \vspace*{-0.9em}
        \caption{Performance about Java250 for Syntax Pair Node prediction~(AST) and Token Syntax Tagging.}
         \Description[Performance about Java250 for Syntax Pair Node prediction~(AST) and Token Syntax Tagging]{}
\end{figure}

\figref{tagging_org_java250} and \figref{tagging_org_poj} present the MCC results of the probing classifiers for the individual token-syntax role (Token Syntax Tagging) on Java250 and POJ-104 respectively. \figref{tagging_org_java250_f1} and \figref{tagging_org_poj_f1} demonstrates the F1 scores. Our observations indicate that CodeT5 outperforms the other models in the middle layers, specifically from $Layer_{4}$ to $Layer_{11}$. CodeBERT exhibits marginally better performance than GraphCodeBERT in these middle layers, while GraphCodeBERT surpasses the other models at the final layer, i.e., $Layer_{12}$. Interestingly, a performance drop is observed for all models except UnixCoder transitioning from $Layer_{11}$ to $Layer_{12}$.
CodeT5 achieves peak performance approximately around $Layer_{10}$. Conversely, UnixCoder lags behind other models, particularly after $Layer_7$. Except for UnixCoder, the performance of other pre-trained models shows an increasing trend with fluctuations from $Layer_0$ to $Layer_{11}$.
\revision{\figref{ast_org_java250_f1_llm} and \figref{ast_org_java250_mcc_llm} shows the F1 score and MCC score for LLMs on Java250 dataset. If we look at all the figures in this task, we find a general trend that the performance improves as the depth of layers increases, and the latter is stable with some drops or ups. Interestingly, at the intersection between the encoder and decoder, the model performance dropped sharply but quickly improved again as the number of layers increased. This phenomenon has also been observed in other experiments. One reasonable explanation is that the encoder and the decoder work with different mechanisms. The encoder encodes information into an abstract representation while the decoder tries to recover the input from the abstract representation. Due to the working difference, the properties in the representation from the interaction layers are challenging to observe when migrating from the encoder to the decoder. As aforementioned, token syntax tagging is a non-trivial task to label the code token with a syntax role, and it is a more difficult task than the previous one. The token syntax role property is more obviously represented in the deeper than the shallow layers. }

Regarding the syntax-relative relationship, \revision{these pre-trained code models} demonstrate comparable proficiency. However, CodeT5~(we consider its encoder) distinguishes itself by more effectively representing the individual syntax-token role within context. Conversely, UnixCoder exhibits a lower degree of syntactic understanding compared to the other models. We speculate this for the following two reasons: 1) CodeBERT, GraphCodeBERT, and CodeT5-encoder all employ masked language modeling (MLM) or similar ways to learn bi-directional token features while UniXCoder incorporates different types of pre-training tasks. Although it also used MLM, the other tasks such as unidirectional language modeling (ULM) and DeNoiSing (DNS) may have a detrimental impact on probing tasks; 2) UnixCoder integrates the encoder and the decoder by sharing weights while the representations from the decoder have a worse ability in information extraction than the encoder representation~\cite{troshin2022probing}. Sharing weights between the encoder and decoder may harm the ability of the model to extract information. \revision{The three studied LLMs show some differences from the four pre-trained models.  LLMs do not have obvious advantages over the pre-trained models, although StarCoder has 6144 hidden dimensions, CodeLlama has 4096 hidden dimensions, and CodeT5p-770m has 1046 hidden dimensions.  Their dimensions are much higher than the pre-trained code models~(768).  All of them have a huge number of parameters.  One possible reason is that the representation space dimension is too large and covers too much information, making the embedded syntax features difficult to observe.}

In summary, from Section~\ref{sec:syntax-pair-node-prediction} and Section~\ref{sec:token-syntax-tagging}, it is evident that the \revision{four} code pre-trained models and \revision{3 large language models} can encode the code syntax, both in terms of syntax relationship between tokens (Syntax Pair Node prediction) and the individual token-syntax role (Token Syntax Tagging). \revision{Different syntax features have different degrees of difficulty in being observed at shallow and deep hidden layers. }

\subsection{Semantic Analysis}
\subsubsection{Semantic Relation Prediction}\label{sec:semantic-relation-prediction}
\begin{figure}[]
     \centering
     \begin{subfigure}[b]{0.45\textwidth}
         \centering
         \includegraphics[width=\textwidth]{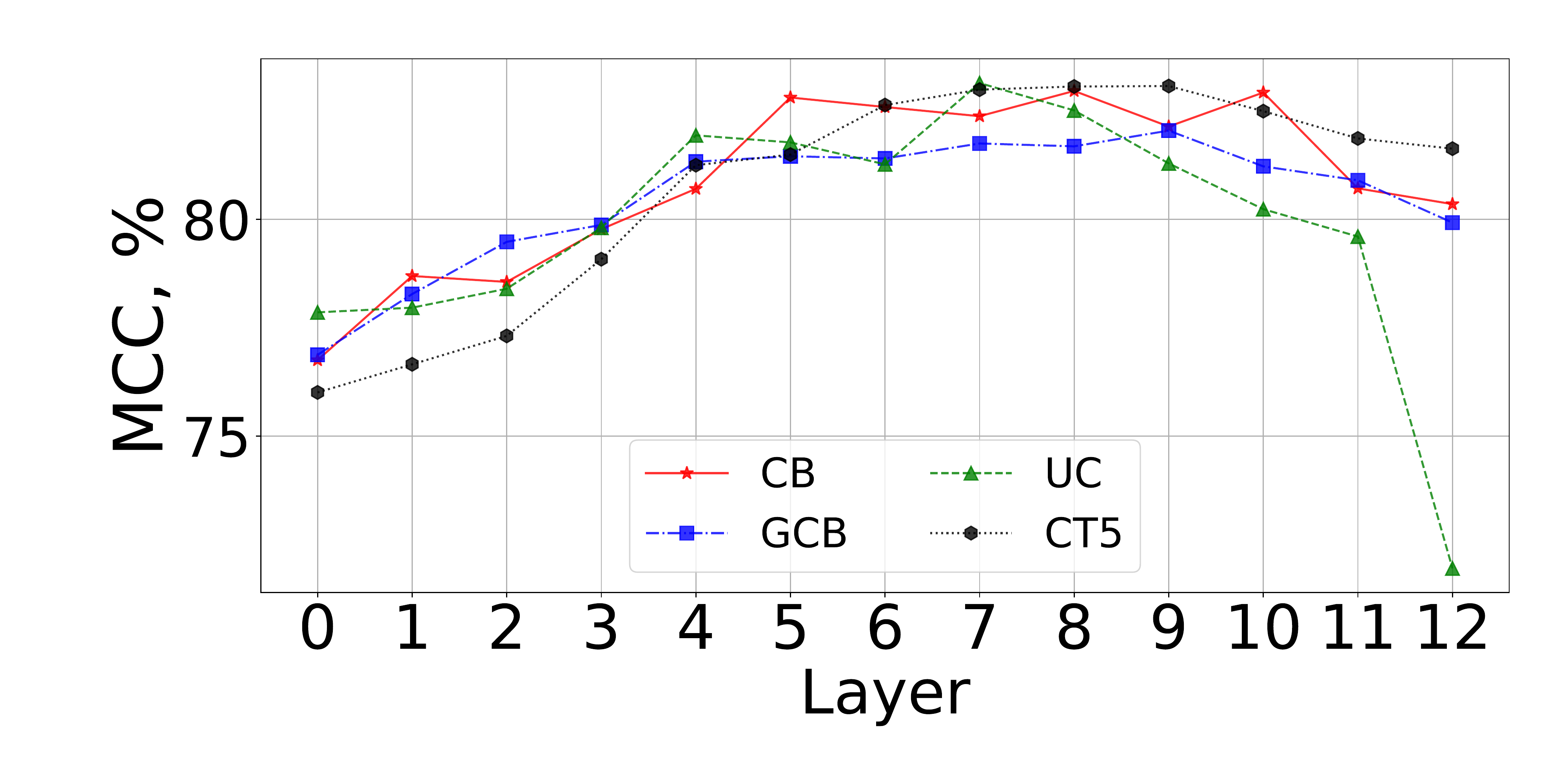}
          \vspace*{-2em}
         \caption{Java250-CDG.}
         \label{fig:cdg_org_java250}
     \end{subfigure}
     \begin{subfigure}[b]{0.45\textwidth}
         \centering
         \includegraphics[width=\textwidth]{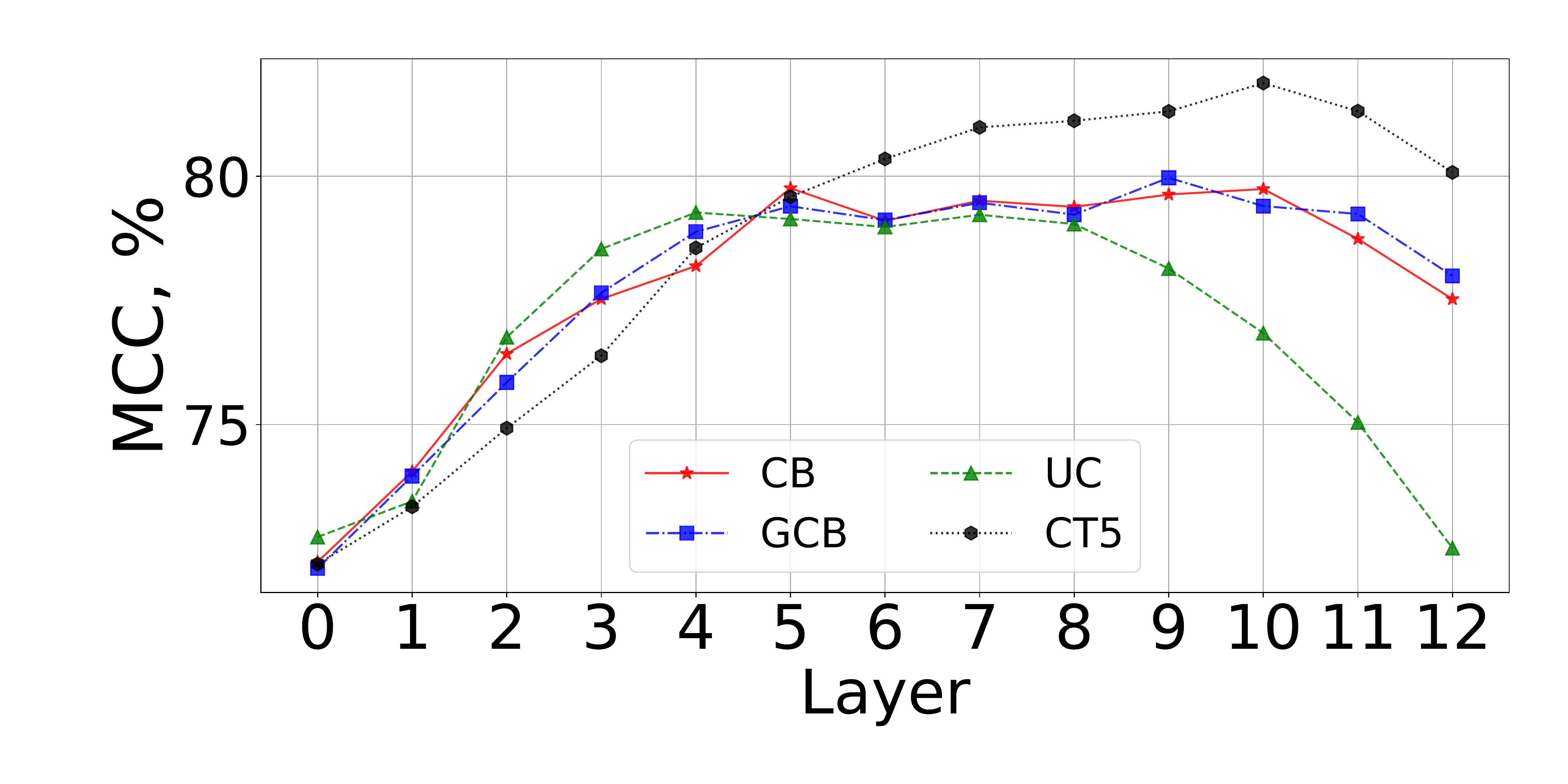}
          \vspace*{-2em}
         \caption{POJ-104-CDG.}
         \label{fig:cdg_org_poj}
     \end{subfigure} 
     \begin{subfigure}[b]{0.45\textwidth}
         \centering
         \includegraphics[width=\textwidth]{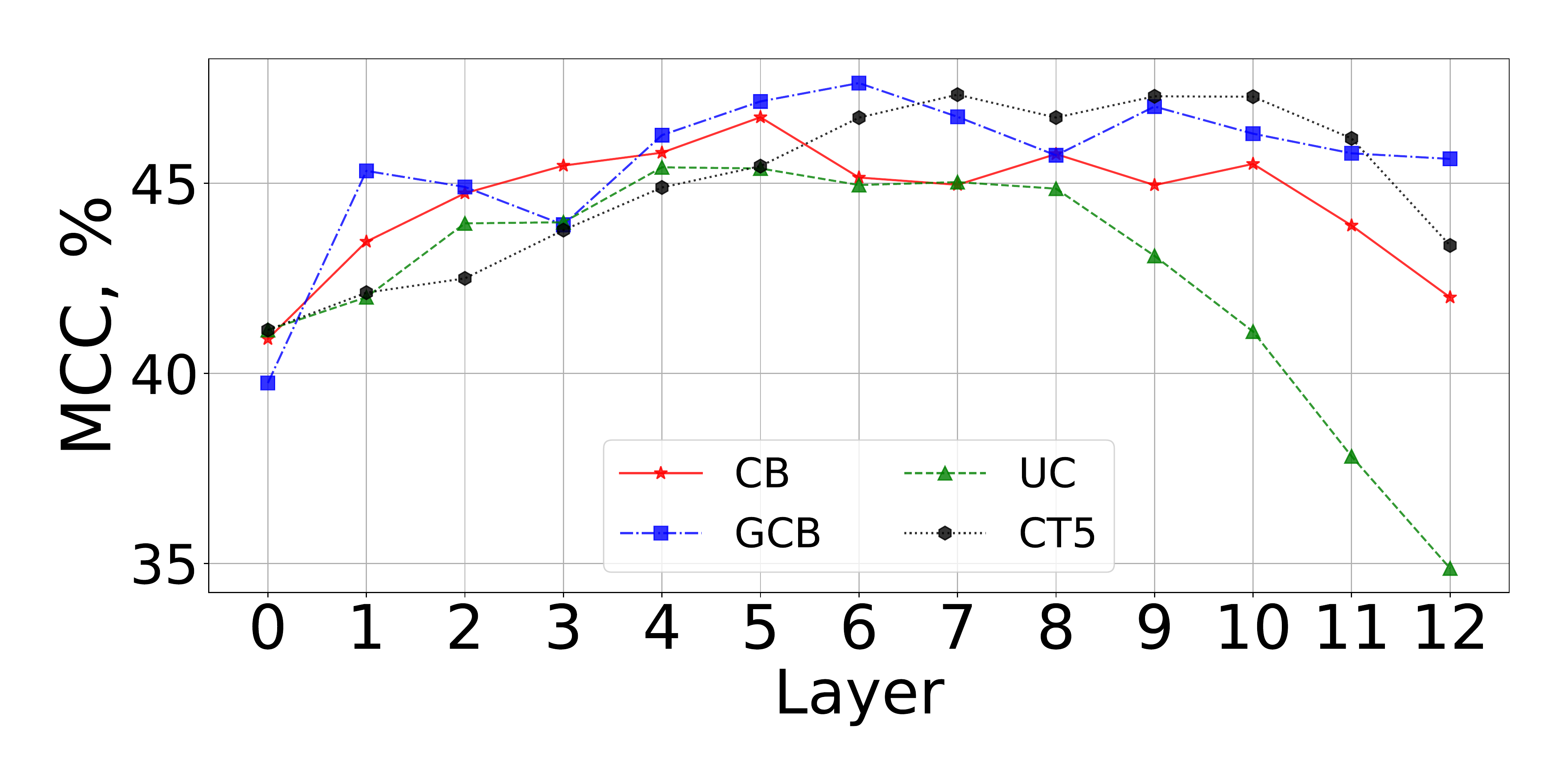}
          \vspace*{-2em}
         \caption{Java250-CFG.}
         \label{fig:cfg_org_java250}
     \end{subfigure} 
     \begin{subfigure}[b]{0.45\textwidth}
         \centering
         \includegraphics[width=\textwidth]{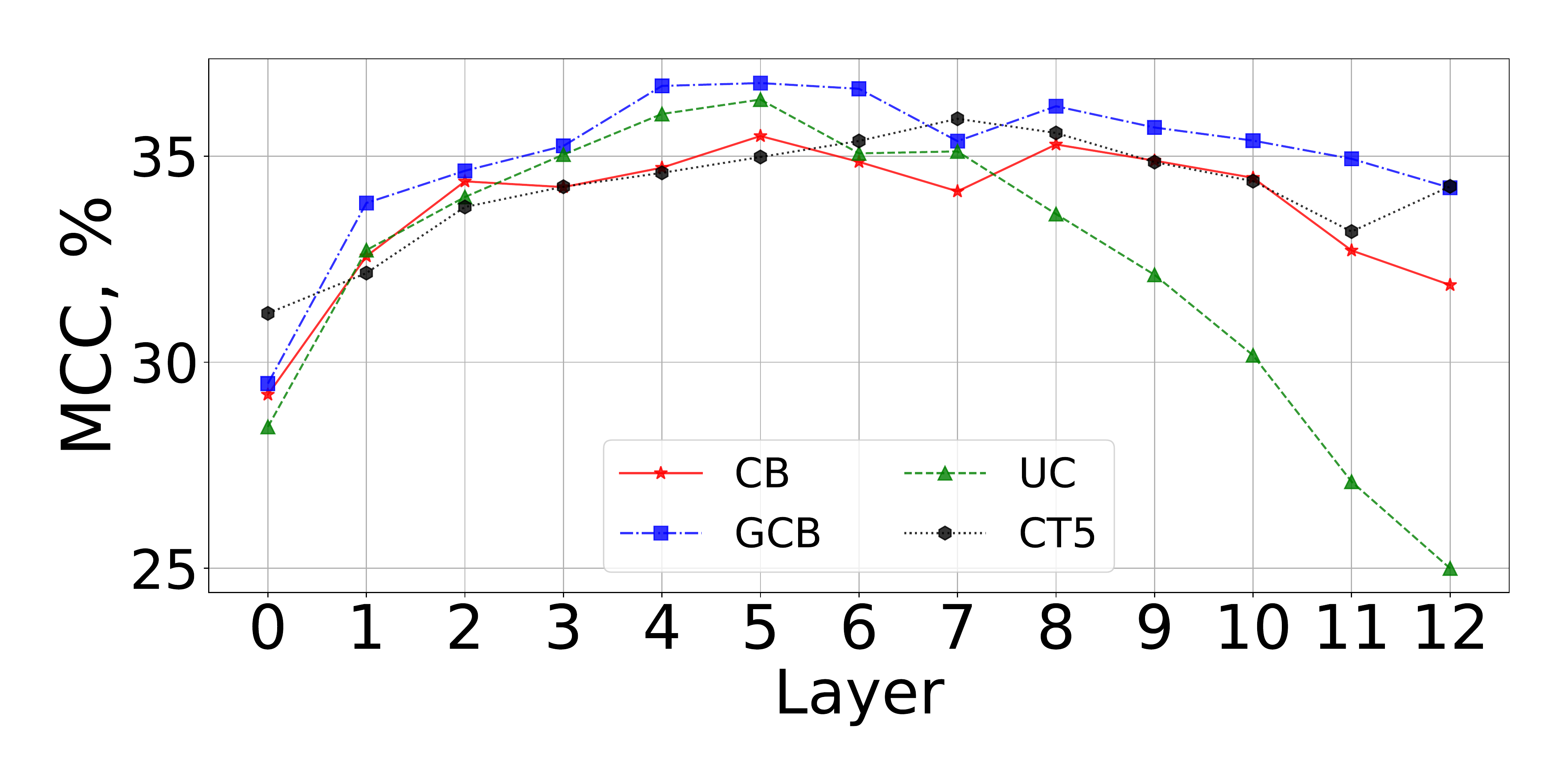}
          \vspace*{-2em}
         \caption{POJ-104-CFG.}
         \label{fig:cfg_org_poj}
     \end{subfigure} 
     \begin{subfigure}[b]{0.45\textwidth}
         \centering
         \includegraphics[width=\textwidth]{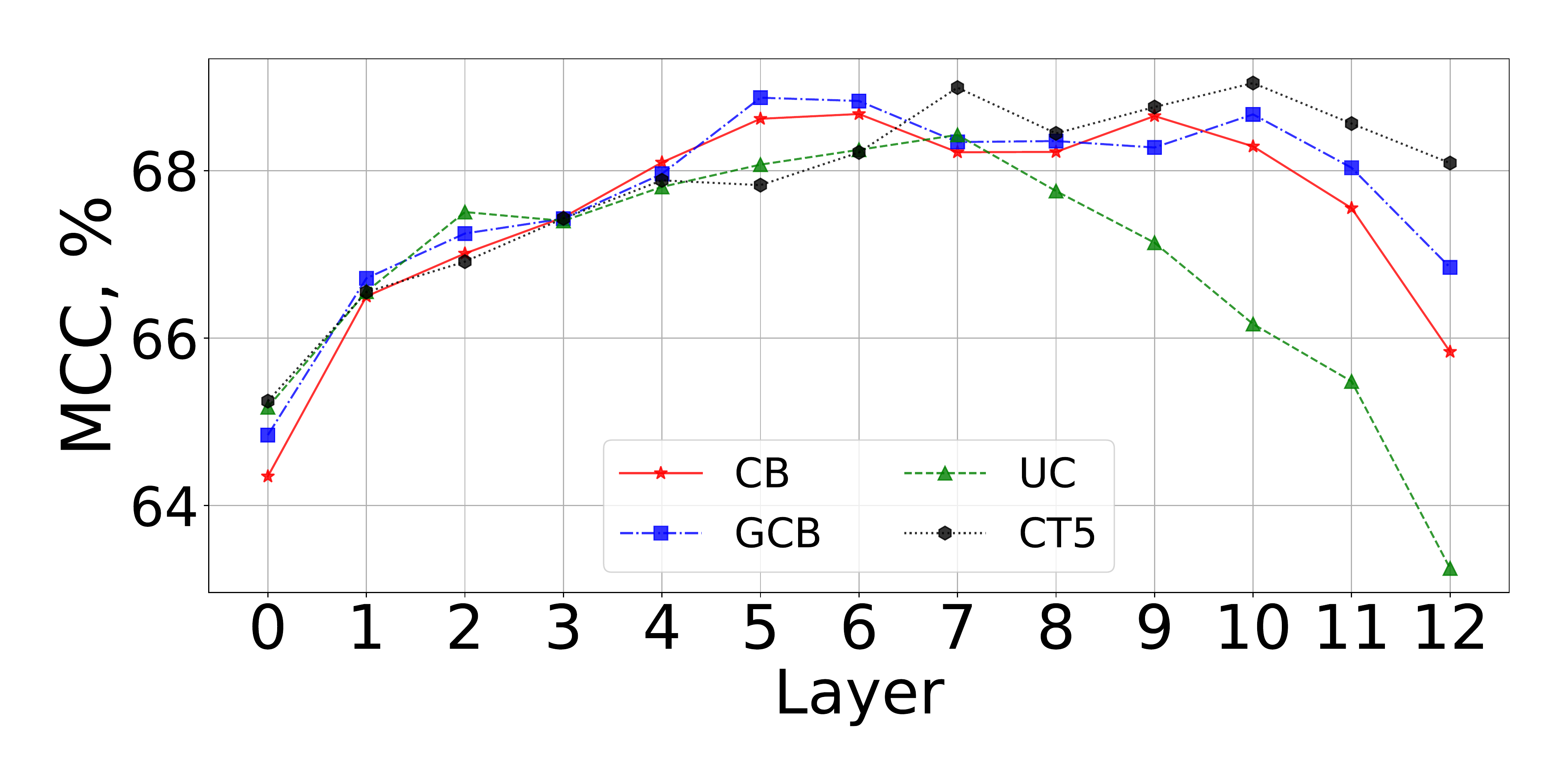}
          \vspace*{-2em}
         \caption{Java250-DDG.}
         \label{fig:ddg_org_java250}
     \end{subfigure} 
     \begin{subfigure}[b]{0.45\textwidth}
         \centering
         \includegraphics[width=\textwidth]{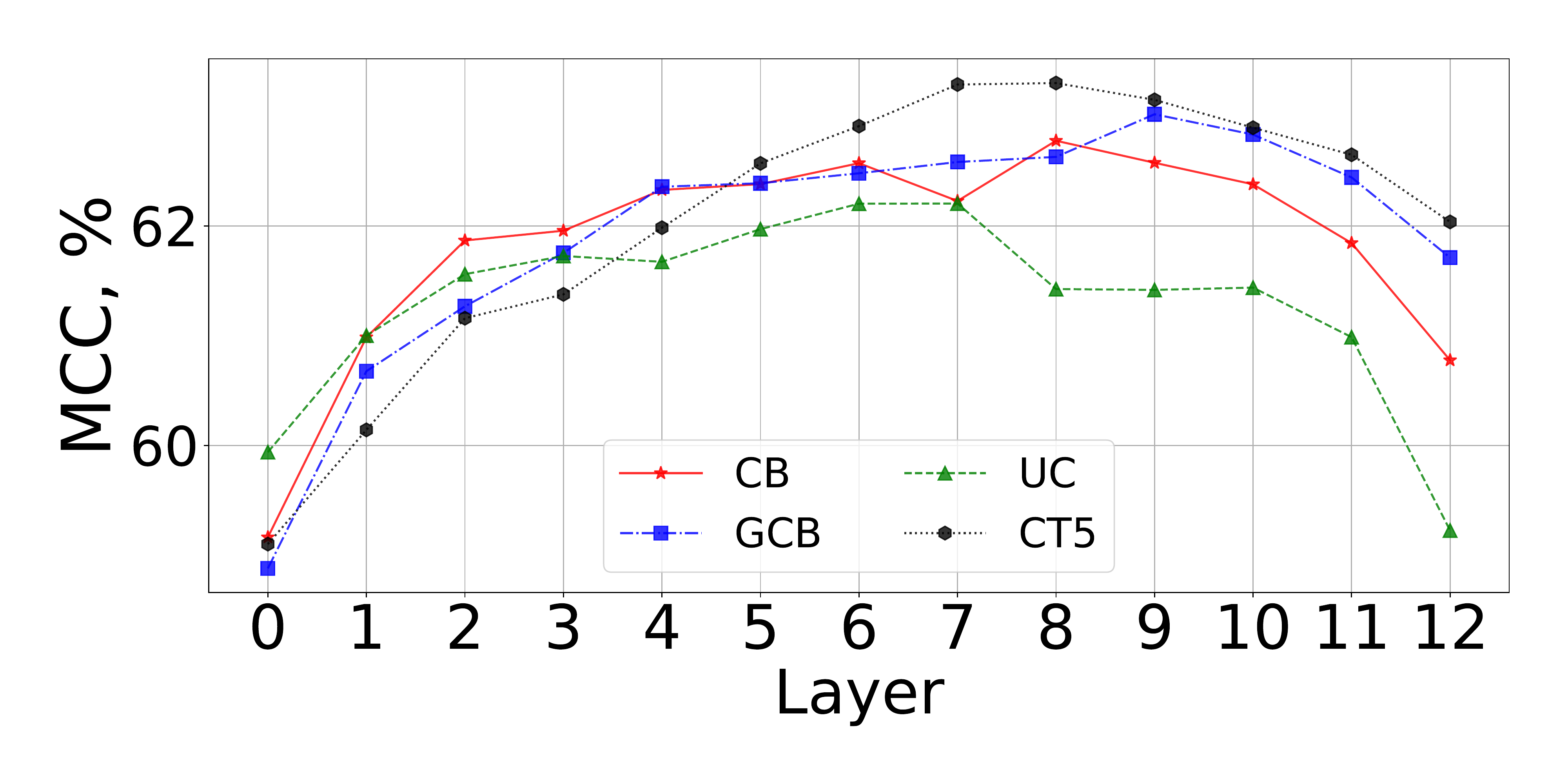}
          \vspace*{-2em}
         \caption{POJ-104-DDG.}
         \label{fig:ddg_org_poj}
     \end{subfigure}
     \vspace*{-0.9em}
        \caption{Performance~(MCC) about Java250 and POJ-104 for Semantic Relation.}
        \Description[Performance about Java250 and POJ-104 for Semantic Relation.]{}
        \label{fig:performance_org_semantics}
\end{figure}

\figref{performance_org_semantics} presents the probing performance of \revision{the four pre-trained code models,} CodeBERT~(CB), GraphCodeBERT~(GCB), UnixCoder~(UC), and CodeT5~(CT5) on the semantic relation prediction task. We conducted a comparison of the probing performance across three different program semantics: control dependency (CDG, \figref{cdg_org_java250} and \figref{cdg_org_poj}), control flow information (CFG, \figref{cfg_org_java250} and \figref{cfg_org_poj}), and data dependency (DDG, \figref{ddg_org_java250} and \figref{ddg_org_poj}). Firstly, it is evident that CodeT5 demonstrates the highest proficiency in understanding program semantics, i.e., CDG and DDG, especially in the last few layers, and GraphCodeBERT proves to be a little better than CodeT5 in terms of CFG. Secondly, CodeBERT is also capable of encoding program semantics, despite not utilizing data flow information during pre-training. For example, the MCC reaches over 60\% in CDG and DDG. It indicates that the pre-training task Masked Language Modeling (MLM) is able to aid the model in learning code semantics. 
UnixCoder performs less effectively in representing code semantics compared to the other three models, particularly after $Layer_{7}$. 

\begin{figure}[]
     \centering
     \begin{subfigure}[b]{0.45\textwidth}
         \centering
         \includegraphics[width=\textwidth]{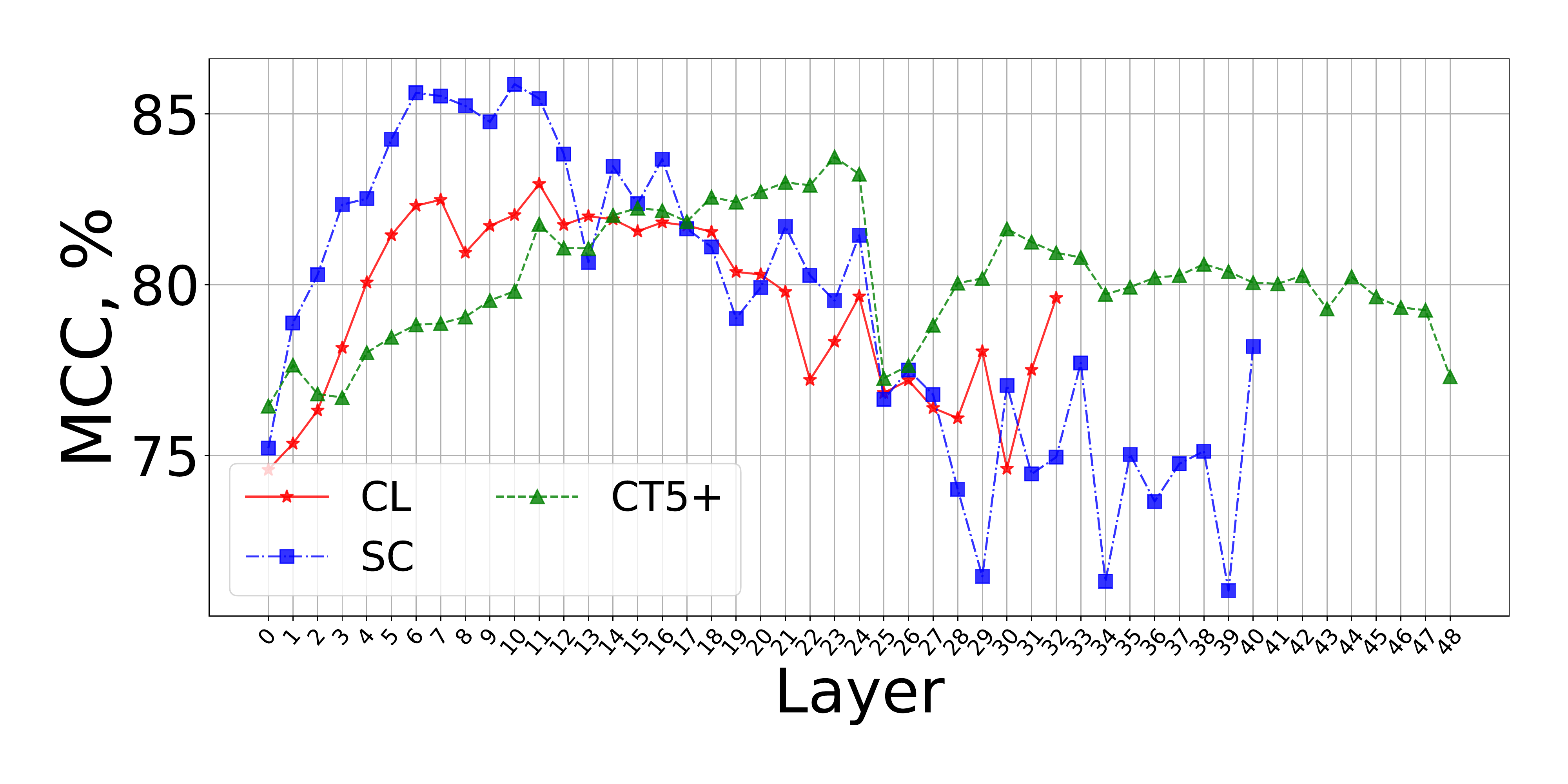}
          \vspace*{-2em}
         \caption{Java250-CDG.}
         \label{fig:cdg_org_java250_llm}
     \end{subfigure}
     \begin{subfigure}[b]{0.45\textwidth}
         \centering
         \includegraphics[width=\textwidth]{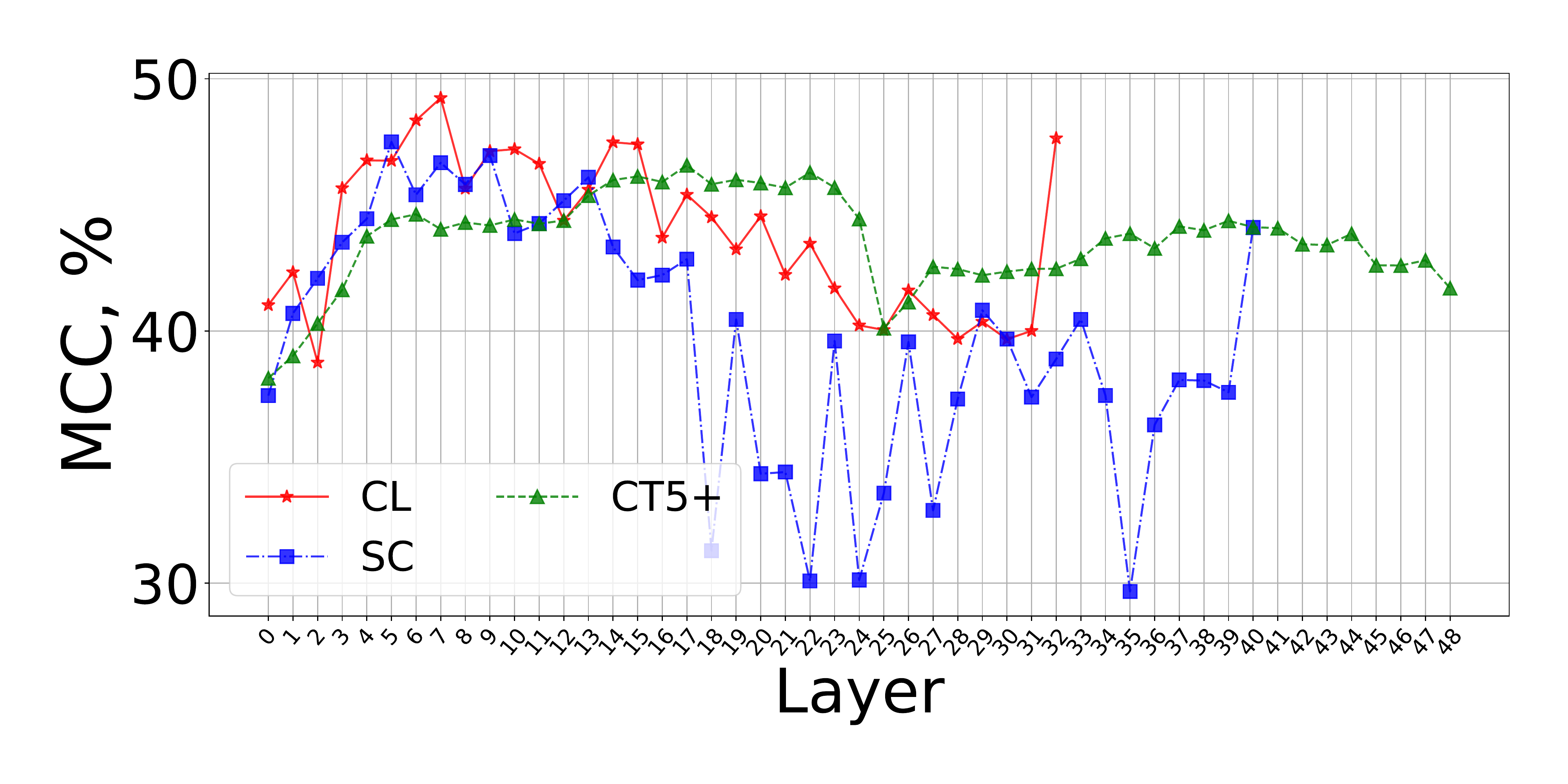}
          \vspace*{-2em}
         \caption{Java250-CFG.}
         \label{fig:cfg_org_java250_llm}
     \end{subfigure} 
     \begin{subfigure}[b]{0.45\textwidth}
         \centering
         \includegraphics[width=\textwidth]{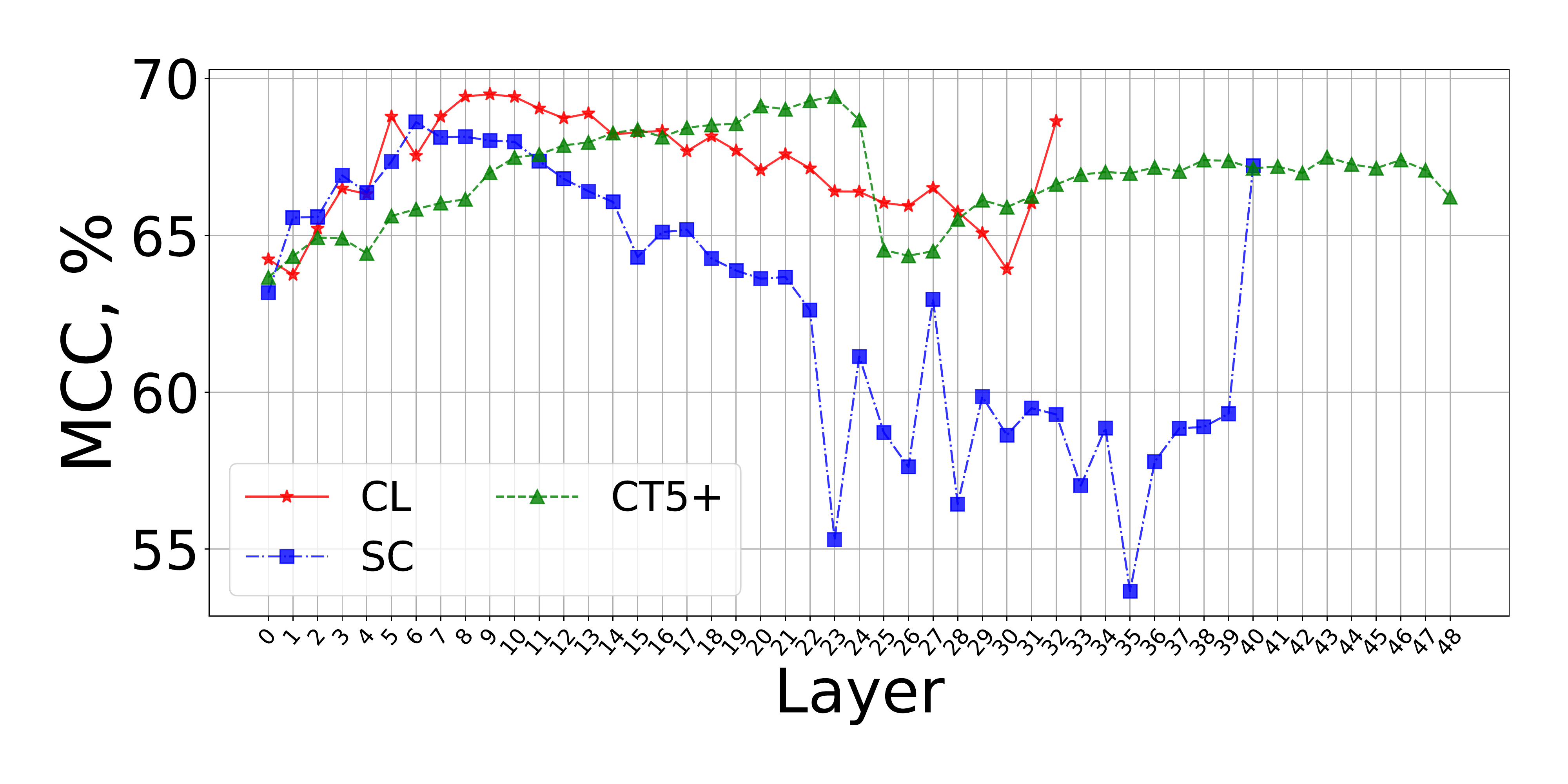}
          \vspace*{-2em}
         \caption{Java250-DDG.}
         \label{fig:ddg_org_java250_llm}
     \end{subfigure} \\
     \vspace*{-0.9em}
        \caption{LLM Performance~(MCC) about Java250 for Semantic Relation.}
        \Description[LLM Performance about Java250 for Semantic Relation.]{}
\label{fig:performance_org_semantics_llm}
\end{figure}

\revision{\figref{performance_org_semantics_llm} demonstrates the probing performance of three LLMs, CodeLlama, StarCoder and CodeT5+, on the three code semantics understanding tasks, CDG, CFG and DDG. Compared with the code pre-trained model, the representation from the shallow layer of LLM makes it easier to observe the three semantic information of the code. When we compared StarCoder with CodeLlama and CodeT5+, we found that the performance of StarCoder dropped significantly for the deep layers. Since the current LLMs have emergency ability, we can think LLMs have learned the three code semantics while they are not easily observed. Furthermore, how to induce the code semantic representation of LLM to be more easily observed in prompting LLM, thereby improving the quality of model answers, is a topic worthy of study in the future.}

When comparing the performance of the code models across CDG, CFG, and DDG, it is clear that the code models exhibit the lowest MCCs on CFG, higher MCCs on DDG, and the highest MCCs on CDG, respectively. It indicates that the code models struggle to capture control flow semantics (CFG) as effectively as the other two semantic types. \revision{CFG is an approximation execution trace of the program. LLMs are shown to lack the ability to handle the tasks related to code execution~\cite{ma2023scope}. Model trainers need to consider how to improve the understanding of code dynamic behaviour of LLMs.} 

\subsubsection{Semantic Propagation (inGraph)}\label{sec:semantic-propagation}
\begin{figure}[]
     \centering
     \begin{subfigure}[b]{0.4\textwidth}
         \centering
         \includegraphics[width=\textwidth]{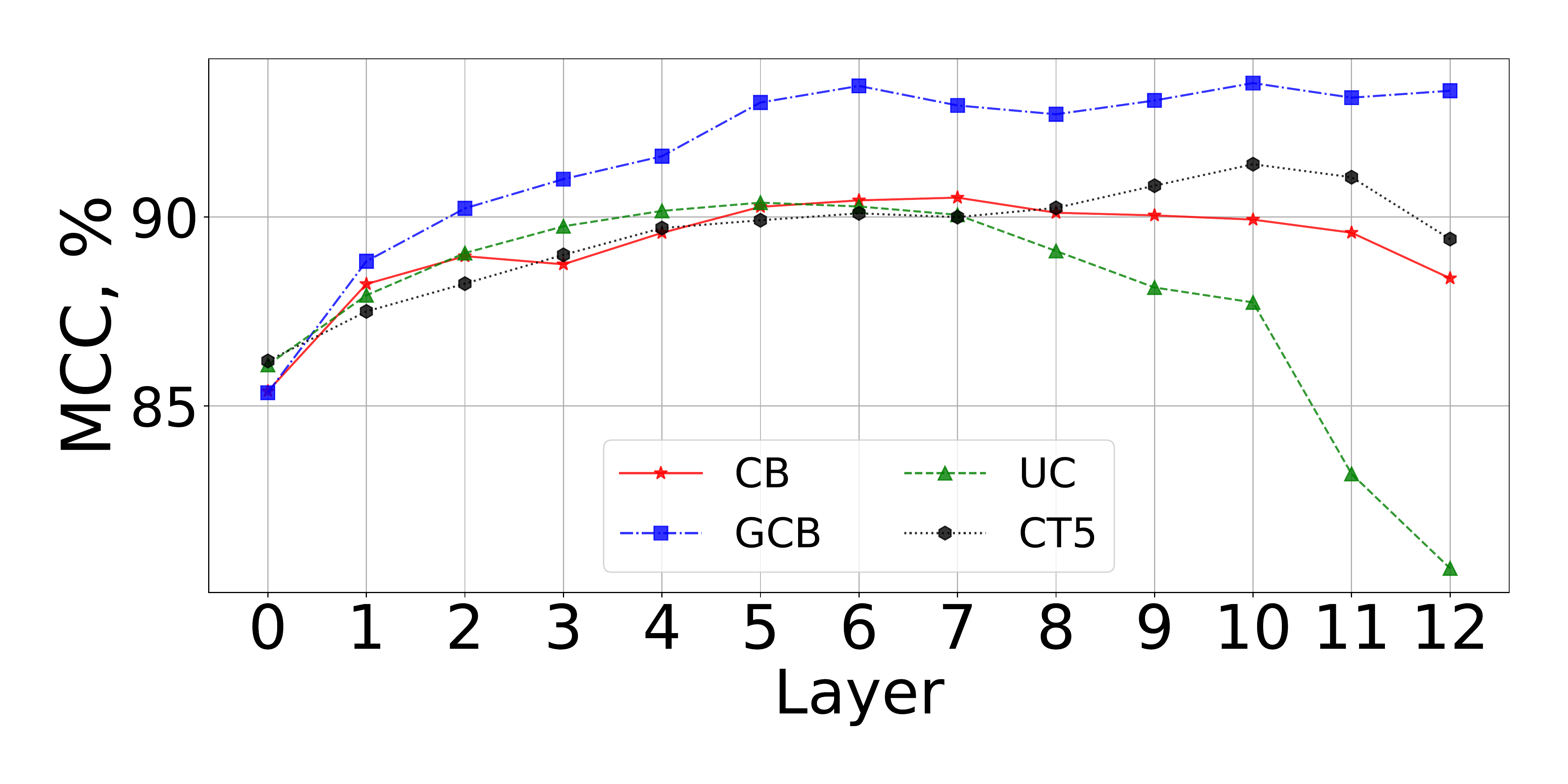}
         \vspace*{-2em}
         \caption{Java250-CDG.}
         \label{fig:ingraph_cdg_java250}
     \end{subfigure} 
     \begin{subfigure}[b]{0.4\textwidth}
         \centering
         \includegraphics[width=\textwidth]{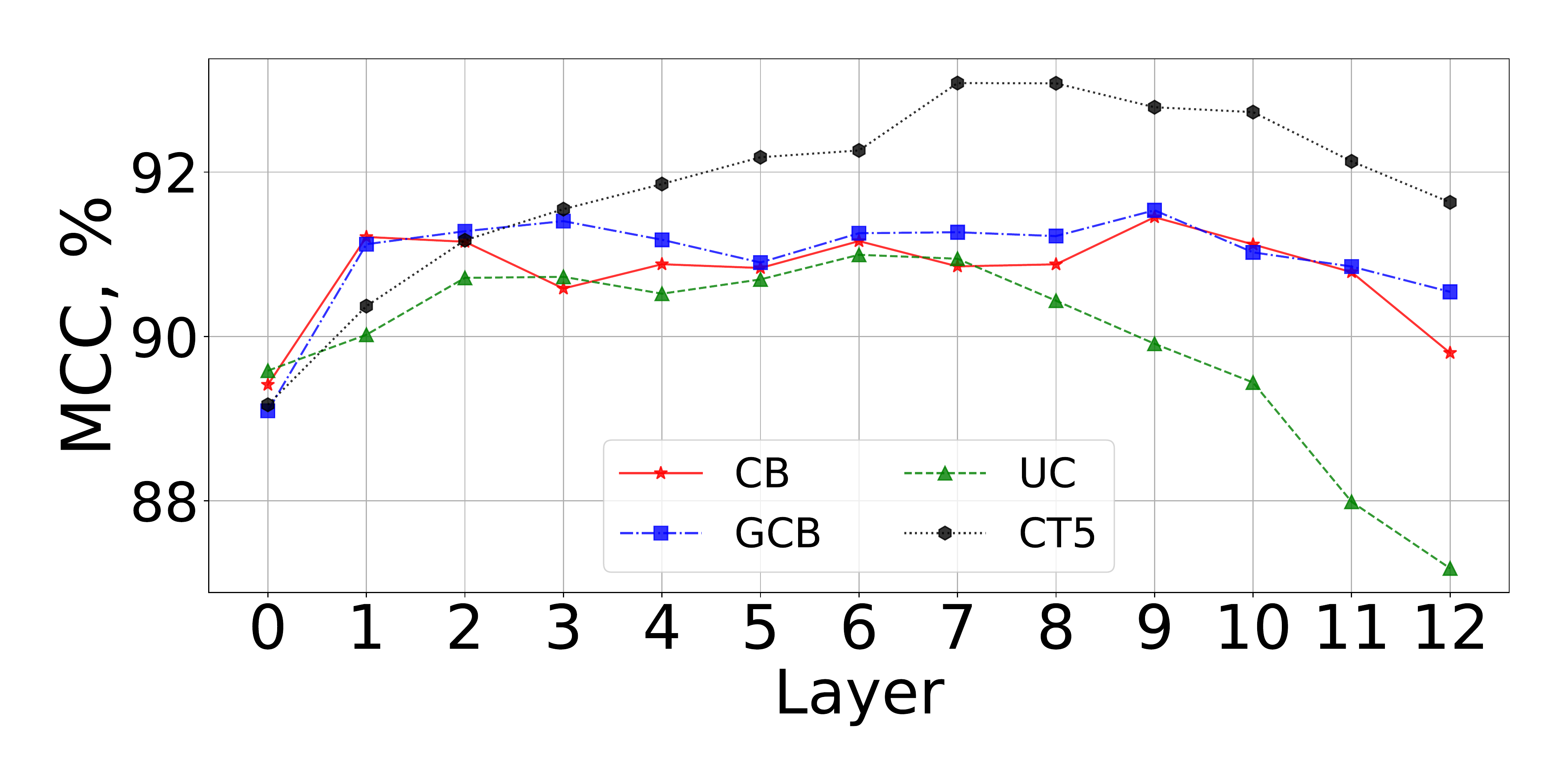}
          \vspace*{-2em}
         \caption{POJ-104-CDG.}
         \label{fig:ingraph_cdg_poj}
     \end{subfigure}  \hspace*{-0.9em} \\
      \begin{subfigure}[b]{0.4\textwidth}
         \centering
         \includegraphics[width=\textwidth]{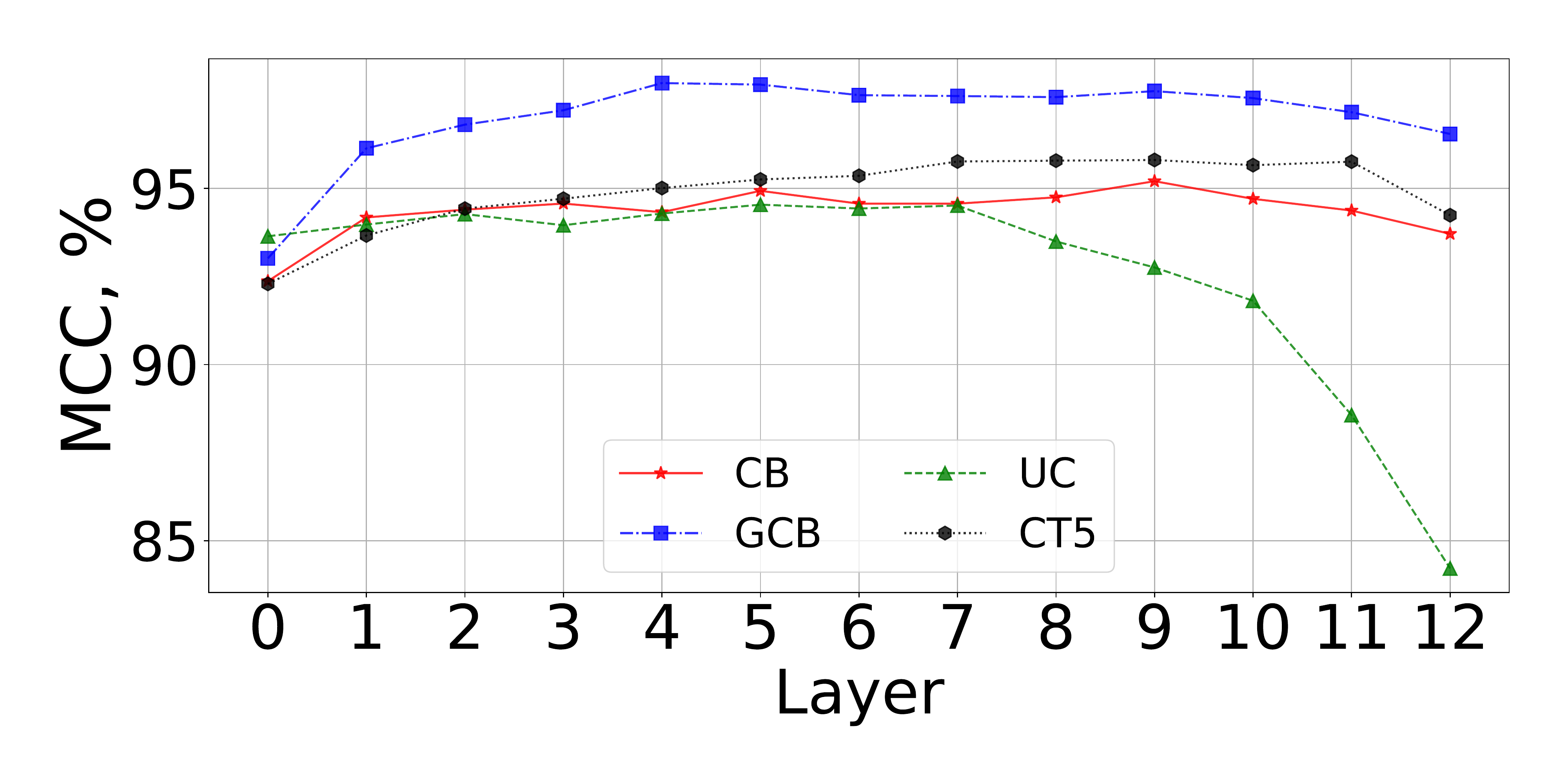}
          \vspace*{-2em}
        \caption{Java250-DDG.}
         \label{fig:ingraph_ddg_java250}
     \end{subfigure}     
     \begin{subfigure}[b]{0.4\textwidth}
         \centering
         \includegraphics[width=\textwidth]{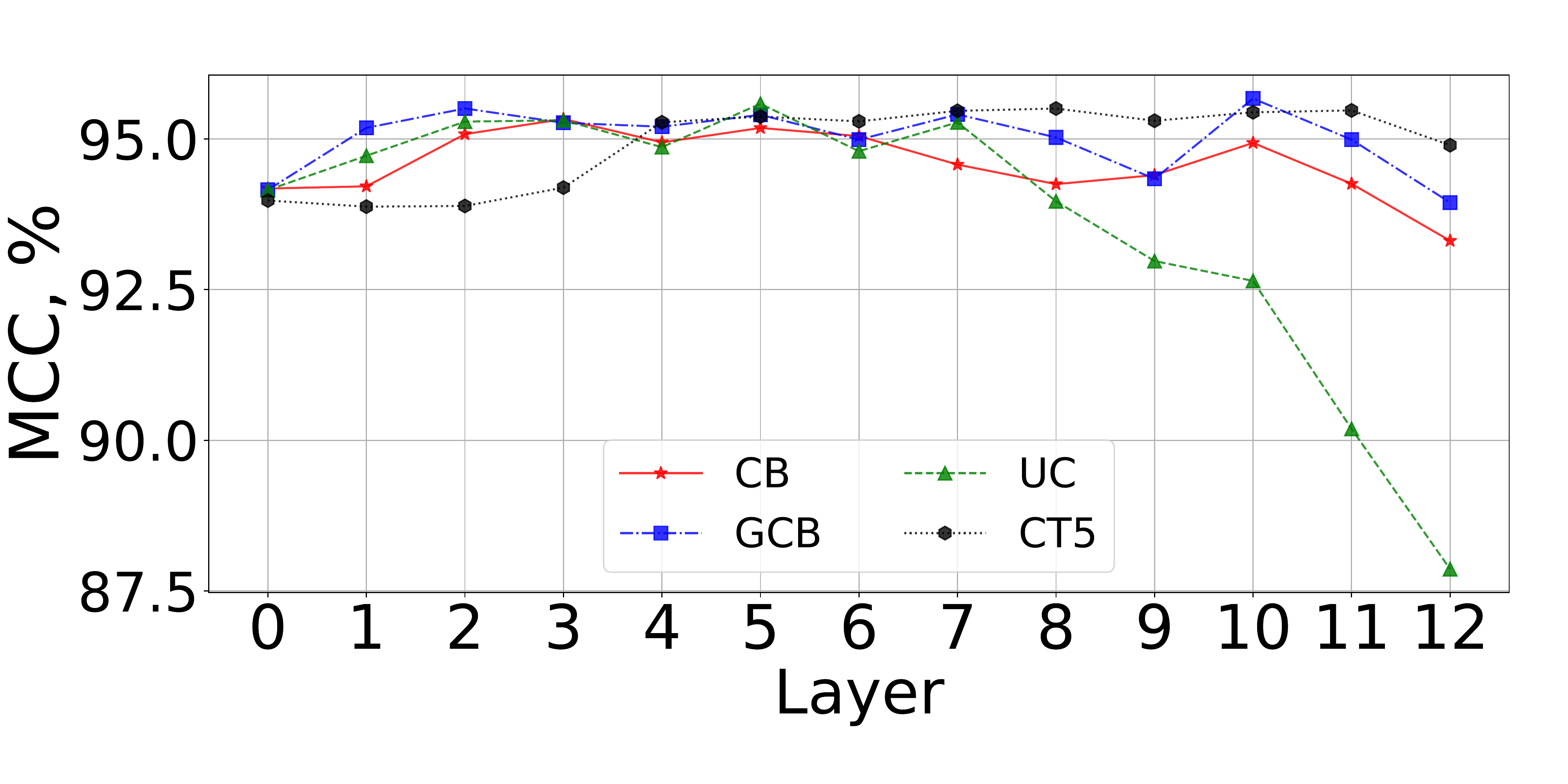}
          \vspace*{-2em}
        \caption{POJ-104-DDG.}
         \label{fig:ingraph_ddg_poj}
     \end{subfigure} \hspace*{-0.9em} \\
     \vspace*{-0.9em}
        \caption{Performance~(MCC) about Java250 and POJ-104 for inGraph.}
        \label{fig:performance_isingraph}
        \Description[Performance about Java250 and POJ-104 for inGraph.]{}
        \vspace*{-2mm}
\end{figure}

\figref{performance_isingraph} illustrates the probing results of four \revision{pre-trained} models on the inGraph task. It is evident that GraphCodeBERT outperforms the other models in terms of semantic propagation for CDG and DDG on the Java250 dataset. CodeT5 is better on POJ-104 about CDG. However, both GraphCodeBERT and CodeT5 exhibit similar performance on POJ-104 about DDG. This suggests that the data propagation employed by GraphCodeBERT proves beneficial in capturing the semantic propagation within the code. Furthermore, we observe that CodeBERT can encode code semantics, albeit with lower performance compared to CodeT5 and GraphCodeBERT. 
However, UnixCoder performs worse than the other models, especially after $Layer_7$. When comparing the performance of the models between CDG and DDG,  we can find that models exhibit higher MCCs on DDG than CDG, which means that data dependency propagation is encoded better than control dependency propagation by these 4 pre-trained code models. \revision{For the LLMs, we observe that the shallow layers have better performance than the deep layers. For CodeT5+, this phenomenon is even more apparent and has a sharp drop between the encoder and the decoder ($Layer_{24}$ and $Layer_{25}$). One possible explanation is that the ability of code representations for the decoder to express long dependencies is diminished because the decoder can only see the previous token information, and the future tokens are masked. If we see \figref{performance_org_semantics_llm} for the short dependency~(constructing the semantic graphs), we can find that there is a performance drop between the encoder and the decoder ($Layer_{24}$ and $Layer_{25}$) of CodeT5+ and the drops gets more evident for the long dependency.}

To summarize the findings from Section~\ref{sec:semantic-relation-prediction} and Section~\ref{sec:semantic-propagation}, all four \revision{pre-trained code} models possess the capability to learn code semantics. However, their abilities to encode semantics vary across different types of semantics. \revision{For the LLM group, the representation from the shallow layers makes more easier to observe the code semantics. The ability to encode the code execution information should be enhanced for LLMs. }

\begin{figure}[]
     \centering
     \begin{subfigure}[b]{0.4\textwidth}
         \centering
         \includegraphics[width=\textwidth]{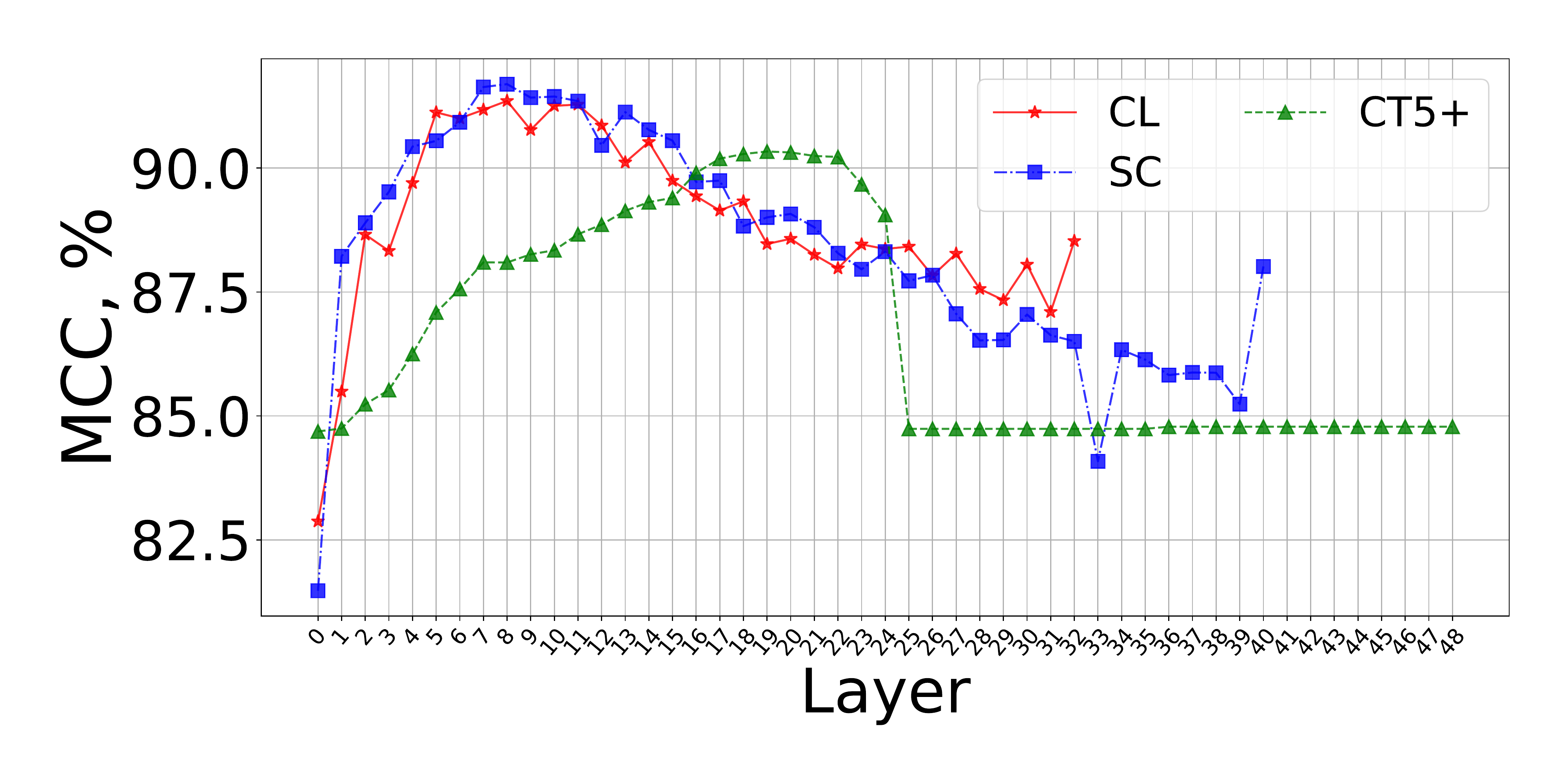}
          \vspace*{-2em}
        \caption{Java250-CDG.}
         \label{fig:ingraph_cdg_java250_llm}
     \end{subfigure}     
     \begin{subfigure}[b]{0.4\textwidth}
         \centering
         \includegraphics[width=\textwidth]{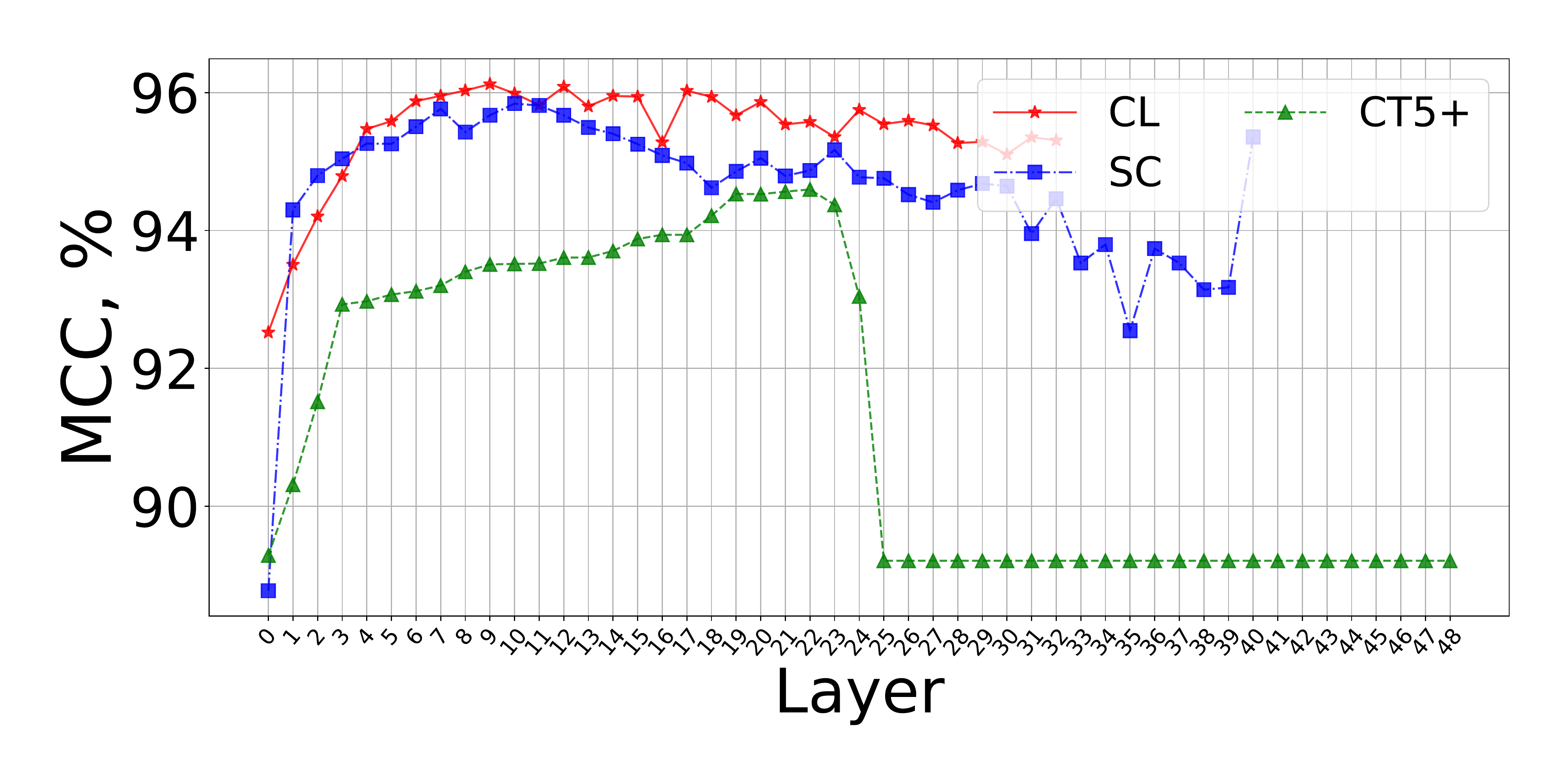}
         \vspace*{-2em}
         \caption{Java250-DDG.}
         \label{fig:ingraph_ddg_java250_llm}
     \end{subfigure} 
     \vspace*{-0.9em}
        \caption{LLM Performance~(MCC) about Java250 and POJ-104 for inGraph.}
        \label{fig:performance_isingraph_llm}
        \Description[LLM Performance about Java250 and POJ-104 for inGraph]{}
        \vspace{-2mm}
\end{figure}

\subsection{Attention Analysis}
\revision{Based on the analyses mentioned above, we have observed that pre-trained code models and large language models~(LLM) demonstrate good or middle-level proficiency in learning code syntax and semantics. To gain further insights into how these models encode semantics, we conducted an investigation into the roles of self-attention heads in learning code semantics, utilizing a dataset of over 10,000 semantic inputs for pre-trained code models and randomly sampling 100 semantic inputs for LLMs. Every one of the pre-trained models has 144 attention heads in total, respectively. StarCoder has 1920 attention heads, CodeLlama has 960 attention heads and CodeT5+ Decoder/Encoder has 384 attention heads.}

\begin{table}[!t]
\centering
\caption{Number of Semantic Attention Heads on Java250 / POJ-104.}
\label{tab:java250_attention}
\begin{tabular}{l|c|c|c|c}
\hline
    & CodeBERT & GraphCodeBERT & UnixCoder & CodeT5 \\ \hline
CDG & 65 / 93       & 71 / 86            & 50 / 44        & 77 / 91     \\ %
CFG & 129 / 141      & 136 / 139           & 99 / 82        & 130 / 128    \\ %
DDG & 133 / 115      & 135 / 124           & 23 / 58        & 111 / 120    \\ \hline
\end{tabular}
\end{table}

\begin{table}[!t]
\centering
\caption{Number of Semantic Attention Heads of LLMs on Java250 / POJ-104.}
\label{tab:java250_attention_llm}
\begin{tabular}{l|c|c|c|c}
\hline
    & StarCoder & CodeLlama & CodeT5+ Decoder & CodeT5+ Encoder \\ \hline
CDG &  0/0        &  0/0       &  0/0         &  199/252      \\ %
CFG &  1372/1844       &  423/830        &  371/340         &  315/319     \\ %
DDG &  1253/1892       &  100/977         &  318/146        &  247/208     \\ \hline
\end{tabular}
\end{table}

In \tabref{java250_attention} and \tabref{java250_attention_llm}, we present the number of attention heads that significantly learn code semantics in terms of CDG, CFG, and DDG (statistically significant at a p-value threshold of 0.01) for the pre-trained models and LLMs about the Java250 and POJ-104 datasets, respectively. In \tabref{java250_attention}, it is evident that UnixCoder has the fewest number of attention heads dedicated to semantics. This is consistent with the previous conclusion that UnixCoder is not better than others in terms of encoding code syntax and semantics. \revision{In \tabref{java250_attention_llm}, we can find that attention heads that have the control dependency relationship do not assign more weights than the attention heads that have no control dependency relationship for the StarCoder, CodeLlama and CodeT5+ decoder who use the transformer decoder architecture. However, for CodeT5+ encoder, there exist 199/252 attention heads that can assign more weights for the control dependency relationship. We think the reason is that different model architectures may affect the distribution of attention mechanisms. Encoder and decoder in the transformer have different functions. Encoder is more inclined to understand the overall structure and context of the input, while the decoder focuses on generating or predicting the next token based on this understanding. Therefore, for control-dependent statements, the encoder may need to pay more attention to these relationships to understand the logical structure of the entire code, while the decoder may pay more attention to local information or other semantic information for generation, resulting in attention weights of different distributions.}

We further investigate the overlap of semantic attention heads that assign more attention weights to semantic tokens between the Java250 and POJ-104 datasets for these four models. The results reveal that one model shares semantic attention heads across different programming languages. 
\revision{To examine the disparities in semantic attention heads between the Java250 and POJ-104 datasets across four employed models, we present the ratios of the overlapping semantic attention heads for each model in both datasets (\tabref{attention_overlapping} and \tabref{attention_overlapping_llm}).}  The rows Java250 and POJ-104 are computed by the following two equations respectively,
\[ r_{java250} = \frac{|S_{java250} \cap \ S_{POJ}|}{|S_{java250}|}\]
\[ r_{poj} = \frac{|S_{java250} \cap \ S_{POJ}|}{|S_{POJ}|}\], where $S_{java250}$ and $S_{POJ}$ are the set of semantic attention heads for Java250 and POJ-104 datasets. 
Surprisingly, despite the distinct programming languages represented in the two datasets, a substantial overlap is observed. \revision{Although the overlapping ratio $r_{poj}$ for DDG in CodeLlama is 8.29\%, it is caused that $S_{POJ}$ is too large and actually majority of attention heads from Java250 are included in $S_{POJ}$ (81.00\%).} This finding suggests that certain semantic attention patterns are shared among code models regardless of the programming language being used.

\begin{table*}[!t]
\centering
 \caption{Overlapping of Semantic Attention Heads on Java250 and POJ-104, Percent \%.}
\label{tab:attention_overlapping}
\scalebox{0.9}{
\begin{tabular}{l|lll|lll|lll|lll}
\hline
\multirow{2}{*}{} & \multicolumn{3}{c|}{CodeBERT}                                   & \multicolumn{3}{c|}{GraphCodeBERT}                              & \multicolumn{3}{c|}{UnixCoder}                                  & \multicolumn{3}{c}{CodeT5}                                     \\ \cline{2-13} 
                  & \multicolumn{1}{l|}{CDG}   & \multicolumn{1}{l|}{CFG}   & DDG   & \multicolumn{1}{l|}{CDG}   & \multicolumn{1}{l|}{CFG}   & DDG   & \multicolumn{1}{l|}{CDG}   & \multicolumn{1}{l|}{CFG}   & DDG   & \multicolumn{1}{l|}{CDG}   & \multicolumn{1}{l|}{CFG}   & DDG   \\ \hline
$r_{Java250}$           & \multicolumn{1}{l|}{96.92} & \multicolumn{1}{l|}{100.0} & 85.71 & \multicolumn{1}{l|}{94.37} & \multicolumn{1}{l|}{97.06} & 91.11 & \multicolumn{1}{l|}{84.62} & \multicolumn{1}{l|}{97.87} & 95.50 & \multicolumn{1}{l|}{84.42} & \multicolumn{1}{l|}{96.92} & 93.69 \\ \hline
$r_{poj}$           & \multicolumn{1}{l|}{67.74} & \multicolumn{1}{l|}{91.49} & 99.13 & \multicolumn{1}{l|}{77.91} & \multicolumn{1}{l|}{94.96} & 99.19 & \multicolumn{1}{l|}{51.16} & \multicolumn{1}{l|}{70.23} & 76.81 & \multicolumn{1}{l|}{71.43} & \multicolumn{1}{l|}{98.44} & 86.67 \\ \hline
\end{tabular}}
\end{table*}

\begin{table*}[!t]
\centering
 \caption{Overlapping of LLM Semantic Attention Heads on Java250 and POJ-104, Percent \%.}
\label{tab:attention_overlapping_llm}
\scalebox{0.9}{
\begin{tabular}{l|lll|lll|lll|lll}
\hline
\multirow{2}{*}{} & \multicolumn{3}{c|}{StarCoder}                                   & \multicolumn{3}{c|}{CodeLlama}                              & \multicolumn{3}{c|}{CodeT5+ Decoder}                                  & \multicolumn{3}{c}{CodeT5+ Encoder}                                     \\ \cline{2-13} 
                  & \multicolumn{1}{l|}{CDG}   & \multicolumn{1}{l|}{CFG}   & DDG   & \multicolumn{1}{l|}{CDG}   & \multicolumn{1}{l|}{CFG}   & DDG   & \multicolumn{1}{l|}{CDG}   & \multicolumn{1}{l|}{CFG}   & DDG   & \multicolumn{1}{l|}{CDG}   & \multicolumn{1}{l|}{CFG}   & DDG   \\ \hline
$r_{Java250}$           & \multicolumn{1}{c|}{-} & \multicolumn{1}{l|}{99.05} & 98.24 & \multicolumn{1}{c|}{-} & \multicolumn{1}{l|}{95.98} & 81.00 & \multicolumn{1}{c|}{-} & \multicolumn{1}{l|}{90.56} & 40.25 & \multicolumn{1}{l|}{98.49} & \multicolumn{1}{l|}{95.87} & 74.09 \\ \hline
$r_{poj}$           & \multicolumn{1}{c|}{-} & \multicolumn{1}{l|}{73.69} & 65.06 & \multicolumn{1}{c|}{-} & \multicolumn{1}{l|}{48.92} & 8.29 & \multicolumn{1}{c|}{-} & \multicolumn{1}{l|}{98.82} & 87.67 & \multicolumn{1}{l|}{77.78} & \multicolumn{1}{l|}{94.67} & 87.98 \\ \hline
\end{tabular}}
\end{table*}

\section{Related Work}
\label{sec:related_work}
\subsection{Code Models} Pre-trained models have been used to support many tasks due to their excellent generalization ability in natural language processing tasks. Recently, researchers pre-train transformers~\cite{niu2023empirical} using code data to solve programming tasks. According to the pre-training strategies and model architectures,
we can group the pre-trained models into three (3) types:
auto-encoding models, auto-regressive models, and sequence-to-sequence (Seq2Seq) models. 
Auto-encoding models utilize Transformer encoders and are pre-trained with objectives such as Masked Language Modelling (MLM). MLM masks some tokens in the code sequence and expects the model to predict the masked tokens using bidirectional context information, which in fact enables the model to use future tokens to predict current mask tokens. 
CodeBERT~\cite{feng2020codebert} is pre-trained on CodeSearchNet dataset~\cite{husain2019codesearchnet}.
GraphCodeBERT~\cite{guo2020graphcodebert} includes one additional input type, data flow sequence, compared with CodeBERT.  CodeBERT and GraphCodeBERT use the encoder of Transformer.
Auto-regressive models use Causal Language Modelling (CLM) or its variants to pre-train the transformers in a left-to-right manner. CodeGPT~\cite{lu2021codexglue} uses this pre-training strategy and keeps the transformer decoder. 
Seq2Seq models, e.g., CodeT5~\cite{wang2021codet5}, use both an encoder and a decoder in the Transformer. CommitBART~\cite{liu2022commitbart} uses BART~\cite{lewis2019bart} architecture to pre-train a model for GitHub commits. \\
\revision{Recently, ChatGPT and other large language models have gained significant attention. As a result, several LLMs specifically designed for coding have emerged. StarCoder~\cite{li2023starcoder} with 15B parameters trained using 1 trillion tokens is released. CodeLlama~\cite{roziere2023code} is tuned based on Llama2~\cite{touvron2023llama} using the code dataset. WizardCoder~\cite{luo2023wizardcoder} uses the code evolutionary instruction to tune the mode weights. All of them are based on the transformer decoder. Differently from them, CodeT5+~\cite{wang2023codet5+} uses the encoder-decoder architecture like CodeT5~\cite{wang2021codet5}. LLMs show powerful ability in program repair, code generation and summarization. \citet{hou2023large} and \citet{zhang2023survey} comprehensively reviews how large the applications of language models in software engineering.}\\
\revision{These code models are widely used to solve various software engineering tasks~\cite{hou2023large,zhang2023survey,sun2024llm4vuln,ma2024combining}, such as defect detection, code summarization, vulnerability repair, and bug localization. Before the emergence of Language Model Learners (LLMs), these deep learning models were used in two ways in software engineering~\cite{10.1145/3505243,li2018deep}: 1) Adding a task model on top of these code models and fine-tuning the weights of the entire model. 2) These models are used as feature extractors, and other algorithms are applied after extracting features. Compared to the first method, the second does not require fine-tuning the weights of the model. After the emergence of LLMs, learning methods based on in-context learning~\cite{dong2022survey} are gradually accepted and used. These models are usually not used alone but as a step in the workflow. Moreover, researchers will use domain knowledge to determine how to use the code model for specific scenarios. Despite the many methods for various software engineering tasks based on code models available today and researchers trying to explain each step of their methods as much as possible, the use of the code model itself as a black box might discount these explanations to some extent. Our work aims to help software engineers understand how models understand code syntax and semantics as much as possible.}

\subsection{Probing Analysis for Code Models} 
The impressive performance of pre-trained models stimulates loads of work trying to interpret and understand these newly invented large-scale black box models. These analysis works can help users understand and apply pre-trained models.
\emph{Probing} \citep{bertology, conneau-etal-2018-cram, pse} is one of the most prominent techniques widely leveraged for interpretability.
Probing analysis aims at diagnosing which types of regularities are encoded in a representation extracted from data. The basis of probing is that if a \emph{simple} classifier, e.g., a linear classifier, built upon the representations can solve a task sufficiently well, then the representations should contain informative features about the task already. \\
Recent works strive to analyze code pre-trained models via probing. \citet{10.1145/3510003.3510050} evaluated if pre-trained models learn programming language syntax, and they measured the number of edges among node tokens at AST, and tried to learn this distance in the vector space. \revision{ This approach cannot recover the AST structure, given the distances among all nodes. Although the number of edges between nodes can reflect the syntax information to some degree, it still has some problems. First, it cannot reconstruct AST structures in the vector space, which means that it partially checks the code syntax. Second, two tokens with similar syntax have a small number of edges, but the small number of edges does not imply syntax closeness as shown in the motivation section. }
\citet{lopez2022ast} analyzed pre-trained models in a global-level AST by projecting AST into a subspace. \revision{This work converts AST to a middle-format binary tree and then learns a subspace utilizing the syntax distance~\cite{shen-etal-2018-straight} that is designed for natural language. However, the syntax distance for natural language may not be suitable for code data since \citet{allamanis2018a} list the difference between the code and the natural language, including the differences in syntax tree between the natural language and the code. In contrast, our approach is concise and efficient, and we directly recover AST structures from the vector space. In addition, we conduct the semantic analysis for the code.}
\citet{troshin2022probing} developed a group of probing tasks to check if pre-trained models learn the code syntax structure and data flow information. \revision{First, this work does not consider the whole structure of AST and considers partially code syntax. Second, this work does not consider control-flow and control-dependency semantics. \citet{shen2022benchmarking} extract the syntax sub-structure and predict their syntax relation. However, they lack semantic analysis and only include four syntax types: assignment, function calling, if statements, and loop structures. 
The latest work~\cite{ma2024lms} observes syntax and semantic relationships and also studies how LLMs understand code behaviour based on the generated outputs via in context learning for code analysis. Our work uses the inner states of code models and comprehensively studies how code models encode code syntax and semantics by targeting the whole structure of AST, control and data dependency graphs, and the control-flow graph. } 

\subsection{Deep Learning Testing}
\revision{
Deep learning models are treated as a system. Although we can know the weights of the model and its internal calculation method, we do not know its intrinsic logic. \citet{10.1145/3643678} comprehensively study deep learning testing. Testing such a blackbox system is quite challenging. Early software engineering researchers mainly focused on detecting defects in deep learning systems~\cite{10.1145/3417330}. For example, DeepMutation~\cite{ma2018deepmutation}  and DeepGini~\cite{10.1145/3395363.3397357} are all used to find defects in deep learning systems. These defects are usually defined as inputs that cause the system to produce incorrect outputs. Once the system defects are discovered, researchers try to repair these discovered defects. Also, in situations where multiple models are used for the same task, researchers have proposed ways to select high-quality models with few defects~\cite{10.1145/3611666, 9710827, meng2021measuring}. The models being tested are usually models that solve specific tasks in a specific scenario. Besides, some research focuses on testing the model performance in unknown scenarios. For example, ATC (Leveraging Unlabeled Data to Predict Out-of-Distribution Performance)~\cite{garg2022leveraging}, Aries~\cite{10.1109/ICSE48619.2023.00152}, and OODRobustBench~\cite{li2023oodrobustbench} all study the performance of models on out-of-distribution data. 
It is important to emphasize that our work is not to detect defects in the model; we aim to explain the understanding of the code syntax and semantics for code models. Probing analysis and testing are different. Probing analysis focuses on understanding the internal mechanisms and knowledge representation of models while testing code models aims to evaluate the actual performance of models and identify potential defects. Both are indispensable steps in the development and evaluation process of code models, but they focus on different aspects.}

\section{Conclusion and Discussion}
\label{sec:conclusion}
\subsection{Conclusion}
\revision{In this work, we aim to gain a deeper understanding of how code models handle and comprehend complex code syntax and semantic structures. Specifically, we investigate whether these code models can accurately capture the syntax trees (AST), control dependencies (CDG), control flow (CFG), and data dependencies (DDG) in code by reconstructing these syntax and semantic structures in vector space. These are all fundamental aspects of program understanding. 
We design a series of probing tasks to evaluate the ability of code models to handle code syntax and semantics. We explore four popular pre-trained code models: CodeBERT, GraphCodeBERT, UnixCoder, and CodeT5, and introduce three large-scale language models (LLMs): CodeLlama, StarCoder, and CodeT5+, to assess their performance in understanding code syntax and semantics. Additionally, we indirectly observe how the models handle dependencies in code through attention analysis. From our experiments, we obtain several interesting findings. Pre-trained code models and LLMs are able to express code syntax quite well and capture code semantics to some extent, especially in handling data dependencies and control dependencies. By comparing the performance of different models, we observe that specific models have different advantages in probing tasks, indicating the varied effects of model strategies and training datasets on the understanding of the models in code syntax and semantics. Though LLMs demonstrate the ability to learn code syntax and semantics, in-depth performance analysis reveals differences in the presentation of this understanding in the different hidden layers, suggesting the possibility of optimizing pre-training strategies to improve deep semantic processing. When we employ the SOTA code models like CodeLlama to solve software engineering tasks, we still need to carefully design downstream-task models because the code features extracted by large models are hidden deeply and are not easily observed.}

\subsection{Discussion}
\revision{
Our research provides valuable insights into understanding and improving the capabilities of code models and raises challenging questions and opportunities for future research directions. Based on our findings, we believe that future research should focus on several key areas:
1). Further optimizing the architecture and training strategies of code models to enhance their ability to understand complex code semantics. This could involve exploring new neural network architectures, training objectives, and data augmentation techniques. One possible solution is to use the graph transformer~\cite{li2019graph, yun2019graph} to directly learn the syntax and semantic structures of the code. \citet{10.1145/3524842.3528456} have demonstrated that the graph neural network can perform as well as the transformer but with fewer parameters. The graph transformer combines the advantages of the graph neural network and the transformer: encoding graph structure and learning from a large corpus. 
2). Exploring new probing tasks and evaluation methods to further unveil the inner workings of code models, especially the ability of decoder-based models to understand code. This could involve developing new metrics for measuring code understanding, as well as designing adversarial examples to test the robustness of code models.
3). Investigating the relationship between the robustness of code models and their ability to understand code semantics. This could involve studying how different types of noise and adversarial attacks affect the performance of code models and developing methods to improve their robustness while maintaining their ability to understand code.
Looking ahead, code models are increasingly being used for software development. Our work has highlighted that code models still have significant shortcomings in understanding code semantics. However, code semantics is closely related to code security. Therefore, it is crucial to perform security checks on code generated by code models.}

\section{THREATS TO VALIDITY}
\label{sec:limitation}
Firstly, the results of the detection analysis are influenced by the random seed and the dataset used. There is a certain level of randomness that may affect the performance of the final detection classifier. To mitigate the impact of randomness, we conducted multiple experiments and performed statistical analyses of the results. \\
Secondly, we used static analysis tools to build a high-quality dataset for analysis. Although this approach provides valuable information, it also introduces certain biases. Static analysis tools may not fully capture subtle variations in dynamic semantics and may be influenced by the syntax rules of specific programming languages. To reduce this bias, we choose the tools that are used by many researchers to avoid mistakes in the data. In the process of cross-task comparisons, we extracted syntax and semantic tasks from the same data source to minimize the bias in cross-task comparisons. \\
Third, different performance metrics may exhibit different biases. To evaluate the models more accurately, we used the Matthews correlation coefficient (MCC) for binary classification and the F1 score for multi-class classification. In fact, we evaluated the models using both MCC and F1 for all tasks, and the conclusions drawn from both metrics were consistent. \\
Fourth, although the probing analysis can reveal the learning capability of code models in terms of syntax and semantics, it does not guarantee good downstream task performance in practical applications. Code models may perform well in probing tasks but poorly in actual applications, especially tasks requiring complex reasoning or multimodal interactions. The performance of downstream task models is related not only to the expressive power of the base model but also to the quality of the downstream task dataset and the robustness of the model. \\
\revision{Fifth, probing analysis is to observe if the representation contains one specific property that exists in the input. However, the poor performance of the probing analysis does not mean that the representation does not contain such information. The representation space is a high-dimension space and mixes all learned things together. Some information is more observable and is easy to probe. Our probing approach can only be used to answer if this property is more observable or less. It is limited to answer `yes' or `no'.}\\
\revision{Last but not least, many code models have been proposed. To ensure that our evaluation approach and conclusion are generalized, we consider three different architectures of code models: encoder, decoder and encoder-decoder. We include the traditional pre-trained models and large language models from large companies or organizations, and all of these models are widely used as the based models to solve the downstream tasks. When we design our approach, we do not consider any specific model and programming language, and we only consider the syntax and semantics of code. AST represents all syntax information of code. CFG represents the logic/action of the code. DDG and CDG represent the semantic relationship between different code parts. Almost all SE tasks related to the code need this information. However, different models have different abilities, and our conclusion may not be suitable for one model with very different training strategies and datasets.}

\begin{acks}
This research is supported by the National Research Foundation, Singapore, and the Cyber Security Agency under its National Cybersecurity R\&D Programme (NCRP25-P04-TAICeN), the National Research Foundation, Singapore, and DSO National Laboratories under the AI Singapore Programme (AISG Award No: AISG2-GC-2023-008), and NRF Investigatorship NRF-NRFI06-2020-0001. Any opinions, findings and conclusions or recommendations expressed in this material are those of the author(s) and do not reflect the views of National Research Foundation, Singapore and Cyber Security Agency of Singapore.
\end{acks}

\bibliographystyle{ACM-Reference-Format}
\bibliography{sample-base}


\begin{thebibliography}{81}


\ifx \showCODEN    \undefined \def \showCODEN     #1{\unskip}     \fi
\ifx \showDOI      \undefined \def \showDOI       #1{#1}\fi
\ifx \showISBNx    \undefined \def \showISBNx     #1{\unskip}     \fi
\ifx \showISBNxiii \undefined \def \showISBNxiii  #1{\unskip}     \fi
\ifx \showISSN     \undefined \def \showISSN      #1{\unskip}     \fi
\ifx \showLCCN     \undefined \def \showLCCN      #1{\unskip}     \fi
\ifx \shownote     \undefined \def \shownote      #1{#1}          \fi
\ifx \showarticletitle \undefined \def \showarticletitle #1{#1}   \fi
\ifx \showURL      \undefined \def \showURL       {\relax}        \fi
\providecommand\bibfield[2]{#2}
\providecommand\bibinfo[2]{#2}
\providecommand\natexlab[1]{#1}
\providecommand\showeprint[2][]{arXiv:#2}

\bibitem[Aggarwal et~al\mbox{.}(2001)]%
        {10.5555/645504.656414}
\bibfield{author}{\bibinfo{person}{Charu~C. Aggarwal}, \bibinfo{person}{Alexander Hinneburg}, {and} \bibinfo{person}{Daniel~A. Keim}.} \bibinfo{year}{2001}\natexlab{}.
\newblock \showarticletitle{On the Surprising Behavior of Distance Metrics in High Dimensional Spaces}. In \bibinfo{booktitle}{\emph{Proceedings of the 8th International Conference on Database Theory}} \emph{(\bibinfo{series}{ICDT '01})}. \bibinfo{publisher}{Springer-Verlag}, \bibinfo{address}{Berlin, Heidelberg}, \bibinfo{pages}{420–434}.
\newblock
\showISBNx{3540414568}


\bibitem[Ahmad et~al\mbox{.}(2021)]%
        {ahmad2021unified}
\bibfield{author}{\bibinfo{person}{Wasi~Uddin Ahmad}, \bibinfo{person}{Saikat Chakraborty}, \bibinfo{person}{Baishakhi Ray}, {and} \bibinfo{person}{Kai-Wei Chang}.} \bibinfo{year}{2021}\natexlab{}.
\newblock \showarticletitle{Unified pre-training for program understanding and generation}.
\newblock \bibinfo{journal}{\emph{arXiv preprint arXiv:2103.06333}} (\bibinfo{year}{2021}).
\newblock


\bibitem[Allamanis et~al\mbox{.}(2018)]%
        {allamanis2018a}
\bibfield{author}{\bibinfo{person}{Miltos Allamanis}, \bibinfo{person}{Earl T.~Barr}, \bibinfo{person}{Premkumar Devanbu}, {and} \bibinfo{person}{Charles Sutton}.} \bibinfo{year}{2018}\natexlab{}.
\newblock \showarticletitle{A Survey of Machine Learning for Big Code and Naturalness}.
\newblock \bibinfo{journal}{\emph{Comput. Surveys}} \bibinfo{volume}{51}, \bibinfo{number}{4} (\bibinfo{date}{July} \bibinfo{year}{2018}), \bibinfo{pages}{81}.
\newblock
\urldef\tempurl%
\url{https://www.microsoft.com/en-us/research/publication/a-survey-of-machine-learning-for-big-code-and-naturalness/}
\showURL{%
\tempurl}


\bibitem[Buratti et~al\mbox{.}(2020)]%
        {buratti2020exploring}
\bibfield{author}{\bibinfo{person}{Luca Buratti}, \bibinfo{person}{Saurabh Pujar}, \bibinfo{person}{Mihaela Bornea}, \bibinfo{person}{Scott McCarley}, \bibinfo{person}{Yunhui Zheng}, \bibinfo{person}{Gaetano Rossiello}, \bibinfo{person}{Alessandro Morari}, \bibinfo{person}{Jim Laredo}, \bibinfo{person}{Veronika Thost}, \bibinfo{person}{Yufan Zhuang}, {et~al\mbox{.}}} \bibinfo{year}{2020}\natexlab{}.
\newblock \showarticletitle{Exploring software naturalness through neural language models}.
\newblock \bibinfo{journal}{\emph{arXiv preprint arXiv:2006.12641}} (\bibinfo{year}{2020}).
\newblock


\bibitem[Chicco and Jurman(2020)]%
        {chicco2020advantages}
\bibfield{author}{\bibinfo{person}{Davide Chicco} {and} \bibinfo{person}{Giuseppe Jurman}.} \bibinfo{year}{2020}\natexlab{}.
\newblock \showarticletitle{The advantages of the Matthews correlation coefficient (MCC) over F1 score and accuracy in binary classification evaluation}.
\newblock \bibinfo{journal}{\emph{BMC genomics}} \bibinfo{volume}{21}, \bibinfo{number}{1} (\bibinfo{year}{2020}), \bibinfo{pages}{1--13}.
\newblock


\bibitem[Conneau et~al\mbox{.}(2018)]%
        {conneau-etal-2018-cram}
\bibfield{author}{\bibinfo{person}{Alexis Conneau}, \bibinfo{person}{German Kruszewski}, \bibinfo{person}{Guillaume Lample}, \bibinfo{person}{Lo{\"\i}c Barrault}, {and} \bibinfo{person}{Marco Baroni}.} \bibinfo{year}{2018}\natexlab{}.
\newblock \showarticletitle{What you can cram into a single {\$}{\&}!{\#}* vector: Probing sentence embeddings for linguistic properties}. In \bibinfo{booktitle}{\emph{Proceedings of the 56th Annual Meeting of the Association for Computational Linguistics (Volume 1: Long Papers)}}. \bibinfo{publisher}{Association for Computational Linguistics}, \bibinfo{address}{Melbourne, Australia}, \bibinfo{pages}{2126--2136}.
\newblock
\urldef\tempurl%
\url{https://doi.org/10.18653/v1/P18-1198}
\showDOI{\tempurl}


\bibitem[Dong et~al\mbox{.}(2022)]%
        {dong2022survey}
\bibfield{author}{\bibinfo{person}{Qingxiu Dong}, \bibinfo{person}{Lei Li}, \bibinfo{person}{Damai Dai}, \bibinfo{person}{Ce Zheng}, \bibinfo{person}{Zhiyong Wu}, \bibinfo{person}{Baobao Chang}, \bibinfo{person}{Xu Sun}, \bibinfo{person}{Jingjing Xu}, {and} \bibinfo{person}{Zhifang Sui}.} \bibinfo{year}{2022}\natexlab{}.
\newblock \showarticletitle{A survey for in-context learning}.
\newblock \bibinfo{journal}{\emph{arXiv preprint arXiv:2301.00234}} (\bibinfo{year}{2022}).
\newblock


\bibitem[Dou et~al\mbox{.}(2023)]%
        {dou2023towards}
\bibfield{author}{\bibinfo{person}{Shihan Dou}, \bibinfo{person}{Junjie Shan}, \bibinfo{person}{Haoxiang Jia}, \bibinfo{person}{Wenhao Deng}, \bibinfo{person}{Zhiheng Xi}, \bibinfo{person}{Wei He}, \bibinfo{person}{Yueming Wu}, \bibinfo{person}{Tao Gui}, \bibinfo{person}{Yang Liu}, {and} \bibinfo{person}{Xuanjing Huang}.} \bibinfo{year}{2023}\natexlab{}.
\newblock \showarticletitle{Towards understanding the capability of large language models on code clone detection: a survey}.
\newblock \bibinfo{journal}{\emph{arXiv preprint arXiv:2308.01191}} (\bibinfo{year}{2023}).
\newblock


\bibitem[Drain et~al\mbox{.}(2021)]%
        {drain2021generating}
\bibfield{author}{\bibinfo{person}{Dawn Drain}, \bibinfo{person}{Chen Wu}, \bibinfo{person}{Alexey Svyatkovskiy}, {and} \bibinfo{person}{Neel Sundaresan}.} \bibinfo{year}{2021}\natexlab{}.
\newblock \showarticletitle{Generating bug-fixes using pretrained transformers}. In \bibinfo{booktitle}{\emph{Proceedings of the 5th ACM SIGPLAN International Symposium on Machine Programming}}. \bibinfo{pages}{1--8}.
\newblock


\bibitem[Du et~al\mbox{.}(2023)]%
        {du2023classeval}
\bibfield{author}{\bibinfo{person}{Xueying Du}, \bibinfo{person}{Mingwei Liu}, \bibinfo{person}{Kaixin Wang}, \bibinfo{person}{Hanlin Wang}, \bibinfo{person}{Junwei Liu}, \bibinfo{person}{Yixuan Chen}, \bibinfo{person}{Jiayi Feng}, \bibinfo{person}{Chaofeng Sha}, \bibinfo{person}{Xin Peng}, {and} \bibinfo{person}{Yiling Lou}.} \bibinfo{year}{2023}\natexlab{}.
\newblock \showarticletitle{Classeval: A manually-crafted benchmark for evaluating llms on class-level code generation}.
\newblock \bibinfo{journal}{\emph{arXiv preprint arXiv:2308.01861}} (\bibinfo{year}{2023}).
\newblock


\bibitem[Fan et~al\mbox{.}(2023)]%
        {fan2023large}
\bibfield{author}{\bibinfo{person}{Angela Fan}, \bibinfo{person}{Beliz Gokkaya}, \bibinfo{person}{Mark Harman}, \bibinfo{person}{Mitya Lyubarskiy}, \bibinfo{person}{Shubho Sengupta}, \bibinfo{person}{Shin Yoo}, {and} \bibinfo{person}{Jie~M Zhang}.} \bibinfo{year}{2023}\natexlab{}.
\newblock \showarticletitle{Large language models for software engineering: Survey and open problems}.
\newblock \bibinfo{journal}{\emph{arXiv preprint arXiv:2310.03533}} (\bibinfo{year}{2023}).
\newblock


\bibitem[Feng et~al\mbox{.}(2020b)]%
        {10.1145/3395363.3397357}
\bibfield{author}{\bibinfo{person}{Yang Feng}, \bibinfo{person}{Qingkai Shi}, \bibinfo{person}{Xinyu Gao}, \bibinfo{person}{Jun Wan}, \bibinfo{person}{Chunrong Fang}, {and} \bibinfo{person}{Zhenyu Chen}.} \bibinfo{year}{2020}\natexlab{b}.
\newblock \showarticletitle{DeepGini: prioritizing massive tests to enhance the robustness of deep neural networks}. In \bibinfo{booktitle}{\emph{Proceedings of the 29th ACM SIGSOFT International Symposium on Software Testing and Analysis}} (Virtual Event, USA) \emph{(\bibinfo{series}{ISSTA 2020})}. \bibinfo{publisher}{Association for Computing Machinery}, \bibinfo{address}{New York, NY, USA}, \bibinfo{pages}{177–188}.
\newblock
\showISBNx{9781450380089}
\urldef\tempurl%
\url{https://doi.org/10.1145/3395363.3397357}
\showDOI{\tempurl}


\bibitem[Feng et~al\mbox{.}(2020a)]%
        {feng2020codebert}
\bibfield{author}{\bibinfo{person}{Zhangyin Feng}, \bibinfo{person}{Daya Guo}, \bibinfo{person}{Duyu Tang}, \bibinfo{person}{Nan Duan}, \bibinfo{person}{Xiaocheng Feng}, \bibinfo{person}{Ming Gong}, \bibinfo{person}{Linjun Shou}, \bibinfo{person}{Bing Qin}, \bibinfo{person}{Ting Liu}, \bibinfo{person}{Daxin Jiang}, {et~al\mbox{.}}} \bibinfo{year}{2020}\natexlab{a}.
\newblock \showarticletitle{Codebert: A pre-trained model for programming and natural languages}.
\newblock \bibinfo{journal}{\emph{arXiv preprint arXiv:2002.08155}} (\bibinfo{year}{2020}).
\newblock


\bibitem[Gao et~al\mbox{.}(2023)]%
        {10.1145/3591227}
\bibfield{author}{\bibinfo{person}{Fengjuan Gao}, \bibinfo{person}{Yu Wang}, {and} \bibinfo{person}{Ke Wang}.} \bibinfo{year}{2023}\natexlab{}.
\newblock \showarticletitle{Discrete Adversarial Attack to Models of Code}.
\newblock \bibinfo{journal}{\emph{Proc. ACM Program. Lang.}} \bibinfo{volume}{7}, \bibinfo{number}{PLDI}, Article \bibinfo{articleno}{113} (\bibinfo{date}{jun} \bibinfo{year}{2023}), \bibinfo{numpages}{24}~pages.
\newblock
\urldef\tempurl%
\url{https://doi.org/10.1145/3591227}
\showDOI{\tempurl}


\bibitem[Garg et~al\mbox{.}(2022)]%
        {garg2022leveraging}
\bibfield{author}{\bibinfo{person}{Saurabh Garg}, \bibinfo{person}{Sivaraman Balakrishnan}, \bibinfo{person}{Zachary~C Lipton}, \bibinfo{person}{Behnam Neyshabur}, {and} \bibinfo{person}{Hanie Sedghi}.} \bibinfo{year}{2022}\natexlab{}.
\newblock \showarticletitle{Leveraging unlabeled data to predict out-of-distribution performance}.
\newblock \bibinfo{journal}{\emph{arXiv preprint arXiv:2201.04234}} (\bibinfo{year}{2022}).
\newblock


\bibitem[Guo et~al\mbox{.}(2022)]%
        {guo2022unixcoder}
\bibfield{author}{\bibinfo{person}{Daya Guo}, \bibinfo{person}{Shuai Lu}, \bibinfo{person}{Nan Duan}, \bibinfo{person}{Yanlin Wang}, \bibinfo{person}{Ming Zhou}, {and} \bibinfo{person}{Jian Yin}.} \bibinfo{year}{2022}\natexlab{}.
\newblock \showarticletitle{UniXcoder: Unified Cross-Modal Pre-training for Code Representation}.
\newblock \bibinfo{journal}{\emph{arXiv preprint arXiv:2203.03850}} (\bibinfo{year}{2022}).
\newblock


\bibitem[Guo et~al\mbox{.}(2020)]%
        {guo2020graphcodebert}
\bibfield{author}{\bibinfo{person}{Daya Guo}, \bibinfo{person}{Shuo Ren}, \bibinfo{person}{Shuai Lu}, \bibinfo{person}{Zhangyin Feng}, \bibinfo{person}{Duyu Tang}, \bibinfo{person}{Shujie Liu}, \bibinfo{person}{Long Zhou}, \bibinfo{person}{Nan Duan}, \bibinfo{person}{Alexey Svyatkovskiy}, \bibinfo{person}{Shengyu Fu}, {et~al\mbox{.}}} \bibinfo{year}{2020}\natexlab{}.
\newblock \showarticletitle{Graphcodebert: Pre-training code representations with data flow}.
\newblock \bibinfo{journal}{\emph{arXiv preprint arXiv:2009.08366}} (\bibinfo{year}{2020}).
\newblock


\bibitem[Hern\'{a}ndez~L\'{o}pez et~al\mbox{.}(2023)]%
        {lopez2022ast}
\bibfield{author}{\bibinfo{person}{Jos\'{e}~Antonio Hern\'{a}ndez~L\'{o}pez}, \bibinfo{person}{Martin Weyssow}, \bibinfo{person}{Jes\'{u}s~S\'{a}nchez Cuadrado}, {and} \bibinfo{person}{Houari Sahraoui}.} \bibinfo{year}{2023}\natexlab{}.
\newblock \showarticletitle{AST-Probe: Recovering abstract syntax trees from hidden representations of pre-trained language models}. In \bibinfo{booktitle}{\emph{Proceedings of the 37th IEEE/ACM International Conference on Automated Software Engineering}} (<conf-loc>, <city>Rochester</city>, <state>MI</state>, <country>USA</country>, </conf-loc>) \emph{(\bibinfo{series}{ASE '22})}. \bibinfo{publisher}{Association for Computing Machinery}, \bibinfo{address}{New York, NY, USA}, Article \bibinfo{articleno}{11}, \bibinfo{numpages}{11}~pages.
\newblock
\showISBNx{9781450394758}
\urldef\tempurl%
\url{https://doi.org/10.1145/3551349.3556900}
\showDOI{\tempurl}


\bibitem[Hong et~al\mbox{.}(2023)]%
        {hong2023metagpt}
\bibfield{author}{\bibinfo{person}{Sirui Hong}, \bibinfo{person}{Mingchen Zhuge}, \bibinfo{person}{Jonathan Chen}, \bibinfo{person}{Xiawu Zheng}, \bibinfo{person}{Yuheng Cheng}, \bibinfo{person}{Ceyao Zhang}, \bibinfo{person}{Jinlin Wang}, \bibinfo{person}{Zili Wang}, \bibinfo{person}{Steven Ka~Shing Yau}, \bibinfo{person}{Zijuan Lin}, \bibinfo{person}{Liyang Zhou}, \bibinfo{person}{Chenyu Ran}, \bibinfo{person}{Lingfeng Xiao}, \bibinfo{person}{Chenglin Wu}, {and} \bibinfo{person}{Jürgen Schmidhuber}.} \bibinfo{year}{2023}\natexlab{}.
\newblock \bibinfo{title}{MetaGPT: Meta Programming for A Multi-Agent Collaborative Framework}.
\newblock
\newblock
\showeprint[arxiv]{2308.00352}~[cs.AI]


\bibitem[Horwitz and Reps(1992)]%
        {10.1145/143062.143156}
\bibfield{author}{\bibinfo{person}{Susan Horwitz} {and} \bibinfo{person}{Thomas Reps}.} \bibinfo{year}{1992}\natexlab{}.
\newblock \showarticletitle{The Use of Program Dependence Graphs in Software Engineering}. In \bibinfo{booktitle}{\emph{Proceedings of the 14th International Conference on Software Engineering}} (Melbourne, Australia) \emph{(\bibinfo{series}{ICSE '92})}. \bibinfo{publisher}{Association for Computing Machinery}, \bibinfo{address}{New York, NY, USA}, \bibinfo{pages}{392–411}.
\newblock
\showISBNx{0897915046}
\urldef\tempurl%
\url{https://doi.org/10.1145/143062.143156}
\showDOI{\tempurl}


\bibitem[Hou et~al\mbox{.}(2023)]%
        {hou2023large}
\bibfield{author}{\bibinfo{person}{Xinyi Hou}, \bibinfo{person}{Yanjie Zhao}, \bibinfo{person}{Yue Liu}, \bibinfo{person}{Zhou Yang}, \bibinfo{person}{Kailong Wang}, \bibinfo{person}{Li Li}, \bibinfo{person}{Xiapu Luo}, \bibinfo{person}{David Lo}, \bibinfo{person}{John Grundy}, {and} \bibinfo{person}{Haoyu Wang}.} \bibinfo{year}{2023}\natexlab{}.
\newblock \showarticletitle{Large language models for software engineering: A systematic literature review}.
\newblock \bibinfo{journal}{\emph{arXiv preprint arXiv:2308.10620}} (\bibinfo{year}{2023}).
\newblock


\bibitem[Hu et~al\mbox{.}(2022)]%
        {hu2022lora}
\bibfield{author}{\bibinfo{person}{Edward~J Hu}, \bibinfo{person}{yelong shen}, \bibinfo{person}{Phillip Wallis}, \bibinfo{person}{Zeyuan Allen-Zhu}, \bibinfo{person}{Yuanzhi Li}, \bibinfo{person}{Shean Wang}, \bibinfo{person}{Lu Wang}, {and} \bibinfo{person}{Weizhu Chen}.} \bibinfo{year}{2022}\natexlab{}.
\newblock \showarticletitle{Lo{RA}: Low-Rank Adaptation of Large Language Models}. In \bibinfo{booktitle}{\emph{International Conference on Learning Representations}}.
\newblock
\urldef\tempurl%
\url{https://openreview.net/forum?id=nZeVKeeFYf9}
\showURL{%
\tempurl}


\bibitem[Hu et~al\mbox{.}(2024)]%
        {10.1145/3643678}
\bibfield{author}{\bibinfo{person}{Qiang Hu}, \bibinfo{person}{Yuejun Guo}, \bibinfo{person}{Xiaofei Xie}, \bibinfo{person}{Maxime Cordy}, \bibinfo{person}{Lei Ma}, \bibinfo{person}{Mike Papadakis}, {and} \bibinfo{person}{Yves Le~Traon}.} \bibinfo{year}{2024}\natexlab{}.
\newblock \showarticletitle{Test Optimization in DNN Testing: A Survey}.
\newblock \bibinfo{journal}{\emph{ACM Trans. Softw. Eng. Methodol.}} (\bibinfo{date}{jan} \bibinfo{year}{2024}).
\newblock
\showISSN{1049-331X}
\urldef\tempurl%
\url{https://doi.org/10.1145/3643678}
\showDOI{\tempurl}
\newblock
\shownote{Just Accepted}.


\bibitem[Hu et~al\mbox{.}(2023a)]%
        {10.1145/3611666}
\bibfield{author}{\bibinfo{person}{Qiang Hu}, \bibinfo{person}{Yuejun Guo}, \bibinfo{person}{Xiaofei Xie}, \bibinfo{person}{Maxime Cordy}, \bibinfo{person}{Mike Papadakis}, {and} \bibinfo{person}{Yves Le~Traon}.} \bibinfo{year}{2023}\natexlab{a}.
\newblock \showarticletitle{LaF: Labeling-free Model Selection for Automated Deep Neural Network Reusing}.
\newblock \bibinfo{journal}{\emph{ACM Trans. Softw. Eng. Methodol.}} \bibinfo{volume}{33}, \bibinfo{number}{1}, Article \bibinfo{articleno}{25} (\bibinfo{date}{nov} \bibinfo{year}{2023}), \bibinfo{numpages}{28}~pages.
\newblock
\showISSN{1049-331X}
\urldef\tempurl%
\url{https://doi.org/10.1145/3611666}
\showDOI{\tempurl}


\bibitem[Hu et~al\mbox{.}(2023b)]%
        {10.1109/ICSE48619.2023.00152}
\bibfield{author}{\bibinfo{person}{Qiang Hu}, \bibinfo{person}{Yuejun Guo}, \bibinfo{person}{Xiaofei Xie}, \bibinfo{person}{Maxime Cordy}, \bibinfo{person}{Mike Papadakis}, \bibinfo{person}{Lei Ma}, {and} \bibinfo{person}{Yves~Le Traon}.} \bibinfo{year}{2023}\natexlab{b}.
\newblock \showarticletitle{Aries: Efficient Testing of Deep Neural Networks via Labeling-Free Accuracy Estimation}. In \bibinfo{booktitle}{\emph{Proceedings of the 45th International Conference on Software Engineering}} (Melbourne, Victoria, Australia) \emph{(\bibinfo{series}{ICSE '23})}. \bibinfo{publisher}{IEEE Press}, \bibinfo{pages}{1776–1787}.
\newblock
\showISBNx{9781665457019}
\urldef\tempurl%
\url{https://doi.org/10.1109/ICSE48619.2023.00152}
\showDOI{\tempurl}


\bibitem[Husain et~al\mbox{.}(2019)]%
        {husain2019codesearchnet}
\bibfield{author}{\bibinfo{person}{Hamel Husain}, \bibinfo{person}{Ho-Hsiang Wu}, \bibinfo{person}{Tiferet Gazit}, \bibinfo{person}{Miltiadis Allamanis}, {and} \bibinfo{person}{Marc Brockschmidt}.} \bibinfo{year}{2019}\natexlab{}.
\newblock \showarticletitle{Codesearchnet challenge: Evaluating the state of semantic code search}.
\newblock \bibinfo{journal}{\emph{arXiv preprint arXiv:1909.09436}} (\bibinfo{year}{2019}).
\newblock


\bibitem[Iraqi({[n.\,d.]})]%
        {huging_llm_comparing}
\bibfield{author}{\bibinfo{person}{Mehdi Iraqi}.} \bibinfo{year}{[n.\,d.]}\natexlab{}.
\newblock \bibinfo{booktitle}{\emph{Comparing the Performance of LLMs: A Deep Dive into Roberta, Llama 2, and Mistral for Disaster Tweets Analysis with Lora}}.
\newblock
\urldef\tempurl%
\url{https://huggingface.co/blog/Lora-for-sequence-classification-with-Roberta-Llama-Mistral#comparing-the-performance-of-llms-a-deep-dive-into-roberta-llama-2-and-mistral-for-disaster-tweets-analysis-with-lora}
\showURL{%
\tempurl}


\bibitem[Jha and Reddy(2023)]%
        {10.1609/aaai.v37i12.26739}
\bibfield{author}{\bibinfo{person}{Akshita Jha} {and} \bibinfo{person}{Chandan~K. Reddy}.} \bibinfo{year}{2023}\natexlab{}.
\newblock \showarticletitle{CodeAttack: code-based adversarial attacks for pre-trained programming language models}. In \bibinfo{booktitle}{\emph{Proceedings of the Thirty-Seventh AAAI Conference on Artificial Intelligence and Thirty-Fifth Conference on Innovative Applications of Artificial Intelligence and Thirteenth Symposium on Educational Advances in Artificial Intelligence}} \emph{(\bibinfo{series}{AAAI'23/IAAI'23/EAAI'23})}. \bibinfo{publisher}{AAAI Press}, Article \bibinfo{articleno}{1670}, \bibinfo{numpages}{9}~pages.
\newblock
\showISBNx{978-1-57735-880-0}
\urldef\tempurl%
\url{https://doi.org/10.1609/aaai.v37i12.26739}
\showDOI{\tempurl}


\bibitem[Kanade et~al\mbox{.}(2019)]%
        {kanade2019pre}
\bibfield{author}{\bibinfo{person}{Aditya Kanade}, \bibinfo{person}{Petros Maniatis}, \bibinfo{person}{Gogul Balakrishnan}, {and} \bibinfo{person}{Kensen Shi}.} \bibinfo{year}{2019}\natexlab{}.
\newblock \showarticletitle{Pre-trained contextual embedding of source code}.
\newblock  (\bibinfo{year}{2019}).
\newblock


\bibitem[Kaplan et~al\mbox{.}(2020)]%
        {kaplan2020scaling}
\bibfield{author}{\bibinfo{person}{Jared Kaplan}, \bibinfo{person}{Sam McCandlish}, \bibinfo{person}{Tom Henighan}, \bibinfo{person}{Tom~B Brown}, \bibinfo{person}{Benjamin Chess}, \bibinfo{person}{Rewon Child}, \bibinfo{person}{Scott Gray}, \bibinfo{person}{Alec Radford}, \bibinfo{person}{Jeffrey Wu}, {and} \bibinfo{person}{Dario Amodei}.} \bibinfo{year}{2020}\natexlab{}.
\newblock \showarticletitle{Scaling laws for neural language models}.
\newblock \bibinfo{journal}{\emph{arXiv preprint arXiv:2001.08361}} (\bibinfo{year}{2020}).
\newblock


\bibitem[Karampatsis and Sutton(2020)]%
        {karampatsis2020scelmo}
\bibfield{author}{\bibinfo{person}{Rafael-Michael Karampatsis} {and} \bibinfo{person}{Charles Sutton}.} \bibinfo{year}{2020}\natexlab{}.
\newblock \showarticletitle{Scelmo: Source code embeddings from language models}.
\newblock \bibinfo{journal}{\emph{arXiv preprint arXiv:2004.13214}} (\bibinfo{year}{2020}).
\newblock


\bibitem[Lewis et~al\mbox{.}(2019)]%
        {lewis2019bart}
\bibfield{author}{\bibinfo{person}{Mike Lewis}, \bibinfo{person}{Yinhan Liu}, \bibinfo{person}{Naman Goyal}, \bibinfo{person}{Marjan Ghazvininejad}, \bibinfo{person}{Abdelrahman Mohamed}, \bibinfo{person}{Omer Levy}, \bibinfo{person}{Ves Stoyanov}, {and} \bibinfo{person}{Luke Zettlemoyer}.} \bibinfo{year}{2019}\natexlab{}.
\newblock \showarticletitle{Bart: Denoising sequence-to-sequence pre-training for natural language generation, translation, and comprehension}.
\newblock \bibinfo{journal}{\emph{arXiv preprint arXiv:1910.13461}} (\bibinfo{year}{2019}).
\newblock


\bibitem[Li et~al\mbox{.}(2023b)]%
        {li2023oodrobustbench}
\bibfield{author}{\bibinfo{person}{Lin Li}, \bibinfo{person}{Yifei Wang}, \bibinfo{person}{Chawin Sitawarin}, {and} \bibinfo{person}{Michael Spratling}.} \bibinfo{year}{2023}\natexlab{b}.
\newblock \showarticletitle{OODRobustBench: benchmarking and analyzing adversarial robustness under distribution shift}.
\newblock \bibinfo{journal}{\emph{arXiv preprint arXiv:2310.12793}} (\bibinfo{year}{2023}).
\newblock


\bibitem[Li et~al\mbox{.}(2023a)]%
        {li2023starcoder}
\bibfield{author}{\bibinfo{person}{Raymond Li}, \bibinfo{person}{Loubna~Ben Allal}, \bibinfo{person}{Yangtian Zi}, \bibinfo{person}{Niklas Muennighoff}, \bibinfo{person}{Denis Kocetkov}, \bibinfo{person}{Chenghao Mou}, \bibinfo{person}{Marc Marone}, \bibinfo{person}{Christopher Akiki}, \bibinfo{person}{Jia Li}, \bibinfo{person}{Jenny Chim}, {et~al\mbox{.}}} \bibinfo{year}{2023}\natexlab{a}.
\newblock \showarticletitle{StarCoder: may the source be with you!}
\newblock \bibinfo{journal}{\emph{arXiv preprint arXiv:2305.06161}} (\bibinfo{year}{2023}).
\newblock


\bibitem[Li et~al\mbox{.}(2018)]%
        {li2018deep}
\bibfield{author}{\bibinfo{person}{Xiaochen Li}, \bibinfo{person}{He Jiang}, \bibinfo{person}{Zhilei Ren}, \bibinfo{person}{Ge Li}, {and} \bibinfo{person}{Jingxuan Zhang}.} \bibinfo{year}{2018}\natexlab{}.
\newblock \showarticletitle{Deep learning in software engineering}.
\newblock \bibinfo{journal}{\emph{arXiv preprint arXiv:1805.04825}} (\bibinfo{year}{2018}).
\newblock


\bibitem[Li et~al\mbox{.}(2022)]%
        {li2022transrepair}
\bibfield{author}{\bibinfo{person}{Xueyang Li}, \bibinfo{person}{Shangqing Liu}, \bibinfo{person}{Ruitao Feng}, \bibinfo{person}{Guozhu Meng}, \bibinfo{person}{Xiaofei Xie}, \bibinfo{person}{Kai Chen}, {and} \bibinfo{person}{Yang Liu}.} \bibinfo{year}{2022}\natexlab{}.
\newblock \showarticletitle{TransRepair: Context-aware Program Repair for Compilation Errors}.
\newblock \bibinfo{journal}{\emph{arXiv preprint arXiv:2210.03986}} (\bibinfo{year}{2022}).
\newblock


\bibitem[Li et~al\mbox{.}(2019)]%
        {li2019graph}
\bibfield{author}{\bibinfo{person}{Yuan Li}, \bibinfo{person}{Xiaodan Liang}, \bibinfo{person}{Zhiting Hu}, \bibinfo{person}{Yinbo Chen}, {and} \bibinfo{person}{Eric~P. Xing}.} \bibinfo{year}{2019}\natexlab{}.
\newblock \bibinfo{title}{Graph Transformer}.
\newblock
\newblock
\urldef\tempurl%
\url{https://openreview.net/forum?id=HJei-2RcK7}
\showURL{%
\tempurl}


\bibitem[Liu et~al\mbox{.}(2020a)]%
        {liu2020retrieval}
\bibfield{author}{\bibinfo{person}{Shangqing Liu}, \bibinfo{person}{Yu Chen}, \bibinfo{person}{Xiaofei Xie}, \bibinfo{person}{Jingkai Siow}, {and} \bibinfo{person}{Yang Liu}.} \bibinfo{year}{2020}\natexlab{a}.
\newblock \showarticletitle{Retrieval-augmented generation for code summarization via hybrid gnn}.
\newblock \bibinfo{journal}{\emph{arXiv preprint arXiv:2006.05405}} (\bibinfo{year}{2020}).
\newblock


\bibitem[Liu et~al\mbox{.}(2020b)]%
        {liu2020atom}
\bibfield{author}{\bibinfo{person}{Shangqing Liu}, \bibinfo{person}{Cuiyun Gao}, \bibinfo{person}{Sen Chen}, \bibinfo{person}{Nie~Lun Yiu}, {and} \bibinfo{person}{Yang Liu}.} \bibinfo{year}{2020}\natexlab{b}.
\newblock \showarticletitle{ATOM: Commit message generation based on abstract syntax tree and hybrid ranking}.
\newblock \bibinfo{journal}{\emph{IEEE Transactions on Software Engineering}} (\bibinfo{year}{2020}).
\newblock


\bibitem[Liu et~al\mbox{.}(2022)]%
        {liu2022commitbart}
\bibfield{author}{\bibinfo{person}{Shangqing Liu}, \bibinfo{person}{Yanzhou Li}, {and} \bibinfo{person}{Yang Liu}.} \bibinfo{year}{2022}\natexlab{}.
\newblock \showarticletitle{CommitBART: A Large Pre-trained Model for GitHub Commits}.
\newblock \bibinfo{journal}{\emph{arXiv preprint arXiv:2208.08100}} (\bibinfo{year}{2022}).
\newblock


\bibitem[Liu et~al\mbox{.}(2023)]%
        {liu2023contrabert}
\bibfield{author}{\bibinfo{person}{Shangqing Liu}, \bibinfo{person}{Bozhi Wu}, \bibinfo{person}{Xiaofei Xie}, \bibinfo{person}{Guozhu Meng}, {and} \bibinfo{person}{Yang Liu}.} \bibinfo{year}{2023}\natexlab{}.
\newblock \showarticletitle{ContraBERT: Enhancing Code Pre-trained Models via Contrastive Learning}.
\newblock \bibinfo{journal}{\emph{arXiv preprint arXiv:2301.09072}} (\bibinfo{year}{2023}).
\newblock


\bibitem[Lu et~al\mbox{.}(2021)]%
        {lu2021codexglue}
\bibfield{author}{\bibinfo{person}{Shuai Lu}, \bibinfo{person}{Daya Guo}, \bibinfo{person}{Shuo Ren}, \bibinfo{person}{Junjie Huang}, \bibinfo{person}{Alexey Svyatkovskiy}, \bibinfo{person}{Ambrosio Blanco}, \bibinfo{person}{Colin Clement}, \bibinfo{person}{Dawn Drain}, \bibinfo{person}{Daxin Jiang}, \bibinfo{person}{Duyu Tang}, {et~al\mbox{.}}} \bibinfo{year}{2021}\natexlab{}.
\newblock \showarticletitle{Codexglue: A machine learning benchmark dataset for code understanding and generation}.
\newblock \bibinfo{journal}{\emph{arXiv preprint arXiv:2102.04664}} (\bibinfo{year}{2021}).
\newblock


\bibitem[Luo et~al\mbox{.}(2023)]%
        {luo2023wizardcoder}
\bibfield{author}{\bibinfo{person}{Ziyang Luo}, \bibinfo{person}{Can Xu}, \bibinfo{person}{Pu Zhao}, \bibinfo{person}{Qingfeng Sun}, \bibinfo{person}{Xiubo Geng}, \bibinfo{person}{Wenxiang Hu}, \bibinfo{person}{Chongyang Tao}, \bibinfo{person}{Jing Ma}, \bibinfo{person}{Qingwei Lin}, {and} \bibinfo{person}{Daxin Jiang}.} \bibinfo{year}{2023}\natexlab{}.
\newblock \showarticletitle{WizardCoder: Empowering Code Large Language Models with Evol-Instruct}.
\newblock \bibinfo{journal}{\emph{arXiv preprint arXiv:2306.08568}} (\bibinfo{year}{2023}).
\newblock


\bibitem[Ma et~al\mbox{.}(2018)]%
        {ma2018deepmutation}
\bibfield{author}{\bibinfo{person}{Lei Ma}, \bibinfo{person}{Fuyuan Zhang}, \bibinfo{person}{Jiyuan Sun}, \bibinfo{person}{Minhui Xue}, \bibinfo{person}{Bo Li}, \bibinfo{person}{Felix Juefei-Xu}, \bibinfo{person}{Chao Xie}, \bibinfo{person}{Li Li}, \bibinfo{person}{Yang Liu}, \bibinfo{person}{Jianjun Zhao}, {et~al\mbox{.}}} \bibinfo{year}{2018}\natexlab{}.
\newblock \showarticletitle{Deepmutation: Mutation testing of deep learning systems}. In \bibinfo{booktitle}{\emph{2018 IEEE 29th international symposium on software reliability engineering (ISSRE)}}. IEEE, \bibinfo{pages}{100--111}.
\newblock


\bibitem[Ma et~al\mbox{.}(2024a)]%
        {ma2024lms}
\bibfield{author}{\bibinfo{person}{Wei Ma}, \bibinfo{person}{Shangqing Liu}, \bibinfo{person}{Zhihao Lin}, \bibinfo{person}{Wenhan Wang}, \bibinfo{person}{Qiang Hu}, \bibinfo{person}{Ye Liu}, \bibinfo{person}{Cen Zhang}, \bibinfo{person}{Liming Nie}, \bibinfo{person}{Li Li}, {and} \bibinfo{person}{Yang Liu}.} \bibinfo{year}{2024}\natexlab{a}.
\newblock \bibinfo{title}{LMs: Understanding Code Syntax and Semantics for Code Analysis}.
\newblock
\newblock
\showeprint[arxiv]{2305.12138}~[cs.SE]


\bibitem[Ma et~al\mbox{.}(2023)]%
        {ma2023scope}
\bibfield{author}{\bibinfo{person}{Wei Ma}, \bibinfo{person}{Shangqing Liu}, \bibinfo{person}{Wenhan Wang}, \bibinfo{person}{Qiang Hu}, \bibinfo{person}{Ye Liu}, \bibinfo{person}{Cen Zhang}, \bibinfo{person}{Liming Nie}, {and} \bibinfo{person}{Yang Liu}.} \bibinfo{year}{2023}\natexlab{}.
\newblock \showarticletitle{The Scope of ChatGPT in Software Engineering: A Thorough Investigation}.
\newblock \bibinfo{journal}{\emph{arXiv preprint arXiv:2305.12138}} (\bibinfo{year}{2023}).
\newblock


\bibitem[Ma et~al\mbox{.}(2021)]%
        {10.1145/3417330}
\bibfield{author}{\bibinfo{person}{Wei Ma}, \bibinfo{person}{Mike Papadakis}, \bibinfo{person}{Anestis Tsakmalis}, \bibinfo{person}{Maxime Cordy}, {and} \bibinfo{person}{Yves~Le Traon}.} \bibinfo{year}{2021}\natexlab{}.
\newblock \showarticletitle{Test Selection for Deep Learning Systems}.
\newblock \bibinfo{journal}{\emph{ACM Trans. Softw. Eng. Methodol.}} \bibinfo{volume}{30}, \bibinfo{number}{2}, Article \bibinfo{articleno}{13} (\bibinfo{date}{jan} \bibinfo{year}{2021}), \bibinfo{numpages}{22}~pages.
\newblock
\showISSN{1049-331X}
\urldef\tempurl%
\url{https://doi.org/10.1145/3417330}
\showDOI{\tempurl}


\bibitem[Ma et~al\mbox{.}(2024b)]%
        {ma2024combining}
\bibfield{author}{\bibinfo{person}{Wei Ma}, \bibinfo{person}{Daoyuan Wu}, \bibinfo{person}{Yuqiang Sun}, \bibinfo{person}{Tianwen Wang}, \bibinfo{person}{Shangqing Liu}, \bibinfo{person}{Jian Zhang}, \bibinfo{person}{Yue Xue}, {and} \bibinfo{person}{Yang Liu}.} \bibinfo{year}{2024}\natexlab{b}.
\newblock \bibinfo{title}{Combining Fine-Tuning and LLM-based Agents for Intuitive Smart Contract Auditing with Justifications}.
\newblock
\newblock
\showeprint[arxiv]{2403.16073}~[cs.SE]


\bibitem[Ma et~al\mbox{.}(2022)]%
        {10.1145/3524842.3528456}
\bibfield{author}{\bibinfo{person}{Wei Ma}, \bibinfo{person}{Mengjie Zhao}, \bibinfo{person}{Ezekiel Soremekun}, \bibinfo{person}{Qiang Hu}, \bibinfo{person}{Jie~M. Zhang}, \bibinfo{person}{Mike Papadakis}, \bibinfo{person}{Maxime Cordy}, \bibinfo{person}{Xiaofei Xie}, {and} \bibinfo{person}{Yves~Le Traon}.} \bibinfo{year}{2022}\natexlab{}.
\newblock \showarticletitle{GraphCode2Vec: generic code embedding via lexical and program dependence analyses}. In \bibinfo{booktitle}{\emph{Proceedings of the 19th International Conference on Mining Software Repositories}} (Pittsburgh, Pennsylvania) \emph{(\bibinfo{series}{MSR '22})}. \bibinfo{publisher}{Association for Computing Machinery}, \bibinfo{address}{New York, NY, USA}, \bibinfo{pages}{524–536}.
\newblock
\showISBNx{9781450393034}
\urldef\tempurl%
\url{https://doi.org/10.1145/3524842.3528456}
\showDOI{\tempurl}


\bibitem[Matthews(1975)]%
        {MATTHEWS1975442}
\bibfield{author}{\bibinfo{person}{B.W. Matthews}.} \bibinfo{year}{1975}\natexlab{}.
\newblock \showarticletitle{Comparison of the predicted and observed secondary structure of T4 phage lysozyme}.
\newblock \bibinfo{journal}{\emph{Biochimica et Biophysica Acta (BBA) - Protein Structure}} \bibinfo{volume}{405}, \bibinfo{number}{2} (\bibinfo{year}{1975}), \bibinfo{pages}{442--451}.
\newblock
\showISSN{0005-2795}
\urldef\tempurl%
\url{https://doi.org/10.1016/0005-2795(75)90109-9}
\showDOI{\tempurl}


\bibitem[Meng et~al\mbox{.}(2021)]%
        {meng2021measuring}
\bibfield{author}{\bibinfo{person}{Linghan Meng}, \bibinfo{person}{Yanhui Li}, \bibinfo{person}{Lin Chen}, \bibinfo{person}{Zhi Wang}, \bibinfo{person}{Di Wu}, \bibinfo{person}{Yuming Zhou}, {and} \bibinfo{person}{Baowen Xu}.} \bibinfo{year}{2021}\natexlab{}.
\newblock \showarticletitle{Measuring discrimination to boost comparative testing for multiple deep learning models}. In \bibinfo{booktitle}{\emph{2021 IEEE/ACM 43rd International Conference on Software Engineering (ICSE)}}. IEEE, \bibinfo{pages}{385--396}.
\newblock


\bibitem[Mirkes et~al\mbox{.}(2020)]%
        {mirkes2020fractional}
\bibfield{author}{\bibinfo{person}{Evgeny~M Mirkes}, \bibinfo{person}{Jeza Allohibi}, {and} \bibinfo{person}{Alexander Gorban}.} \bibinfo{year}{2020}\natexlab{}.
\newblock \showarticletitle{Fractional norms and quasinorms do not help to overcome the curse of dimensionality}.
\newblock \bibinfo{journal}{\emph{Entropy}} \bibinfo{volume}{22}, \bibinfo{number}{10} (\bibinfo{year}{2020}), \bibinfo{pages}{1105}.
\newblock


\bibitem[Mou et~al\mbox{.}(2016)]%
        {mou2016convolutional}
\bibfield{author}{\bibinfo{person}{Lili Mou}, \bibinfo{person}{Ge Li}, \bibinfo{person}{Lu Zhang}, \bibinfo{person}{Tao Wang}, {and} \bibinfo{person}{Zhi Jin}.} \bibinfo{year}{2016}\natexlab{}.
\newblock \showarticletitle{Convolutional neural networks over tree structures for programming language processing}. In \bibinfo{booktitle}{\emph{Proceedings of the Thirtieth AAAI Conference on Artificial Intelligence}}. \bibinfo{pages}{1287--1293}.
\newblock


\bibitem[Niu et~al\mbox{.}(2023)]%
        {niu2023empirical}
\bibfield{author}{\bibinfo{person}{Changan Niu}, \bibinfo{person}{Chuanyi Li}, \bibinfo{person}{Vincent Ng}, \bibinfo{person}{Dongxiao Chen}, \bibinfo{person}{Jidong Ge}, {and} \bibinfo{person}{Bin Luo}.} \bibinfo{year}{2023}\natexlab{}.
\newblock \showarticletitle{An empirical comparison of pre-trained models of source code}.
\newblock \bibinfo{journal}{\emph{arXiv preprint arXiv:2302.04026}} (\bibinfo{year}{2023}).
\newblock


\bibitem[Penha and Hauff(2020)]%
        {10.1145/3383313.3412249}
\bibfield{author}{\bibinfo{person}{Gustavo Penha} {and} \bibinfo{person}{Claudia Hauff}.} \bibinfo{year}{2020}\natexlab{}.
\newblock \showarticletitle{What does BERT know about books, movies and music? Probing BERT for Conversational Recommendation}. In \bibinfo{booktitle}{\emph{Proceedings of the 14th ACM Conference on Recommender Systems}} (Virtual Event, Brazil) \emph{(\bibinfo{series}{RecSys '20})}. \bibinfo{publisher}{Association for Computing Machinery}, \bibinfo{address}{New York, NY, USA}, \bibinfo{pages}{388–397}.
\newblock
\showISBNx{9781450375832}
\urldef\tempurl%
\url{https://doi.org/10.1145/3383313.3412249}
\showDOI{\tempurl}


\bibitem[Puri et~al\mbox{.}(2021)]%
        {puri2021codenet}
\bibfield{author}{\bibinfo{person}{Ruchir Puri}, \bibinfo{person}{David~S Kung}, \bibinfo{person}{Geert Janssen}, \bibinfo{person}{Wei Zhang}, \bibinfo{person}{Giacomo Domeniconi}, \bibinfo{person}{Vladimir Zolotov}, \bibinfo{person}{Julian Dolby}, \bibinfo{person}{Jie Chen}, \bibinfo{person}{Mihir Choudhury}, \bibinfo{person}{Lindsey Decker}, {et~al\mbox{.}}} \bibinfo{year}{2021}\natexlab{}.
\newblock \showarticletitle{CodeNet: A large-scale AI for code dataset for learning a diversity of coding tasks}.
\newblock \bibinfo{journal}{\emph{arXiv preprint arXiv:2105.12655}} (\bibinfo{year}{2021}).
\newblock


\bibitem[Rogers et~al\mbox{.}(2020)]%
        {bertology}
\bibfield{author}{\bibinfo{person}{Anna Rogers}, \bibinfo{person}{Olga Kovaleva}, {and} \bibinfo{person}{Anna Rumshisky}.} \bibinfo{year}{2020}\natexlab{}.
\newblock \showarticletitle{A Primer in {BERT}ology: What We Know About How {BERT} Works}.
\newblock \bibinfo{journal}{\emph{Transactions of the Association for Computational Linguistics}}  \bibinfo{volume}{8} (\bibinfo{year}{2020}), \bibinfo{pages}{842--866}.
\newblock
\urldef\tempurl%
\url{https://doi.org/10.1162/tacl_a_00349}
\showDOI{\tempurl}


\bibitem[Roziere et~al\mbox{.}(2023)]%
        {roziere2023code}
\bibfield{author}{\bibinfo{person}{Baptiste Roziere}, \bibinfo{person}{Jonas Gehring}, \bibinfo{person}{Fabian Gloeckle}, \bibinfo{person}{Sten Sootla}, \bibinfo{person}{Itai Gat}, \bibinfo{person}{Xiaoqing~Ellen Tan}, \bibinfo{person}{Yossi Adi}, \bibinfo{person}{Jingyu Liu}, \bibinfo{person}{Tal Remez}, \bibinfo{person}{J{\'e}r{\'e}my Rapin}, {et~al\mbox{.}}} \bibinfo{year}{2023}\natexlab{}.
\newblock \showarticletitle{Code llama: Open foundation models for code}.
\newblock \bibinfo{journal}{\emph{arXiv preprint arXiv:2308.12950}} (\bibinfo{year}{2023}).
\newblock


\bibitem[Shen et~al\mbox{.}(2022)]%
        {shen2022benchmarking}
\bibfield{author}{\bibinfo{person}{Da Shen}, \bibinfo{person}{Xinyun Chen}, \bibinfo{person}{Chenguang Wang}, \bibinfo{person}{Koushik Sen}, {and} \bibinfo{person}{Dawn Song}.} \bibinfo{year}{2022}\natexlab{}.
\newblock \showarticletitle{Benchmarking Language Models for Code Syntax Understanding}. In \bibinfo{booktitle}{\emph{Findings of the Association for Computational Linguistics: EMNLP 2022}}, \bibfield{editor}{\bibinfo{person}{Yoav Goldberg}, \bibinfo{person}{Zornitsa Kozareva}, {and} \bibinfo{person}{Yue Zhang}} (Eds.). \bibinfo{publisher}{Association for Computational Linguistics}, \bibinfo{address}{Abu Dhabi, United Arab Emirates}, \bibinfo{pages}{3071--3093}.
\newblock
\urldef\tempurl%
\url{https://doi.org/10.18653/v1/2022.findings-emnlp.224}
\showDOI{\tempurl}


\bibitem[Shen et~al\mbox{.}(2018)]%
        {shen-etal-2018-straight}
\bibfield{author}{\bibinfo{person}{Yikang Shen}, \bibinfo{person}{Zhouhan Lin}, \bibinfo{person}{Athul~Paul Jacob}, \bibinfo{person}{Alessandro Sordoni}, \bibinfo{person}{Aaron Courville}, {and} \bibinfo{person}{Yoshua Bengio}.} \bibinfo{year}{2018}\natexlab{}.
\newblock \showarticletitle{Straight to the Tree: Constituency Parsing with Neural Syntactic Distance}. In \bibinfo{booktitle}{\emph{Proceedings of the 56th Annual Meeting of the Association for Computational Linguistics (Volume 1: Long Papers)}}, \bibfield{editor}{\bibinfo{person}{Iryna Gurevych} {and} \bibinfo{person}{Yusuke Miyao}} (Eds.). \bibinfo{publisher}{Association for Computational Linguistics}, \bibinfo{address}{Melbourne, Australia}, \bibinfo{pages}{1171--1180}.
\newblock
\urldef\tempurl%
\url{https://doi.org/10.18653/v1/P18-1108}
\showDOI{\tempurl}


\bibitem[Sun et~al\mbox{.}(2021)]%
        {9710827}
\bibfield{author}{\bibinfo{person}{Xiaoxiao Sun}, \bibinfo{person}{Yunzhong Hou}, \bibinfo{person}{Weijian Deng}, \bibinfo{person}{Hongdong Li}, {and} \bibinfo{person}{Liang Zheng}.} \bibinfo{year}{2021}\natexlab{}.
\newblock \showarticletitle{Ranking Models in Unlabeled New Environments}. In \bibinfo{booktitle}{\emph{2021 IEEE/CVF International Conference on Computer Vision (ICCV)}}. \bibinfo{pages}{11741--11751}.
\newblock
\urldef\tempurl%
\url{https://doi.org/10.1109/ICCV48922.2021.01155}
\showDOI{\tempurl}


\bibitem[Sun et~al\mbox{.}(2024)]%
        {sun2024llm4vuln}
\bibfield{author}{\bibinfo{person}{Yuqiang Sun}, \bibinfo{person}{Daoyuan Wu}, \bibinfo{person}{Yue Xue}, \bibinfo{person}{Han Liu}, \bibinfo{person}{Wei Ma}, \bibinfo{person}{Lyuye Zhang}, \bibinfo{person}{Miaolei Shi}, {and} \bibinfo{person}{Yang Liu}.} \bibinfo{year}{2024}\natexlab{}.
\newblock \bibinfo{title}{LLM4Vuln: A Unified Evaluation Framework for Decoupling and Enhancing LLMs' Vulnerability Reasoning}.
\newblock
\newblock
\showeprint[arxiv]{2401.16185}~[cs.CR]


\bibitem[Svyatkovskiy et~al\mbox{.}(2020)]%
        {svyatkovskiy2020intellicode}
\bibfield{author}{\bibinfo{person}{Alexey Svyatkovskiy}, \bibinfo{person}{Shao~Kun Deng}, \bibinfo{person}{Shengyu Fu}, {and} \bibinfo{person}{Neel Sundaresan}.} \bibinfo{year}{2020}\natexlab{}.
\newblock \showarticletitle{Intellicode compose: Code generation using transformer}. In \bibinfo{booktitle}{\emph{Proceedings of the 28th ACM Joint Meeting on European Software Engineering Conference and Symposium on the Foundations of Software Engineering}}. \bibinfo{pages}{1433--1443}.
\newblock


\bibitem[Tenney et~al\mbox{.}(2019)]%
        {tenney2019you}
\bibfield{author}{\bibinfo{person}{Ian Tenney}, \bibinfo{person}{Patrick Xia}, \bibinfo{person}{Berlin Chen}, \bibinfo{person}{Alex Wang}, \bibinfo{person}{Adam Poliak}, \bibinfo{person}{R~Thomas McCoy}, \bibinfo{person}{Najoung Kim}, \bibinfo{person}{Benjamin Van~Durme}, \bibinfo{person}{Samuel~R Bowman}, \bibinfo{person}{Dipanjan Das}, {et~al\mbox{.}}} \bibinfo{year}{2019}\natexlab{}.
\newblock \showarticletitle{What do you learn from context? probing for sentence structure in contextualized word representations}.
\newblock \bibinfo{journal}{\emph{arXiv preprint arXiv:1905.06316}} (\bibinfo{year}{2019}).
\newblock


\bibitem[Touvron et~al\mbox{.}(2023)]%
        {touvron2023llama}
\bibfield{author}{\bibinfo{person}{Hugo Touvron}, \bibinfo{person}{Louis Martin}, \bibinfo{person}{Kevin Stone}, \bibinfo{person}{Peter Albert}, \bibinfo{person}{Amjad Almahairi}, \bibinfo{person}{Yasmine Babaei}, \bibinfo{person}{Nikolay Bashlykov}, \bibinfo{person}{Soumya Batra}, \bibinfo{person}{Prajjwal Bhargava}, \bibinfo{person}{Shruti Bhosale}, {et~al\mbox{.}}} \bibinfo{year}{2023}\natexlab{}.
\newblock \showarticletitle{Llama 2: Open foundation and fine-tuned chat models}.
\newblock \bibinfo{journal}{\emph{arXiv preprint arXiv:2307.09288}} (\bibinfo{year}{2023}).
\newblock


\bibitem[Troshin and Chirkova(2022)]%
        {troshin2022probing}
\bibfield{author}{\bibinfo{person}{Sergey Troshin} {and} \bibinfo{person}{Nadezhda Chirkova}.} \bibinfo{year}{2022}\natexlab{}.
\newblock \showarticletitle{Probing Pretrained Models of Source Code}.
\newblock \bibinfo{journal}{\emph{arXiv preprint arXiv:2202.08975}} (\bibinfo{year}{2022}).
\newblock


\bibitem[van Aken et~al\mbox{.}(2019)]%
        {10.1145/3357384.3358028}
\bibfield{author}{\bibinfo{person}{Betty van Aken}, \bibinfo{person}{Benjamin Winter}, \bibinfo{person}{Alexander L\"{o}ser}, {and} \bibinfo{person}{Felix~A. Gers}.} \bibinfo{year}{2019}\natexlab{}.
\newblock \showarticletitle{How Does BERT Answer Questions? A Layer-Wise Analysis of Transformer Representations}. In \bibinfo{booktitle}{\emph{Proceedings of the 28th ACM International Conference on Information and Knowledge Management}} (Beijing, China) \emph{(\bibinfo{series}{CIKM '19})}. \bibinfo{publisher}{Association for Computing Machinery}, \bibinfo{address}{New York, NY, USA}, \bibinfo{pages}{1823–1832}.
\newblock
\showISBNx{9781450369763}
\urldef\tempurl%
\url{https://doi.org/10.1145/3357384.3358028}
\showDOI{\tempurl}


\bibitem[Vaswani et~al\mbox{.}(2017)]%
        {vaswani2017attention}
\bibfield{author}{\bibinfo{person}{Ashish Vaswani}, \bibinfo{person}{Noam Shazeer}, \bibinfo{person}{Niki Parmar}, \bibinfo{person}{Jakob Uszkoreit}, \bibinfo{person}{Llion Jones}, \bibinfo{person}{Aidan~N Gomez}, \bibinfo{person}{{\L}ukasz Kaiser}, {and} \bibinfo{person}{Illia Polosukhin}.} \bibinfo{year}{2017}\natexlab{}.
\newblock \showarticletitle{Attention is all you need}.
\newblock \bibinfo{journal}{\emph{Advances in neural information processing systems}}  \bibinfo{volume}{30} (\bibinfo{year}{2017}).
\newblock


\bibitem[Wan et~al\mbox{.}(2022)]%
        {10.1145/3510003.3510050}
\bibfield{author}{\bibinfo{person}{Yao Wan}, \bibinfo{person}{Wei Zhao}, \bibinfo{person}{Hongyu Zhang}, \bibinfo{person}{Yulei Sui}, \bibinfo{person}{Guandong Xu}, {and} \bibinfo{person}{Hai Jin}.} \bibinfo{year}{2022}\natexlab{}.
\newblock \showarticletitle{What Do They Capture? A Structural Analysis of Pre-Trained Language Models for Source Code}. In \bibinfo{booktitle}{\emph{Proceedings of the 44th International Conference on Software Engineering}} (Pittsburgh, Pennsylvania) \emph{(\bibinfo{series}{ICSE '22})}. \bibinfo{publisher}{Association for Computing Machinery}, \bibinfo{address}{New York, NY, USA}, \bibinfo{pages}{2377–2388}.
\newblock
\showISBNx{9781450392211}
\urldef\tempurl%
\url{https://doi.org/10.1145/3510003.3510050}
\showDOI{\tempurl}


\bibitem[Wang et~al\mbox{.}(2023)]%
        {wang2023codet5+}
\bibfield{author}{\bibinfo{person}{Yue Wang}, \bibinfo{person}{Hung Le}, \bibinfo{person}{Akhilesh~Deepak Gotmare}, \bibinfo{person}{Nghi~DQ Bui}, \bibinfo{person}{Junnan Li}, {and} \bibinfo{person}{Steven~CH Hoi}.} \bibinfo{year}{2023}\natexlab{}.
\newblock \showarticletitle{Codet5+: Open code large language models for code understanding and generation}.
\newblock \bibinfo{journal}{\emph{arXiv preprint arXiv:2305.07922}} (\bibinfo{year}{2023}).
\newblock


\bibitem[Wang et~al\mbox{.}(2021)]%
        {wang2021codet5}
\bibfield{author}{\bibinfo{person}{Yue Wang}, \bibinfo{person}{Weishi Wang}, \bibinfo{person}{Shafiq Joty}, {and} \bibinfo{person}{Steven~CH Hoi}.} \bibinfo{year}{2021}\natexlab{}.
\newblock \showarticletitle{Codet5: Identifier-aware unified pre-trained encoder-decoder models for code understanding and generation}.
\newblock \bibinfo{journal}{\emph{arXiv preprint arXiv:2109.00859}} (\bibinfo{year}{2021}).
\newblock


\bibitem[Wei et~al\mbox{.}(2022)]%
        {wei2022emergent}
\bibfield{author}{\bibinfo{person}{Jason Wei}, \bibinfo{person}{Yi Tay}, \bibinfo{person}{Rishi Bommasani}, \bibinfo{person}{Colin Raffel}, \bibinfo{person}{Barret Zoph}, \bibinfo{person}{Sebastian Borgeaud}, \bibinfo{person}{Dani Yogatama}, \bibinfo{person}{Maarten Bosma}, \bibinfo{person}{Denny Zhou}, \bibinfo{person}{Donald Metzler}, {et~al\mbox{.}}} \bibinfo{year}{2022}\natexlab{}.
\newblock \showarticletitle{Emergent abilities of large language models}.
\newblock \bibinfo{journal}{\emph{arXiv preprint arXiv:2206.07682}} (\bibinfo{year}{2022}).
\newblock


\bibitem[Yang et~al\mbox{.}(2022)]%
        {10.1145/3505243}
\bibfield{author}{\bibinfo{person}{Yanming Yang}, \bibinfo{person}{Xin Xia}, \bibinfo{person}{David Lo}, {and} \bibinfo{person}{John Grundy}.} \bibinfo{year}{2022}\natexlab{}.
\newblock \showarticletitle{A Survey on Deep Learning for Software Engineering}.
\newblock \bibinfo{journal}{\emph{ACM Comput. Surv.}} \bibinfo{volume}{54}, \bibinfo{number}{10s}, Article \bibinfo{articleno}{206} (\bibinfo{date}{sep} \bibinfo{year}{2022}), \bibinfo{numpages}{73}~pages.
\newblock
\showISSN{0360-0300}
\urldef\tempurl%
\url{https://doi.org/10.1145/3505243}
\showDOI{\tempurl}


\bibitem[Yao and Shepperd(2020)]%
        {10.1145/3383219.3383232}
\bibfield{author}{\bibinfo{person}{Jingxiu Yao} {and} \bibinfo{person}{Martin Shepperd}.} \bibinfo{year}{2020}\natexlab{}.
\newblock \showarticletitle{Assessing Software Defection Prediction Performance: Why Using the Matthews Correlation Coefficient Matters}. In \bibinfo{booktitle}{\emph{Proceedings of the 24th International Conference on Evaluation and Assessment in Software Engineering}} (Trondheim, Norway) \emph{(\bibinfo{series}{EASE '20})}. \bibinfo{publisher}{Association for Computing Machinery}, \bibinfo{address}{New York, NY, USA}, \bibinfo{pages}{120–129}.
\newblock
\showISBNx{9781450377317}
\urldef\tempurl%
\url{https://doi.org/10.1145/3383219.3383232}
\showDOI{\tempurl}


\bibitem[Yefet et~al\mbox{.}(2020)]%
        {10.1145/3428230}
\bibfield{author}{\bibinfo{person}{Noam Yefet}, \bibinfo{person}{Uri Alon}, {and} \bibinfo{person}{Eran Yahav}.} \bibinfo{year}{2020}\natexlab{}.
\newblock \showarticletitle{Adversarial examples for models of code}.
\newblock \bibinfo{journal}{\emph{Proc. ACM Program. Lang.}} \bibinfo{volume}{4}, \bibinfo{number}{OOPSLA}, Article \bibinfo{articleno}{162} (\bibinfo{date}{nov} \bibinfo{year}{2020}), \bibinfo{numpages}{30}~pages.
\newblock
\urldef\tempurl%
\url{https://doi.org/10.1145/3428230}
\showDOI{\tempurl}


\bibitem[Yuan et~al\mbox{.}(2023)]%
        {yuan2023no}
\bibfield{author}{\bibinfo{person}{Zhiqiang Yuan}, \bibinfo{person}{Yiling Lou}, \bibinfo{person}{Mingwei Liu}, \bibinfo{person}{Shiji Ding}, \bibinfo{person}{Kaixin Wang}, \bibinfo{person}{Yixuan Chen}, {and} \bibinfo{person}{Xin Peng}.} \bibinfo{year}{2023}\natexlab{}.
\newblock \showarticletitle{No More Manual Tests? Evaluating and Improving ChatGPT for Unit Test Generation}.
\newblock \bibinfo{journal}{\emph{arXiv preprint arXiv:2305.04207}} (\bibinfo{year}{2023}).
\newblock


\bibitem[Yun et~al\mbox{.}(2019)]%
        {yun2019graph}
\bibfield{author}{\bibinfo{person}{Seongjun Yun}, \bibinfo{person}{Minbyul Jeong}, \bibinfo{person}{Raehyun Kim}, \bibinfo{person}{Jaewoo Kang}, {and} \bibinfo{person}{Hyunwoo~J Kim}.} \bibinfo{year}{2019}\natexlab{}.
\newblock \showarticletitle{Graph transformer networks}.
\newblock \bibinfo{journal}{\emph{Advances in neural information processing systems}}  \bibinfo{volume}{32} (\bibinfo{year}{2019}).
\newblock


\bibitem[Zhang et~al\mbox{.}(2023)]%
        {zhang2023survey}
\bibfield{author}{\bibinfo{person}{Ziyin Zhang}, \bibinfo{person}{Chaoyu Chen}, \bibinfo{person}{Bingchang Liu}, \bibinfo{person}{Cong Liao}, \bibinfo{person}{Zi Gong}, \bibinfo{person}{Hang Yu}, \bibinfo{person}{Jianguo Li}, {and} \bibinfo{person}{Rui Wang}.} \bibinfo{year}{2023}\natexlab{}.
\newblock \showarticletitle{A survey on language models for code}.
\newblock \bibinfo{journal}{\emph{arXiv preprint arXiv:2311.07989}} (\bibinfo{year}{2023}).
\newblock


\bibitem[Zhao et~al\mbox{.}(2020)]%
        {pse}
\bibfield{author}{\bibinfo{person}{Mengjie Zhao}, \bibinfo{person}{Philipp Dufter}, \bibinfo{person}{Yadollah Yaghoobzadeh}, {and} \bibinfo{person}{Hinrich Sch{\"u}tze}.} \bibinfo{year}{2020}\natexlab{}.
\newblock \showarticletitle{Quantifying the Contextualization of Word Representations with Semantic Class Probing}. In \bibinfo{booktitle}{\emph{Findings of the Association for Computational Linguistics: EMNLP 2020}}. \bibinfo{publisher}{Association for Computational Linguistics}, \bibinfo{address}{Online}, \bibinfo{pages}{1219--1234}.
\newblock
\urldef\tempurl%
\url{https://doi.org/10.18653/v1/2020.findings-emnlp.109}
\showDOI{\tempurl}


\bibitem[Zhou et~al\mbox{.}(2019)]%
        {zhou2019devign}
\bibfield{author}{\bibinfo{person}{Yaqin Zhou}, \bibinfo{person}{Shangqing Liu}, \bibinfo{person}{Jingkai Siow}, \bibinfo{person}{Xiaoning Du}, {and} \bibinfo{person}{Yang Liu}.} \bibinfo{year}{2019}\natexlab{}.
\newblock \showarticletitle{Devign: Effective vulnerability identification by learning comprehensive program semantics via graph neural networks}. In \bibinfo{booktitle}{\emph{Advances in Neural Information Processing Systems}}. \bibinfo{pages}{10197--10207}.
\newblock


\bibitem[Zoph et~al\mbox{.}(2022)]%
        {52065}
\bibfield{author}{\bibinfo{person}{Barret Zoph}, \bibinfo{person}{Colin Raffel}, \bibinfo{person}{Dale Schuurmans}, \bibinfo{person}{Dani Yogatama}, \bibinfo{person}{Denny Zhou}, \bibinfo{person}{Don Metzler}, \bibinfo{person}{Ed~H. Chi}, \bibinfo{person}{Jason Wei}, \bibinfo{person}{Jeff Dean}, \bibinfo{person}{Liam~B. Fedus}, \bibinfo{person}{Maarten~Paul Bosma}, \bibinfo{person}{Oriol Vinyals}, \bibinfo{person}{Percy Liang}, \bibinfo{person}{Sebastian Borgeaud}, \bibinfo{person}{Tatsunori~B. Hashimoto}, {and} \bibinfo{person}{Yi Tay}.} \bibinfo{year}{2022}\natexlab{}.
\newblock \showarticletitle{Emergent abilities of large language models}.
\newblock \bibinfo{journal}{\emph{TMLR}} (\bibinfo{year}{2022}).
\newblock


\end{thebibliography}

\end{document}